%% file: main.tex
\iffalse 
\documentclass[11pt, letter, oneside, reqno]{article}
\usepackage[total={7in, 9.25in}]{geometry}
\else 
\documentclass[letterpaper, 10 pt, journal, twoside]{IEEEtran}
\fi 
\input{mystyle}
\input{notations}
\usepackage{lipsum}
\usepackage{graphicx}
\ifCLASSOPTIONcompsoc
    \usepackage[caption=false, font=normalsize, labelfont=sf, textfont=sf]{subfig}
\else
\usepackage[caption=false, font=footnotesize]{subfig}
\fi
\title{\LARGE  \vspace{-6mm} Representer Theorem for Learning Koopman Operators}
\author{Mohammad Khosravi}
\affil{Delft Center for Systems and  Control, TU  Delft  
\authorcr\texttt{mohammad.khosravi@tudelft.nl} \vspace{-7mm}}
\begin{document}
\maketitle
\input{sec_01_abstract.tex}
\input{sec_02_introduction.tex}
\input{sec_03_notations.tex}


\input{sec_04_Koopman.tex}

\input{sec_05_learning.tex}

\input{sec_06_edmd.tex}
\input{sec_07_generalization.tex}
\input{sec_08_stable.tex}

\input{sec_09_numerics.tex}

\input{sec_97_conc}
\appendix
\section{Appendix}
\input{sec_99_myappendix}


\bibliographystyle{IEEEtran}
\bibliography{references2}
\end{document}

%% file: mystyle.tex
\usepackage[noadjust]{cite}
\usepackage{amsmath, amssymb,mathrsfs, dsfont} 
\usepackage{amsmath,amssymb,bm,bbm}
\usepackage{ color, bm,bbm,authblk}
\usepackage{times}
\usepackage{ graphicx,graphics, epstopdf} 
\usepackage{algorithm, algorithmicx,multirow, titlecaps} %
\usepackage{threeparttable}
\usepackage{tikzsymbols}

\DeclareMathAlphabet{\mathbbold}{U}{bbold}{m}{n}

\usepackage[english]{babel}
\usepackage[noend]{algpseudocode} 

\usepackage{tabularx}
\usepackage{booktabs}
\usepackage{multirow}
\usepackage{courier}
\usepackage{float}
\usepackage{makecell}
\setcellgapes{5pt}

\newcommand{\cmm}[1]{{#1}}

\newcommand{\stiny}[1]{{\scalebox{.5}{#1}}}
\newcommand{\smtiny}[1]{{\scalebox{.63}{#1}}}

\newcommand{\argmin}{{\mathrm{argmin}}}

\newcommand*{\argminOp}{\operatornamewithlimits{argmin}\limits}

\newcommand*{\minOp}{\operatornamewithlimits{min}\limits}

\newcommand*{\sumOp}{\operatornamewithlimits{\sum}\limits}

\newcommand{\tr}{{\smtiny{$\mathsf{T}$ }}\!}

\iftrue
\newcommand{\zeromx}{{\mathbbold{0}}} 
\newcommand{\eye}{\mathbb{I}}
\else
\newcommand{\zeromx}{{\mathbf{0}}} 
\newcommand{\eye}{\mathbf{I}}
\fi
\newcommand{\rank}{\mathrm{rank}}
\newcommand{\vc}[1]{{ \mathrm{#1} }}
\newcommand{\mx}[1]{{ \mathrm{#1} }}
\newcommand{\dom}{\mathrm{dom}}	
\newcommand{\diag}{\mathrm{diag}}

\newcommand{\drm}{\mathrm{d}}
\newcommand{\linspan}{\mathrm{span}} 

\newcommand{\inner}[2]{{ \langle {#1,#2} \rangle}}
\newcommand{\nth}{{\text{\tiny{th}}}}
\newcommand{\trace}{\mathrm{tr}}

\newcommand{\Dscr}{{\mathscr{D}}}

\newcommand{\Lscr}{{\mathscr{L}}}

\newcommand{\Acal}{{\mathcal{A}}}

\newcommand{\Ccal}{{\mathcal{C}}}
\newcommand{\Dcal}{{\mathcal{D}}}
\newcommand{\Ecal}{{\mathcal{E}}}
\newcommand{\Fcal}{{\mathcal{F}}}
\newcommand{\Gcal}{{\mathcal{G}}}
\newcommand{\Hcal}{{\mathcal{H}}}
\newcommand{\Ical}{{\mathcal{I}}}
\newcommand{\Jcal}{{\mathcal{J}}}
\newcommand{\Kcal}{{\mathcal{K}}}
\newcommand{\Lcal}{{\mathcal{L}}}
\newcommand{\Mcal}{{\mathcal{M}}}

\newcommand{\Pcal}{{\mathcal{P}}}

\newcommand{\Rcal}{{\mathcal{R}}}

\newcommand{\Ucal}{{\mathcal{U}}}
\newcommand{\Vcal}{{\mathcal{V}}}
\newcommand{\Wcal}{{\mathcal{W}}}
\newcommand{\Xcal}{{\mathcal{X}}}

\newcommand{\Zcal}{{\mathcal{Z}}}

\newcommand{\Nbb}{{\mathbb{N}}}

\newcommand{\Rbb}{{\mathbb{R}}}

\newcommand{\Vbb}{{\mathbb{V}}}

\newcommand{\Zbb}{{\mathbb{Z}}}

\newcommand{\bbh}{\mathds{h}}
\newcommand{\bbk}{\mathds{k}}

\newcommand{\PP}{\mathbbm{P}}
\newcommand{\EE}{\mathbbm{E}}

\usepackage{amsthm}

\newtheorem{theorem}{Theorem}
\newtheorem*{theorem*}{Theorem}
\newtheorem{definition}{Definition}
\newtheorem*{definition*}{Definition}
\newtheorem{assumption}{Assumption}

\newtheorem{corollary}[theorem]{Corollary}
\newtheorem{lemma}[theorem]{Lemma}
\newtheorem{remark}{Remark}
\newtheorem{example}{Example}
\newtheorem*{example*}{Example}

\newtheorem*{claim*}{Claim}

\newtheorem*{problem*}{Problem}
\iffalse
	\usetikzlibrary{arrows}
	\usetikzlibrary{trees}

\else
	\iftrue
	\usepackage{pifont}
\fi

\newcommand\xqed[1]{%
	\leavevmode\unskip\penalty9999 \hbox{}\nobreak\hfill
	\quad\hbox{#1}}
\iftrue 
\usepackage[colorlinks=true]{hyperref}
\usepackage{xcolor}
\hypersetup{
	colorlinks,
	linkcolor={blue!85!black},
	citecolor={red!65!black},
	urlcolor={green!80!black}
}
\allowdisplaybreaks

\fi

%% file: notations.tex
\newcommand{\vcx}{{\vc{x}}}
\newcommand{\vcy}{{\vc{y}}}
\newcommand{\vcd}{{\vc{d}}}
\newcommand{\vce}{{\vc{e}}}

\newcommand{\vcv}{{\vc{v}}}
\newcommand{\vcq}{{\vc{q}}}
\newcommand{\vcp}{{\vc{p}}}

\newcommand{\mxT}{\mx{T}}

\newcommand{\mxW}{{\mx{W}}}
\newcommand{\mxS}{\mx{S}}
\newcommand{\mxB}{\mx{B}}
\newcommand{\mxV}{\mx{V}}
\newcommand{\mxG}{\mx{G}}
\newcommand{\mxA}{\mx{A}}
\newcommand{\mxY}{\mx{Y}}
\newcommand{\mxC}{\mx{C}}
\newcommand{\mxZ}{\mx{Z}}
\newcommand{\mxE}{\mx{E}}
\newcommand{\mxQ}{\mx{Q}}

\newcommand{\mxU}{\mx{U}}
\newcommand{\mxR}{\mx{R}}
\newcommand{\mxM}{\mx{M}}
\newcommand{\mxMstar}{\mx{M}^{\star\!}}
\newcommand{\mxP}{\mx{P}}

\newcommand{\mxK}{\mx{K}}
\newcommand{\hatK}{\hat{\mx{K}}}
\newcommand{\tildeK}{\widetilde{\mx{K}}}
\newcommand{\tildeWcal}{\widetilde{\Wcal}}

\newcommand{\nT}{n_{\mathrm{t}}}
\newcommand{\nS}{n_{\mathrm{s}}}
\newcommand{\nG}{n_{\mathrm{g}}}
\newcommand{\nGb}{n_{\overline{\mathrm{g}}}}
\newcommand{\barnG}{\overline{n}_{\mathrm{g}}}
\newcommand{\nW}{n_{\mathrm{w}}}
\newcommand{\nX}{n_{\mathrm{x}}}
\newcommand{\nZ}{n_{\mathrm{z}}}
\newcommand{\barnZ}{\overline{n}_{\mathrm{z}}}

\newcommand{\itrajs}{{\stiny{\text{$(i)$}}}}
\newcommand{\itrajsOp}[1]{{\stiny{\text{$(#1)$}}}}

\newcommand{\kernel}{\bbk}
\newcommand{\kernelh}{\bbh}

\newcommand{\fro}{{\mathrm{F}}}
\newcommand{\xeq}{{\vc{x}_{\text{eq}}}}
\newcommand{\barJcal}{\overline{\Jcal}}
\newcommand{\barRcal}{\overline{\Rcal}}

%% file: sec_01_abstract.tex
\begin{abstract} 
In this work, the problem of learning Koopman operator of a discrete-time autonomous system is considered. The learning problem is formulated as a constrained regularized empirical loss minimization in the infinite-dimensional space of linear operators. We show that under certain but general conditions, a representer theorem holds for the learning problem. This allows reformulating the problem in a finite-dimensional space without any approximation and loss of precision. Following this, we consider various cases of regularization and constraints in the learning problem, including the operator norm, the Frobenius norm,  rank, nuclear norm, and stability. Subsequently, we derive the corresponding finite-dimensional problem. Furthermore, we discuss the connection between the proposed formulation and the extended dynamic mode decomposition. Finally, we provide an illustrative numerical example.
\end{abstract}

%% file: sec_02_introduction.tex
\section{Introduction}
Learning-based and data-driven approaches for analysis, modeling and control of nonlinear dynamics have received considerable attention in recent years \cite{schoukens2019nonlinear,fisac2018general,mania2020active,kaiser2018sparse,boffi2020regret,boffi2020learning}. In these approaches, various tools are developed towards a systematic analysis of nonlinear dynamical systems based on the collected data, and for exploiting valuable information and features such as stability. The main root of these techniques is in the classical point of view of the dynamical systems in which the system is described based on the state-space models. Indeed, considering that the evolution in nonlinear systems is described by nonlinear maps characterizing the difference or differential equation of the system, regression and classification methods are employed for learning these governing rules using the available measurement or synthetic data. These methodologies include regression techniques in reproducing kernel Hilbert spaces (RKHS), polynomial optimization methods, Gaussian process regression, training neural networks, and many other ones \cite{umlauft2019feedback,khansari2017learning,ijspeert2013dynamical,khosravi2021ROA,ahmadi2020learning,khosravi2021grad}. Working directly with the time-series data generated by the system has the leverage of dealing with finite-dimensional objects, however, the nonlinearity of the dynamics and incomplete information about the state trajectories can be challenging issues. 

Recent efforts in learning nonlinear dynamics have focused on data-driven operator-theoretic methods which lift the dynamics to an infinite-dimensional space via linear embedding.  Indeed, along with each dynamical system, one can introduce a so-called Koopman operator  \cite{koopman1931hamiltonian} which translates the evolution of the system to linear dynamics on a Hilbert space of functions called observables \cite{mauroy2020koopman}. While this approach leads to working in an infinite-dimensional space, we only need to deal with linear maps, and thus, one can leverage the linearity of the resulting objects, thanks to the well-developed tools of functional analysis and operator theory. The main feature of this lifting is the potentials for full representation and global characterization of the underlying dynamical system  \cite{koopman1931hamiltonian, singh1993composition,mezic2005spectral}. This alternative formalism offers applicability in data-driven settings for the analysis of large classes of nonlinear and high-dimensional systems  \cite{mezic2004comparison, mauroy2020koopman}, e.g.,  the framework has been applied to systems with different features such as hyperbolic fixed points, limit cycles, and,  attractors  \cite{mezic2019spectrum}.  In addition to the analysis of the nonlinear dynamical systems, Koopman linear representation has been also used for control, whereby not only the problem simplifies but also, in certain cases, it can outperform feedback policies that are based on the underlying nonlinear dynamics  \cite{brunton_invariant, koopman_kronic}.  In the data-driven settings, Koopman operators have already been utilized in many applications, such as robotics  \cite{Ian_KoopmanMPC, Bruder_Koopman}, human locomotion  \cite{applications_human_locomotion}, neuroscience  \cite{applications_neuroscience}, fluid mechanics  \cite{Applications_fluid_flows}, and climate forecast  \cite{Koopman_longterm_prediction}. Meanwhile, inspired by the physics of these applications or the expected behavior of the underlying system, features, constraints and \textit{a priori} information about the original system has been included in the algorithms for learning Koopman operators \cite{Koopman_EDMD, mamakoukas2020learning, Koopman_dissipativity}. For example, the stability constraints are imposed on the data-driven approximations of Koopman operators in  \cite{mamakoukas2020learning} to address the long-term accuracy issue. In  \cite{Koopman_dissipativity}, the dissipativity constraint which is a physics-based property of the system is imposed in the approximation of the corresponding Koopman operator.

While this operator-theoretic approach has various potentials and benefits, unless one can obtain a suitable finite-dimensional approximation for the Koopman operator  \cite{brunton_invariant, koopman_invariant_Naoya, haseli2019efficient}, its underlying infinite-dimensional nature hinders its practical use. 
To this end, various methods are developed, e.g., the Dynamic Mode Decomposition (DMD)  \cite{DMD}, Hankel-DMD  \cite{Hankel_DMD}, extended DMD (EDMD) \cite{koopman_datadrivenapproximation_edmd, Koopman_EDMD}, generator EDMD (gEDMD) \cite{klus2020data}, to name a few. These linearization methods are either locally accurate or depend strongly on the choice of the observable functions to provide suitable accuracy. 
Moreover, in all of the above approaches, rather than learning directly the infinite-dimensional Koopman operator, they initially enforce a restriction to a finite-dimensional subspace, and then, the data is employed to learn the corresponding finite-dimensional matrix representation. This leads to a systematic loss of precision and inefficient exploitation of the data. On the other hand, an infinite-dimensional learning problem formulation is, in general, ill-posed and intractable. Similar issues arise in nonparametric learning problems which are addressed by the well-known representer theorems  \cite{scholkopf2001generalized,dinuzzo2012representer,unser2019representer}, where certain conditions are introduced under which the solution of the problem is characterized as a finite linear combination of known objects. This highlights the necessity of analogous mathematical results for learning infinite-dimensional Koopman operators.

Motivated by the above discussion, we propose an operator-theoretic approach for learning the Koopman operator, that is formulated directly in the infinite-dimensional space of linear operators. The learning problem, in its most general form, is introduced as a constrained regularized empirical loss minimization, where the constraints enforce attributes of interests and possibly available side-information about the unknown Koopman operator, and the regularization is considered to penalize undesired features and also to avoid overfitting. We address the main concerns of data efficiency and tractability by introducing a set of representer theorems that provide equivalent finite-dimensional optimization problems. First, we consider the case where the operator norm is employed for the regularization. We extend the results to the formulation in which linear constraints are additionally imposed to enforce features of interest. Subsequently, we discuss the connection between the proposed Koopman learning formulation and the well-known EDMD method. Then, we generalize the developed representer theorem and provide conditions under which the results hold. Following this, we consider different cases of regularization and constraints, e.g., the Frobenius norm, the nuclear norm, and rank. For each of these cases, we derive the aforementioned equivalent finite-dimensional version of the learning problem which can be solved in a computationally tractable way by standard techniques of convex optimizations and singular value decomposition. Finally, we provide illustrative numerical examples.

%% file: sec_03_notations.tex
\section{Notation  and Preliminaries}
The set of natural numbers, the set of non-negative integers, the set of real scalars, the set of non-negative real numbers, the $n$-dimensional Euclidean space, and the space of $n$ by $m$ real matrices are denoted by $\Nbb$, $\Zbb_+$,  $\Rbb$, $\Rbb_+$, $\Rbb^n$ and $\Rbb^{n\times m}$, respectively. 
For each $n\in\Nbb$, 
$\{1,\ldots,n\}$ is denoted by $[n]$.
For matrix $\mxA\in\Rbb^{n\times m}$, the entry at $i^\nth$ row and $j^\nth$ column is denoted by $[\mxA]_{(i,j)}$. 
Given  Hilbert spaces $\Hcal,\Wcal$, the space of bounded linear operators $\mxT:\Hcal\to\Wcal$ is denoted by $\Lcal(\Hcal,\Wcal)$, and when $\Wcal$ is the same as $\Hcal$, we simply use $\Lcal(\Hcal)$. 
Given vectors $u,v\in \Hcal$, one can define a rank-one bounded linear operator, denoted by $v\otimes u$, such that $(v\otimes u) w = v \inner{u}{w}$, for any $w\in\Hcal$.
For vectors $v_1,\ldots,v_n\in\Hcal$, the Gram matrix $\mxV\in\Rbb^{n\times n}$ is defined as $[\mxV]_{(i,j)} = \inner{v_i}{v_j}$, for $i,j=1,\ldots,n$.
Given $\Ccal\subset\Hcal$, the indicator function $\delta_{\Ccal}:\Hcal\to\Rbb\cup\{+\infty\}$ is defined as $\delta_{\Ccal}(x) = 0$, when $x\in\Ccal$, and $\delta_{\Ccal}(x) = +\infty$, if $x\notin\Ccal$.
For matrix or operator $\mxA$,
the Frobenius or the Hilbert-Schmidt norm,
the trace, the nuclear norm, and the adjoint are respectively denoted by $\|\mxA\|_{\mathrm{F}}$,  $\trace(\mxA)$, $\|\mxA\|_*$, and $\mxA^*$.
We have $(u\otimes v)^*= v\otimes u$.
Let $\Hcal$ contain $\Vbb$-valued functions defined on $\Xcal$, where $\Xcal$ is a given set and $\Vbb$ is a normed space. Then, for each $x\in\Xcal$,
the \emph{evaluation} operator at $x$, denoted by $\vce_{x}$, is a linear map $\vce_{\vcx}:\Hcal\to\Vbb$ such that $\vce_{x}(g) = g(x)$, for any $g\in\Hcal$.  
\cmm{Given kernel $\kernel:\Xcal\times\Xcal\to\Rbb$ and $\vcx\in\Xcal$, the \emph{section of kernel} at $\vcx$ is denoted by $\kernel_{\vcx}$ and defined as function $\kernel(\vcx,\cdot):\Xcal\to\Rbb$.}
The domain of function $f:\Xcal\to\Rbb\cup\{+\infty\}$, denoted by $\dom(f)$, is defined as $\dom(f):=\{x\in\Xcal\ \!|\ \! f(x)<\infty\}$.
\cmm{The composition operation is denoted by $\circ$, which is dropped when it is clear from the context.}

%% file: sec_04_Koopman.tex
\section{Dynamical Systems and Koopman Operators}
Let $f:\Xcal\to\Xcal$ be a map defined over $\Xcal\subseteq \Rbb^{\nX}$.  
Consider the following autonomous dynamical system 
\begin{equation}\label{eqn:dyn_f}
\vcx_{k+1} = f(\vcx_{k}), \qquad k \in \Zbb_+,  
\end{equation}
which defines a discrete-time flow on the state space $\Xcal$.
Let $\Hcal$ be a Hilbert space of real-valued functions defined over $\Xcal$.  
Suppose  $\Hcal$ is closed under composition with map $f$, i.e., we know that 
$g\circ f\in\Hcal$, for each $\vc{g}\in \Hcal$.
Given a trajectory of the system, with respect to each $g\in\Hcal$, we have a sequence of real numbers transferring information about the system \eqref{eqn:dyn_f} through the lens of $g$. 
Accordingly, each element of $\Hcal$ is called an \emph{observable}. 

\begin{definition}[Koopman operator]
With respect to dynamical system \eqref{eqn:dyn_f}, the \emph{Koopman} operator is defined as a linear map $\mxK:\Hcal\to\Hcal$ such that $\mxK g = g \circ f$, for each $ g \in \Hcal$.
\end{definition}
\begin{figure}[t]
	\centering
	\includegraphics[width =0.485\textwidth]{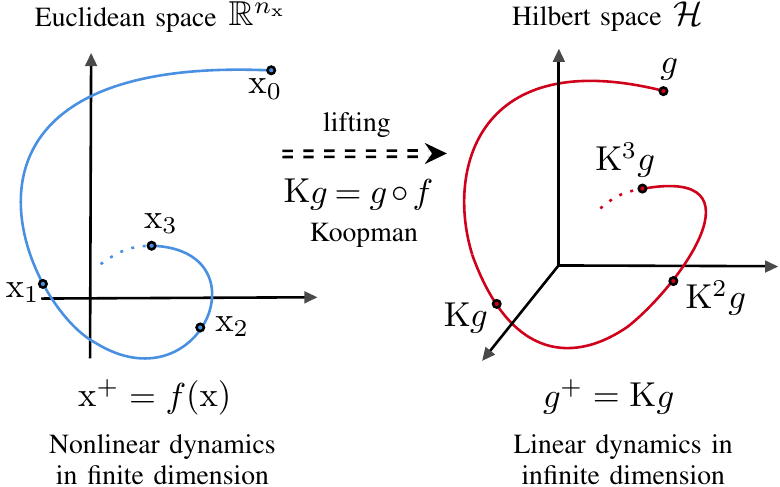}
	\caption{Lifting of the dynamics from finite-dimensional state space to the infinite-dimensional Hilbert space of observables.}
	\label{fig:lifting}
\end{figure}
The Koopman operator induces a lifting of the finite-dimensional nonlinear dynamics \eqref{eqn:dyn_f} to the infinite-dimensional linear dynamics $g^+=\mxK g$, which is in the Hilbert space of observables~$\Hcal$ (see Figure \ref{fig:lifting}).
Accordingly,  the dynamical system \eqref{eqn:dyn_f} can be 
completely described based on the corresponding Koopman operator $\mxK$ \cite{mauroy2020koopman}. Motivated by this fact, when a set of data about the dynamical system \eqref{eqn:dyn_f} is provided through several observables, we can pose the problem of learning the Koopman operator using the given data and possibly available side-information. 
To this end, we propose a general operator-theoretic learning framework formulated as a constrained regularized empirical loss minimization.
In the next section, Theorem \ref{thm:Tikhonov_reg_case} provides a representer theorem for the case of Tikhonov regularization, which also shows that the learning problem is well-posed and tractable.  
Theorem \ref{thm:Tikhonov_reg_case_W_part_1} extends this result to the case where we have additional linear constraints, and following this, the connection to the EDMD method is elaborated in Section \ref{sec:edmd}. 
The most general case is studied by Theorem \ref{thm:gen_reg_case}, and various cases of interest are discussed in Section \ref{sec:gen}.

%% file: sec_05_learning.tex
\section{Learning Koopman Operator} \label{sec:learning}
Let $\vcx_0,\vcx_1,\ldots,\vcx_{\nS}$ be a trajectory of the dynamical system \eqref{eqn:dyn_f} and 
$g_1,g_2,\ldots,g_{\nG}\in\Hcal$ be a set of observable maps. 
Accordingly, the set of data $\Dscr$ is provided as following
\begin{equation}\label{eqn:D}
\Dscr\!:=\!\Big\{
y_{kl}:= g_l(\vcx_k)
\ \!\Big|\ \! k=0,\ldots,\nS,l=1,\ldots,\nG\Big\}.	\!
\end{equation}
Note that one may define $y_{kl}$ as $y_{kl}= g_l(\vcx_k)+\varepsilon_{kl}$, for $k=0,\ldots,\nS$ and $l=1,\ldots,\nG$, where 
$\varepsilon_{kl}$ is introduced for considering the possible uncertainties in the value of observable $g_l$ at $\vcx_k$, which can be potentially due to the imperfect measurements or evaluations.
We can apply similar considerations to the trajectory data. 
In order to learn the Koopman operator of the dynamical system \eqref{eqn:dyn_f}, we need to consider a suitable learning objective function to be minimized over the hypothesis space of candidate operators  $\Lcal(\Hcal)$.
To this end, we define \emph{regularized empirical loss} function, $\Jcal:\Lcal(\Hcal)\to\Rbb_+\cup\{+\infty\}$, as 
\begin{equation}\label{eqn:cost_learning_koopman}
	\Jcal(\mxK) := 
	\Ecal(\mxK) 
	+
	\lambda
	\Rcal(\mxK),
\end{equation}  
where $\Rcal:\Lcal(\Hcal)\to \Rbb_+\cup\{+\infty\}$ is a regularization function, $\lambda>0$ is the weight of regularization, and $\Ecal:\Lcal(\Hcal)\to\Rbb_+$ is the \emph{empirical loss} function
characterized as
\begin{equation}\label{eqn:empirical_loss}
	\Ecal(\mxK) := 
	\sum_{k=1}^{\nS}
	\sum_{l=1}^{\nG}  
	\big(y_{kl}-(\mxK   g_l)(\vcx_{k-1})\big)^2.
\end{equation}
Furthermore, we may consider a constraint set
$\Ccal\subseteq\Lcal(\Hcal)$ in the learning problem to include some possibly available side information on the Koopman operator. 
This side information can be about different aspects of the local or global behavior of the dynamical system \eqref{eqn:dyn_f}, e.g., the stability of an equilibrium point.
Accordingly, the optimization  problem for learning the Koopman operator is defined as following
\begin{equation}\label{eqn:reg_learning_koopman}
\begin{array}{cl}
\minOp_{\mxK\in\Lcal(\Hcal)} &  \Jcal(\mxK)= \Ecal(\mxK) + \lambda \Rcal(\mxK)\\
\mathrm{s.t.} &
\mxK\in\Ccal.	
\end{array}	
\end{equation}
Note that \eqref{eqn:reg_learning_koopman} is an optimization problem over an infinite-dimensional set.
Hence, the main concern is whether this problem admits a solution, and if such solution exists, how one can obtain it in a computationally tractable way.
These issues depend on the choice of the regularization function and the constraint set. 
In the following, we study various  settings and provide conditions under which a representer theorem holds for the learning problem \eqref{eqn:cost_learning_koopman}.

\subsection{Learning Koopman Operator with Tikhonov Regularization}\label{ssec:Tikhonov}
In the statistical learning theory \cite{hastie2009elements}, Tikhonov regularization is the most common choice, 
i.e., $\Rcal$ is defined as the quadratic function $\Rcal(\mxK) := \|\mxK\|^2$.
If there is no additional constraint on the Koopman operator, we have the following learning problem 
\begin{equation}\label{eqn:reg_learning_koopman_Tikhonov}
		\!\!
		\minOp_{\mxK\in\Lcal(\Hcal)}\   
		\sumOp_{k=1}^{\nS}
		\sumOp_{l=1}^{\nG}  
		\big(y_{kl}-(\mxK   g_l)(\vcx_{k-1})\big)^2 + \lambda \|\mxK\|^2\!,\\
\end{equation}
which is a special case of \eqref{eqn:reg_learning_koopman}.
To study this problem we need the assumption given below.
\begin{assumption}\label{ass:e_xk_bounded}
	\normalfont
For $k=0,\ldots,\nS-1$, the evaluation operator $\vce_{\vcx_k}$ is continuous (bounded), i.e., $\vce_{\vcx_k}\in\Lcal(\Hcal,\Rbb)$.
\end{assumption}
This assumption says that, for each $k\in\{0,\ldots,\nS-1\}$, the value of $g(\vcx_k)$ depends continuously on $g$. 
More precisely, there exists a non-negative constant $C$ such that we have $|g(\vcx_k)|\le C\|g\|$, for each $g\in\Hcal$.
Roughly speaking, by a small perturbation of $g$, we expect that the value of $g(\vcx_k)$ does not change significantly, or in other words, when $g,h\in\Hcal$ are almost similar observables, i.e., $\|g-h\|$ is small, we expect that the values $g(\vcx_k)$ and $h(\vcx_k)$ are close to each other.
Accordingly, one can see that the Assumption \ref{ass:e_xk_bounded} is quite natural, otherwise, the observable functions $g\in\Hcal$ do not provide reliable and useful  information on dynamics \eqref{eqn:dyn_f}. 

\begin{remark}
\normalfont
A special case where Assumption \ref{ass:e_xk_bounded} holds is when $\Hcal$ is a reproducing kernel Hilbert space (RKHS) \cite{aronszajn1950theory,berlinet2011reproducing}. 
More precisely, if $\Hcal$ is endowed with \emph{reproducing kernel} $\kernel:\Xcal\times \Xcal\to \Rbb$, 
we know that  
\begin{equation}
|g(\vcx_k)| = |\inner{\kernel(\vcx_k,\cdot)}{g}|\le \|\kernel(\vcx_k,\cdot)\| \|g\|,\quad \forall g\in\Hcal,
\end{equation}
which is due to the \emph{reproducing property} of the kernel and the Cauchy-Schwartz inequality.
Accordingly, if $C$ is defined as 
\begin{equation}
C:=\max\Big\{\|\kernel(\vcx_k,\cdot)\|\ \! \Big|\ \! k=0,\ldots,\nS-1 \Big\},	
\end{equation}
then, for any $g\in\Hcal$, we have $|g(\vcx_k)|\le C\|g\|$. 
\end{remark}
The next theorem characterizes the solution of \eqref{eqn:reg_learning_koopman_Tikhonov}. 	
\begin{theorem}
\label{thm:Tikhonov_reg_case}
Let Assumption \ref{ass:e_xk_bounded} hold and $\lambda>0$.
Then, the optimization problem \eqref{eqn:reg_learning_koopman_Tikhonov} admits a unique solution denoted by $\hatK$.
Moreover, there exists a set of vectors $\{v_1,\ldots,v_{\nS}\}\subset\Hcal$, such that
\begin{equation}\label{eqn:hatK}
\hatK = \sum_{k=1}^{\nS} \sum_{l=1}^{\nG} a_{kl}\ v_k \otimes g_l,	
\end{equation}
where $\mxA=[a_{kl}]_{k=1,l=1}^{\nS,\!\ \nG}\in\Rbb^{\nS\times\nG}$ is the solution of the following optimization problem
\begin{equation}\label{eqn:opt_finite_Tikhonov_reg_case}
\min_{\mxA\in \Rbb^{\nS\times\nG}}\ \|\mxV\mxA\mxG-\mxY\|_{\fro}^2 \ \!+ \ \! \lambda \ \!  \|\mxV^{\frac{1}{2}}\mxA\mxG^{\frac{1}{2}}\|^2,
\end{equation}
given that $\mxV$ and $\mxG$ are respectively the Gramian matrix of $\{v_1,\ldots,v_{\nS}\}$ and $\{g_1,\ldots,g_{\nG}\}$, and $\mxY\in \Rbb^{\nS\times\nG}$ is the matrix defined as  $\mxY:=[y_{kl}]_{k=1,l=1}^{\nS,\!\ \nG}$.
\end{theorem}
\begin{proof}
Since operators $\vce_{\vcx_0},\ldots,\vce_{\vcx_{\nS-1}}$ are bounded, due to the Riesz representation theorem \cite{brezis2010functional}, we know that there exists $v_{k}\in\Hcal$ such that $\vce_{\vcx_{k-1}}(\cdot) = \inner{v_{k}}{\cdot}$, for 
$k\in[\nS]$.
Therefore, we have $ (\mxK  g_l)(\vcx_{k-1}) =  \inner{v_k}{\mxK g_l}$, for each 
$k\in[\nS]$ 
and each 
$l\in[\nG]$.
Subsequently, one can write the objective function in \eqref{eqn:reg_learning_koopman_Tikhonov} as following
\begin{equation}
\Jcal(\mxK):=
\sum_{k=1}^{\nS}
\sum_{l=1}^{\nG}  
(y_{kl}-\inner{v_k}{\mxK g_l})^2 
+ 
\lambda
\|\mxK\|^2.
\end{equation}
Since $\inner{v_k}{\mxK g_l}$ is linear and continuous with respect to $\mxK$, we have that
$\Jcal:\Lcal(\Hcal)\to\Rbb$ is a continuous map. Also, as $\lambda>0$, one can see that 
$\Jcal$ is a strongly convex function. 
Therefore, the optimization problem $\min_{\mxK\in \Lcal(\Hcal)}\Jcal(\mxK)$ admits a unique solution \cite{peypouquet2015convex} denoted by $\hatK$.

Let the linear subspaces  $\Vcal$ and $\Gcal$ be defined respectively as 
\begin{equation}\label{eqn:Vcal}
	\Vcal:=\linspan\{v_1,\ldots,v_{\nS}\},
\end{equation}
and
\begin{equation}\label{eqn:Gcal}
\Gcal:=\linspan\{g_1,\ldots,g_{\nG}\}.
\end{equation}
Also, let $\Pi_{\Vcal}:\Hcal\to\Hcal$  and 
$\Pi_{\Gcal}:\Hcal\to\Hcal$ denote the projection operator on  $\Vcal$ and $\Gcal$, respectively. 
Since the dimension of $\Vcal$ and $\Gcal$ are finite, they are closed subspaces of $\Hcal$, and hence, the projection operators $\Pi_{\Vcal}$ and $\Pi_{\Gcal}$ are well-defined.
Let operator $\mxS$ be defined as $\mxS:=\Pi_{\Vcal} \hatK \Pi_{\Gcal}$.
Due to the definition of $\Pi_{\Gcal}$, we know that $\Pi_{\Gcal}g_l=g_l$, for 
$l\in[\nG]$.
Accordingly, for each $k$ and $l$, we have
\begin{equation}\label{eqn:inner_S_hatK}
\begin{split}
\inner{v_k}{\mxS g_l} 
&=
\inner{v_k}{\Pi_{\Vcal} \hatK \Pi_{\Gcal} g_l}\\ 
&=
\inner{v_k}{\Pi_{\Vcal}\hatK g_l}=\inner{\Pi_{\Vcal}^* v_k}{\hatK g_l},
\end{split}
\end{equation}
where  $\Pi_{\Vcal}^*$ is the adjoint of $\Pi_{\Vcal}$.
Since $\Vcal$ is a closed subspace,  the projection operator $\Pi_{\Vcal}$ is self-adjoint, i.e., $\Pi_{\Vcal}^*=\Pi_{\Vcal}$.
Hence, for each 
$k\in[\nS]$, 
we have $\Pi_{\Vcal}^*v_k=\Pi_{\Vcal}v_k=v_k$, 
where the second equality is due to the definition of operator $\Pi_{\Vcal}$.
Accordingly, from \eqref{eqn:inner_S_hatK}, we know that  
$ 
\inner{v_k}{\mxS g_l}  = \inner{\Pi_{\Vcal} v_k}{\hatK g_l}, 
$ 
for each 
$k\in[\nS]$ and $l\in[\nG]$.
Consequently, it follows that
\begin{equation}\label{eqn:ERM_ERM}
\sum_{k=1}^{\nS}
\sum_{l=1}^{\nG} 
\big(y_{kl}-\inner{v_k}{\mxS g_l}\big)^2
=
\sum_{k=1}^{\nS}
\sum_{l=1}^{\nG}  
\big(y_{kl}-\inner{v_k}{\hatK g_l}\big)^2.	
\end{equation}
Since $\Gcal$ is a closed subspace, we know that $\eye=\Pi_{\Gcal}+\Pi_{\Gcal^\perp}$, where $\eye$ denotes the identity operator on $\Hcal$ and $\Pi_{\Gcal^\perp}$ is the projection operator on $\Gcal^\perp$.
Therefore, for any $h\in\Hcal$, one has
\begin{equation*}\begin{split}
&\frac{\|\mxS h\|^2}{\|h\|^2}
=
\frac{\|\Pi_{\Vcal}\hatK \Pi_{\Gcal} h\|^2}{\|h\|^2}
\\&
\qquad =
\frac{\|\Pi_{\Vcal}\hatK\Pi_{\Gcal} h\|^2}
{\|\Pi_{\Gcal} h\|^2 + \|\Pi_{\Gcal^\perp} h\|^2}
\le
\frac{\|\Pi_{\Vcal} \hatK \Pi_{\Gcal} h\|^2}
{\|\Pi_{\Gcal} h\|^2}
\le
\|\Pi_{\Vcal} \hatK\|^2,
\end{split}\end{equation*}
where the last inequality is due to the definition of \emph{operator norm} for $\Pi_{\Vcal} \hatK$. 
Accordingly, we have 
\begin{equation}\label{eqn:normS_normPVK}
\|\mxS\|^2=\sup_{h\in\Hcal}\frac{\|\mxS\!\ h\|^2}{\|h\|^2}\le\|\Pi_{\Vcal}\ \! \hatK\|^2.
\end{equation}
Since for any $h\in\Hcal$ we have
$\|\Pi_{\Vcal}\ \! \hatK\!\ h\|^2 \le \|\Pi_{\Vcal}\ \! \hatK\!\ h\|^2 +\|\Pi_{\Vcal^\perp}\ \! \hatK\!\ h\|^2  = \|\hatK\!\ h\|^2$, one can see that
\begin{equation}\label{eqn:normPVK_normK}
\|\Pi_{\Vcal}\ \! \hatK\|^2 = 
\sup_{h\in\Hcal}\frac{\|\Pi_{\Vcal}\ \! \hatK\!\ h\|^2}{\|h\|^2} \le
\sup_{h\in\Hcal}\frac{\|\hatK\!\ h\|^2}{\|h\|^2} =
\|\hatK\|^2. 
\end{equation}
Therefore, due to \eqref{eqn:normS_normPVK} and \eqref{eqn:normPVK_normK}, we have $\|\mxS\|^2 \le \|\hatK\|^2$. 
Consequently, from \eqref{eqn:ERM_ERM}, one can see that $\Jcal(\mxS)\le \Jcal(\hatK)$.
Since, the operator $\hatK$ is the unique solution of $\min_{\mxK\in \Lcal(\Hcal)}\!\ \Jcal(\mxK)$, we need to have $\hatK=\mxS=\Pi_{\Vcal}\!\ \hatK\!\ \Pi_{\Gcal}$. 
Due to the linearity of operator $\hatK$, it follows that there exist $a_{kl}\in\Rbb$, 
for $k\in[\nS]$ and $l\in[\nG]$,
such that 
$\hatK = \sum_{k=1}^{\nS} \sum_{l=1}^{\nG} a_{kl}  v_k \otimes g_l$.
To find these values, we replace $\mxK$ in $\min_{\mxK\in \Lcal(\Hcal)}\!\ \Jcal(\mxK)$ with $\hatK$ considering the given parametric form. 
To this end, we need to calculate the value of empirical loss and the regularization term for $\hatK$.
Note that for each 
$k\in[\nS]$ and $l,j\in[\nG]$, 
we have 
$(v_k \otimes g_l) g_j = v_k \!\ \inner{g_l}{g_j}$.
Accordingly, due to the linearity of inner product, for $\hatK = \sum_{k=1}^{\nS}\sum_{l=1}^{\nG} a_{kl}\ v_k \otimes g_l$, we have 
\begin{equation}
\begin{split}
(\hatK   g_j)(\vcx_{i-1})
&=
\sum_{k=1}^{\nS} \sum_{l=1}^{\nG}a_{kl}\ \! \inner{v_i}{(v_k \otimes g_l)g_j}
\\&=
\sum_{k=1}^{\nS} \sum_{l=1}^{\nG}a_{kl}\ \! \inner{v_i}{v_k \inner{g_l}{g_j}}
\\&= 
\sum_{k=1}^{\nS} \sum_{l=1}^{\nG} \inner{v_i}{v_k }\ \! a_{kl}\ \! \inner{g_l}{g_j},
\end{split}
\end{equation}
for each 
$i\in[\nS]$ and $j\in[\nG]$. 
Subsequently, due to the definition of Gramian matrices $\mxV$ and $\mxG$, it follows that
\begin{equation}
(\hatK   g_j)(\vcx_{i-1}) = [\mxV\mxA\mxG]_{(i,j)}. 
\end{equation}
Therefore, given the definition of the matrix $\mxY$, it follows that
\begin{equation}\label{eqn:ERM_Y_VAG}
\begin{split}
&
\sum_{k=1}^{\nS}
\sum_{l=1}^{\nG}  
\big(y_{kl}-(\hatK   g_l)(\vcx_{k-1})\big)^2 
\\&\ \ 
=
\sum_{k=1}^{\nS}
\sum_{l=1}^{\nG}  
\big([\mxY]_{(k,l)}-[\mxV\mxA\mxG]_{(k,l)}\big)^2 
=
\|\mxY-\mxV\mxA\mxG\|_\fro^2,
\end{split}
\end{equation}
where $\mxA$ is the matrix defined as $\mxA=[a_{kl}]_{k=1,l=1}^{\nS,\!\ \nG}$.
We also need to derive the value of $\|\hatK\|^2$. For each $h\in\Hcal$, we have
\begin{equation*}
\hatK h =\bigg(\sum_{k=1}^{\nS} \sum_{l=1}^{\nG} a_{kl}\ \!v_k \otimes g_l\bigg)h
=
\sum_{k=1}^{\nS} \sum_{l=1}^{\nG} a_{kl}\ \! v_k \inner{g_l}{h}.
\end{equation*}
We know that $\inner{g_l}{h} = \inner{g_l}{\Pi_{\Gcal}h}$, for each 
$l\in[\nG]$. 
Accordingly, since $\Pi_{\Gcal}h\in \Gcal=\linspan\{g_l\}_{l=1}^{\nG}$ and due to the definition of operator norm, one has
\begin{equation*}
	\begin{split}
		\|\hatK\|^2 &
		= 
		\sup_{h\in\Hcal}\frac{\|\sum_{k=1}^{\nS} \sum_{l=1}^{\nG} a_{kl}\ v_k 
			\inner{g_l}{\Pi_{\Gcal}\!\ h}\|^2}{\|\Pi_{\Gcal}\!\ h\|^2+ \|\Pi_{\Gcal^\perp}\!\ h\|^2}
		\\&=\!\!
		\sup_{\vc{c}\in\Rbb^{\nG}}
		\frac{\|\sum_{k=1}^{\nS} \sum_{l=1}^{\nG} a_{kl}\ v_k 
			\inner{g_l}{\sum_{j=1}^{\nG} c_jg_j}\|^2}
		{\|\sum_{j=1}^{\nG} c_jg_j\|^2},
	\end{split}
\end{equation*}
where $\vc{c}=[c_1,\ldots,c_{\nG}]^\tr \in\Rbb^{\nG}$ is the vector of coefficients in the expansion of $\Pi_{\Gcal}h$ as in $\Pi_{\Gcal}h = \sum_{j=1}^{\nG} c_jg_j$.
Note that we have 
\begin{equation*}
	\begin{split}
		\Big\|\sum_{j=1}^{\nG} c_jg_j\Big\|^2
		&=
		\sum_{j_1=1}^{\nS}\sum_{j_2=1}^{\nS}c_{j_1}  \inner{g_{j_1}}{g_{j_2}}c_{j_2}
		=\vc{c}^\tr\mxG\vc{c}.
	\end{split}
\end{equation*}
Also, due to the linearity of inner product, one can see that
\begin{equation*}
	\begin{split}
		\sum_{k=1}^{\nS} \sum_{l=1}^{\nG} a_{kl}\ v_k 
		\inner{g_l}{\sum_{j=1}^{\nG} c_jg_j}
		&=
		\sum_{k=1}^{\nS} \sum_{l=1}^{\nG}\sum_{j=1}^{\nG}  a_{kl}v_k 
		c_j\inner{g_l}{g_j}
		\\&=
		\sum_{k=1}^{\nS}v_kd_k,
	\end{split}
\end{equation*}
where $d_k$ is defined as 
$
d_k:=\sum_{l=1}^{\nG}\sum_{j=1}^{\nG}a_{kl}\inner{g_l}{g_j}c_j
$,
for each 
$k\in[\nS]$.
Accordingly, we have
\begin{equation*}
	\begin{split}
		\Big\|\sum_{k=1}^{\nS} \sum_{l=1}^{\nG} a_{kl}\ v_k \inner{g_l}{\sum_{j=1}^{\nG} c_jg_j}\Big\|^2
		=
		\vcd^\tr \mxV \vcd,
	\end{split}
\end{equation*}
where $\mxV$ is the matrix defined as $\mxV=[\inner{v_{k_1}}{v_{k_2}}]_{k_1,k_2=1}^{\nS,\nS}$, and
$\vcd$ is the vector defined as $\vcd:=[d_k]_{k=1}^{\nS}$.
one can 
see that $\vcd=\mxA\mxG \vc{c}$.
Subsequently, we have 
\begin{equation}\label{eqn:norm_hatK}\begin{split}
		\|\hatK\|^2 
		&=
		\sup_{\vc{c}\in\Rbb^{\nG}}
		\frac
		{\vc{c}^\tr\mxG\mxA^\tr\mxV\mxA\mxG\vc{c}}
		{\vc{c}^\tr\mxG\vc{c}}
		\\&=
		\sup_{\vc{c}\in\Rbb^{\nG}}
		\frac
		{\|\mxV^{\frac{1}{2}}\mxA
			\mxG^{\frac{1}{2}}\ 
			\mxG^{\frac{1}{2}}
			\vc{c}\|^2}
		{\|\mxG^{\frac{1}{2}}\vc{c}\|^2}
		=
		\|\mxV^{\frac{1}{2}}\mxA
		\mxG^{\frac{1}{2}}\|^2,
	\end{split}
\end{equation}
where in the last equality we have used the definition of matrix norm and also the change of variable  $\vc{b}=\mxG^{\frac{1}{2}}
\vc{c}$.
Therefore, due to \eqref{eqn:ERM_Y_VAG} and \eqref{eqn:norm_hatK}, one can see that
\begin{equation}
\Jcal(\hatK)= \|\mxY-\mxV\mxA\mxG\|_\fro^2
\!\ +\!\ 
\lambda\!\
\|\mxV^{\frac{1}{2}}\mxA
\mxG^{\frac{1}{2}}\|^2.
\end{equation}
This concludes the proof.
\end{proof}
\begin{remark}\label{rem:RKHS}
	\normalfont
	Let $\Hcal$ be a RKHS endowed with kernel $\kernel:\Xcal\times\Xcal\to\Rbb$.
	Then, from the reproducing property, we know that 
	\begin{equation}
		\vce_{\vcx_{k-1}}(g) = g(\vcx_{k-1}) = \inner{\kernel(\vcx_{k-1},\cdot)}{g},
	\end{equation}
	and subsequently, we have $\vcv_k = \kernel(\vcx_{k-1},\cdot)$, for 
	$k\in[\nS]$. 
    Following this, we can derive the Gramian matrix $\mxV$ as
	\begin{equation}\label{eqn:V_ij}
	\begin{split}
	[\mxV]_{(i,j)} & =\inner{\vcv_i}{\vcv_j}\\
	& = \inner{\kernel(\vcx_{i-1},\cdot)}{\kernel(\vcx_{j-1},\cdot)} = 
	\kernel(\vcx_{i-1},\vcx_{j-1}),		
	\end{split}
	\end{equation}
	for each $i,j\in[\nS]$, 
	where the last equality is due to the reproducing property.
	\cmm{Since the linear span of $\{\kernel_{\vcx}|\vcx\in\Xcal\}$ is dense in $\Hcal$, 
	one may take observables as the sections of kernel at points of a given set $\Pcal:=\{\vcp_1,\ldots,\vcp_{\nG}\}\subset\Xcal$, 
	i.e.,  
	$g_l(\cdot):=\kernel(\vcp_l,\cdot)$, for $l\in[\nG]$.
	Then, similar to \eqref{eqn:V_ij}, one can see that
	$[\mxG]_{(l,j)}= \kernel(\vcp_l,\vcp_j)$, for any $l,j\in[\nG]$.
	Moreover, we have $g_l(\vcx_{k-1}) = \kernel(\vcp_l,\vcx_{k-1})$, for any $k\in[\nS+1]$ and $l\in[\nG]$.
	One can obtain similar expressions when observables are finite linear combinations of the sections of kernel.}
\end{remark}

\subsection{Solving the Optimization Problem} \label{ssec:opt}
Using the change of variables $\mxB=\mxV^{\frac{1}{2}}\mxA\mxG^{\frac{1}{2}}$, one can see that the optimization problem \eqref{eqn:opt_finite_Tikhonov_reg_case} is equivalent to the following program
\begin{equation}\label{eqn:opt_finite_Tikhonov_reg_case_B_01}
\begin{array}{cl}
\minOp_{\mxB\in \Rbb^{\nS\times\nG},\ \! \beta\in\Rbb}
&
\ \|\mxV^{\frac{1}{2}}\mxB\mxG^{\frac{1}{2}}-\mxY\|_{\fro}^2 \ \!+ \ \! \lambda \ \!  \beta^2\\
\mathrm{s.t.}
&
\ \|\mxB\| \le \beta.
\end{array}	
\end{equation}
One can easily see that $\|\mxB\| =\sigma_{\max}(\mxB)\le \beta$ is equivalent to $\beta\eye_{\nG}-\mxB^\tr(\beta\eye_{\nS})^{-1}\mxB\succ \zeromx$ when $\beta$ is positive, or equivalently $\beta\eye_{\nG}\succ \zeromx$.
Accordingly, due to the Schur complement, we can reformulate \eqref{eqn:opt_finite_Tikhonov_reg_case_B_01}
as following
\begin{equation}\label{eqn:opt_finite_Tikhonov_reg_case_B_02}
	\begin{array}{cl}
		\minOp_{\mxB\in \Rbb^{\nS\times\nG},\ \! \beta\in\Rbb}
		&
		\ \|\mxV^{\frac{1}{2}}\mxB\mxG^{\frac{1}{2}}-\mxY\|_{\fro}^2 \ \!+ \ \! \lambda \ \!  \beta^2\\
		\mathrm{s.t.}
		&
		\ \begin{bmatrix}
		  \beta \eye_{\nG} & \mxB\\\mxB^\tr &  \beta \eye_{\nS}
		  \end{bmatrix} \succeq 0,
	\end{array}	
\end{equation}
which is a convex program with quadratic cost and a linear matrix inequality (LMI) constraint. Hence, the learning problem can be solved efficiently using commonly available solvers.

\begin{remark}
\normalfont
The results introduced in Theorem~\ref{thm:Tikhonov_reg_case} and the equivalent finite-dimensional optimization problem can be extended to the case where we have multiple trajectories of the system.
More details are provided in Appendix \ref{ssec:multi_traj}.
\end{remark}

\section{Learning Koopman Operator with Image in a Subspace of Interest}
Let $\Wcal$ be a closed linear subspace of $\Hcal$, and define $\Lcal_{\Wcal}$ as the set of bounded linear operators mapping $\Gcal=\linspan\{g_l\}_{l=1}^{\nG}$ into $\Wcal$, i.e., 
\begin{equation}\label{eqn:L_W}
\Lcal_{\Wcal}:=\{\mxS\in\Lcal(\Hcal)\ |\ \mxS(\Gcal)\subseteq\Wcal\}.
\end{equation}
Since we may encode some specific form of prior knowledge through employing $\Wcal$, it might be of particular interest to learn the Koopman operator as an element of $\Lcal_{\Wcal}$. 
For example, when $g_1,\ldots,g_{\nG}$ and $f$ are polynomials respectively with maximum degree of $d_g$ and $d_f$, we know that $\mxK g_l = g_l \circ f$ is a polynomial with degree maximally equal to $d_g d_f$, for each $l=1,\ldots,\nG$. Accordingly, one may introduce $\Wcal$ as the set of polynomials with degree less than or equal to $d_g d_f$.
The next theorem provides the closest approximation of learned operator $\hatK$, introduced in \eqref{eqn:hatK}, 
in closed subspace $\Lcal_{\Wcal}$.

\begin{theorem}\label{thm:Tikhonov_reg_case_approx}
Define operator $\tildeK_{\Wcal}$ as
\begin{equation}\label{eqn:tildeK_W}
\tildeK_{\Wcal} := \sum_{k=1}^{\nS} \sum_{l=1}^{\nG} a_{kl}\ \!(\Pi_{\Wcal}v_k) \otimes g_l,
\end{equation}
where $\mxA=[a_{kl}]_{k=1,l=1}^{\nS,\!\ \nG}\in\Rbb^{\nS\times\nG}$ is the solution of \eqref{eqn:opt_finite_Tikhonov_reg_case}
and $v_1,\ldots,v_{\nS}\in\Hcal$ are the vectors defined in Theorem \ref{thm:Tikhonov_reg_case}.
Then, we have $\tildeK_{\Wcal}\in \Lcal_{\Wcal}$ and
$ 
\tildeK_{\Wcal}\in \argmin_{\mxS\in\Lcal_{\Wcal}}\ \|\hatK-\mxS\|.
$ 
\end{theorem}
\begin{proof}
Being $\Wcal$ a closed linear subspace of $\Hcal$, the projection operator $\Pi_{\Wcal}$  is well-defined. One can see that $\Pi_{\Wcal}(v\otimes g)=(\Pi_{\Wcal}v)\otimes g$, for any $v,g\in\Hcal$. Accordingly, we have $\tildeK_{\Wcal}=\Pi_{\Wcal}\hatK$, from which it follows that $\tildeK_{\Wcal}\in\Lcal_{\Wcal}$.	
Since $\Pi_{\Gcal}$ is a projection operator, we know that $\|\Pi_{\Gcal}\|=1$. Accordingly, for any $\mxS \in \Lcal_{\Wcal}$, we have
\begin{equation}\label{eqn:pf_thm2_1}
	\|\hatK-\mxS\Pi_{\Gcal}\|  = \|(\hatK-\mxS)\Pi_{\Gcal}\|\le\|\hatK-\mxS\|\|\Pi_{\Gcal}\|=\|\hatK-\mxS\|,
\end{equation} 
where the first equality is a result of $\hatK\Pi_{\Gcal}=\hatK$ which is according to the definition of $\hatK$ in \eqref{eqn:hatK}.
Also, due to the definition of $\Lcal_{\Wcal}$,  we know that $\mxS(\Gcal)\subseteq\Wcal$, and consequently, one has 
$\Pi_{\Wcal^\perp}\mxS\Pi_{\Gcal}=0$.
Accordingly, it follows that 
\begin{equation}\label{eqn:pf_thm2_2}
\begin{split}
\|\Pi_{\Wcal^\perp}\hatK\|  
&= 
\|\Pi_{\Wcal^\perp}\hatK-\Pi_{\Wcal^\perp}\mxS\Pi_{\Gcal}\| 
\\&\le 
\|\Pi_{\Wcal^\perp}\|\|\hatK-\mxS\Pi_{\Gcal}\| 
= 
\|\hatK-\mxS\Pi_{\Gcal}\|,
\end{split}
\end{equation} 
where the last equality is concluded from $\|\Pi_{\Wcal^\perp}\|=1$ 
that is due to the fact that $\Pi_{\Wcal^\perp}$ is a projection operator.
Furthermore, we know that $\Pi_{\Wcal^\perp} = \eye-\Pi_{\Wcal}$, where $\eye$ is the identity operator on $\Hcal$.
Hence, due to $\tildeK_{\Wcal}=\Pi_{\Wcal}\hatK$, we have $\Pi_{\Wcal^\perp}\hatK=\hatK-\tildeK_{\Wcal}$.	
Therefore,  from \eqref{eqn:pf_thm2_1} and \eqref{eqn:pf_thm2_2}, it follows that
\begin{equation}
	\|\hatK-\tildeK_{\Wcal}\|\le \|\hatK-\mxS\|, \quad 
	\forall \mxS \in \Lcal_{\Wcal},
\end{equation}
which concludes the proof.
\end{proof}
According to Theorem \ref{thm:Tikhonov_reg_case_approx}, to obtain 
the closest approximation of operator $\hatK$ in $\Lcal_{\Wcal}$, i.e.,
$\tildeK_{\Wcal}$, we need to solve \eqref{eqn:opt_finite_Tikhonov_reg_case} and also derive $\Pi_{\Wcal}v_k=\argmin_{w\in\Wcal}\|v_k-w\|$, for $k=1,\ldots,\nS$.
On the other hand, one may propose to learn the Koopman operator in $\Lcal_{\Wcal}$ via a direct approach by finding the solution of the following learning problem
\begin{equation}\label{eqn:reg_learning_koopman_Tikhonov_W}
	\!\!
	\minOp_{\mxK\in\Lcal_{\Wcal}}\   
	\Ecal(\mxK)
	+ \lambda \|\mxK\|^2\!,
\end{equation}
where $\Ecal:\Hcal\to\Rbb$ is the empirical loss defined in \eqref{eqn:empirical_loss}.
The next theorem characterizes the solution of \eqref{eqn:reg_learning_koopman_Tikhonov_W}. 	
\begin{theorem}	\label{thm:Tikhonov_reg_case_W_part_1}
	Let Assumption \ref{ass:e_xk_bounded} hold, $\lambda>0$, and $v_1,\ldots,v_{\nS}\in\Hcal$ be the vectors defined in Theorem \ref{thm:Tikhonov_reg_case}.
	Then, the optimization problem \eqref{eqn:reg_learning_koopman_Tikhonov_W} has a unique solution denoted by $\hatK_{\Wcal}$. Moreover, there exists  $\mxA=[a_{kl}]_{k=1,l=1}^{\nS,\!\ \nG}\in\Rbb^{\nS\times\nG}$ which characterizes $\hatK_{\Wcal}$ as following
	\begin{equation}\label{eqn:hatK_PiW}
		\hatK_{\Wcal} = \sum_{k=1}^{\nS} \sum_{l=1}^{\nG} a_{kl}\ \!(\Pi_{\Wcal}v_k) \otimes g_l.	
	\end{equation}
	Define matrices $\mxG$ and $\mxY$ as in Theorem \ref{thm:Tikhonov_reg_case}, and $\mxW_{\Vcal}$ as the Gramian matrix of vectors $\Pi_{\Wcal}v_1,\ldots,\Pi_{\Wcal}v_{\nS}$. 
	Then, matrix $\mxA$ in \eqref{eqn:hatK_PiW} is the solution of the following optimization problem
	\begin{equation}\label{eqn:opt_finite_Tikhonov_reg_case_PiW}
		\min_{\mxA\in \Rbb^{\nS\times\nG}}\ 
		\|\mxW_{\Vcal}\mxA\mxG-\mxY\|_{\fro}^2 \ \!+ \ \! \lambda \ \!  \|\mxW_{\Vcal}^{\frac{1}{2}}\mxA\mxG^{\frac{1}{2}}\|^2.
	\end{equation}
	\end{theorem}
\iftrue
\begin{proof}
See Appendix \ref{sec:appendix_proof_thm_3}.
\end{proof}
\else
\begin{proof}
The existence and uniqueness of the solution $\hatK_{\Wcal}$ follows from the same lines of arguments as in the proof of Theorem~\ref{thm:Tikhonov_reg_case}.
Define the linear subspaces  $\tildeWcal$ as 
\begin{equation}\label{eqn:tildeWcal}
	\tildeWcal:=\linspan\{\Pi_{\Wcal}v_1,\ldots,\Pi_{\Wcal}v_{\nS}\},
\end{equation}
and, let $\Pi_{\tildeWcal}$ be the projection operator on $\tildeWcal$.
For $k=1,\ldots,\nS$, we know that $\Pi_{\Wcal}v_k \in\Wcal$, and therefore, $\tildeWcal$ is a subspace of $\Wcal$.
Define operator $\mxS$ as $\mxS:=\Pi_{\tildeWcal}\ \!\hatK_{\Wcal}\ \!\Pi_{\Gcal}$ which belongs to $\Lcal_{\Wcal}$ due to $\tildeWcal\subseteq\Wcal$.
From the definition of $\Pi_{\Gcal}$, we know that $\Pi_{\Gcal}g_l=g_l$, for $l=1,\ldots,\nG$.
Accordingly, for any $k$ and $l$, we have
\begin{equation}\label{eqn:inner_S_hatK_W}
	\begin{split}
		\inner{v_k}{\mxS g_l} 
		&=
		\inner{v_k}{\Pi_{\tildeWcal} \hatK_{\Wcal} \Pi_{\Gcal} g_l}\\ 
		&
		=
		\inner{v_k}{\Pi_{\tildeWcal}\hatK_{\Wcal} g_l}
		=\inner{\Pi_{\tildeWcal}^* v_k}{\hatK_{\Wcal} g_l},
	\end{split}
\end{equation}
where  $\Pi_{\tildeWcal}^*$ is the adjoint of $\Pi_{\tildeWcal}$.
Since $\tildeWcal$ is a finite dimensional subspace, it is closed and the projection operator $\Pi_{\tildeWcal}$ is self-adjoint, i.e., $\Pi_{\tildeWcal}^*=\Pi_{\tildeWcal}$.
Also, due to $\tildeWcal\subseteq\Wcal$, we know that  $\Wcal^\perp\subseteq\tildeWcal^\perp$, and subsequently, we have $\Pi_{\tildeWcal}\Pi_{\Wcal^\perp} = 0$.
Accordingly, one can see that
\begin{equation}
	\Pi_{\tildeWcal}-\Pi_{\tildeWcal}\Pi_{\Wcal} 
	= 
	\Pi_{\tildeWcal}(\eye-\Pi_{\Wcal})
	=
	\Pi_{\tildeWcal}\Pi_{\Wcal^\perp} = 0.
\end{equation}
Consequently, for $k=1,\ldots,\nS$, we have 
\begin{equation}\label{eqn:Pi_tildeW*_vk}
\Pi_{\tildeWcal}^*v_k = \Pi_{\tildeWcal}v_k=\Pi_{\tildeWcal}\ \!\Pi_{\Wcal}v_k= \Pi_{\Wcal}v_k,	
\end{equation}
where the last equality is due to 
$\Pi_{\Wcal}v_k\in\tildeWcal$. 
Note that $\Wcal$ is a closed subspace, and subsequently, $\Pi_{\Wcal}$ is a self-adjoint operator, i.e., $\Pi_{\Wcal}^*=\Pi_{\Wcal}$.
Accordingly,  from \eqref{eqn:inner_S_hatK_W} and \eqref{eqn:Pi_tildeW*_vk},  we can see that 
\begin{equation}\label{eqn:inner_S_hatK_W_02}
\begin{split}
	\inner{v_k}{\mxS g_l}  &= 
	\inner{\Pi_{\Wcal} v_k}{\hatK_{\Wcal} g_l}\\ 
	&= 
	\inner{v_k}{\Pi_{\Wcal}^*\hatK_{\Wcal} g_l}
	=
	\inner{v_k}{\Pi_{\Wcal}\hatK_{\Wcal} g_l}, 
\end{split}
\end{equation}
for any $k$ and $l$.
Due to the definition of $\Lcal_{\Wcal}$ in \eqref{eqn:L_W} and since $\hatK_{\Wcal}\in\Lcal_{\Wcal}$, we know that $\hatK_{\Wcal}g_l\in\Wcal$, and subsequently, one has $\Pi_{\Wcal}\hatK_{\Wcal} g_l=\hatK_{\Wcal} g_l$, for $l=1,\ldots,\nG$.
Therefore, from \eqref{eqn:inner_S_hatK_W_02}, it follows that
\begin{equation}\label{eqn:ERM_ERM_W}
	\begin{split}
		\Ecal(\mxS) &= \sum_{k=1}^{\nS}
		\sum_{l=1}^{\nG} 
		\big(  y_{kl}-\inner{v_k}{\mxS g_l}\big)^2
		\\&=
		\sum_{k=1}^{\nS}
		\sum_{l=1}^{\nG}  
		\big(y_{kl}-\inner{v_k}{\hatK_{\Wcal} g_l}\big)^2=\Ecal(\hatK_{\Wcal}).	
	\end{split}
\end{equation}
Similar to the proof of Theorem \ref{thm:Tikhonov_reg_case}, one can show that $\|\mxS\|^2 \le \|\hatK_{\Wcal}\|^2$, and subsequently, one can see that $\Ecal(\mxS)+\lambda\|\mxS\|^2\le \Ecal(\hatK_{\Wcal}) +\lambda\|\hatK_{\Wcal}\|^2$.
From the uniqueness of the solution of \eqref{eqn:reg_learning_koopman_Tikhonov_W}, we  have $\hatK_{\Wcal}=\mxS=\Pi_{\Vcal}\!\ \hatK_{\Wcal}\!\ \Pi_{\Gcal}$. 
Due to the linearity of operator $\hatK_{\Wcal}$, it follows that there exist $a_{kl}\in\Rbb$, for $k=1,\ldots,\nS$ and $l=1,\ldots,\nG$, such that 
$\hatK_{\Wcal} = \sum_{k=1}^{\nS} \sum_{l=1}^{\nG} a_{kl}  (\Pi_{\Wcal}v_k) \otimes g_l$, i.e., we have \eqref{eqn:hatK_PiW}.
Considering $\hatK_{\Wcal}$ in this parametric form and 
due to the linearity of inner product, for any $i=1,\ldots,\nS$ and $j=1,\ldots,\nG$, it follows that 
\begin{equation}
	\begin{split}
		(\hatK   g_j)(\vcx_{i-1})
		&=
		\sum_{k=1}^{\nS} \sum_{l=1}^{\nG}a_{kl}\ \! \inner{v_i}{\big((\Pi_{\Wcal}v_k) \otimes g_l\big)g_j}
		\\&=
		\sum_{k=1}^{\nS} \sum_{l=1}^{\nG}a_{kl}\ \! \inner{v_i}{\Pi_{\Wcal}v_k \inner{g_l}{g_j}}
		\\&= 
		\sum_{k=1}^{\nS} \sum_{l=1}^{\nG} \inner{\Pi_{\Wcal}v_i}{\Pi_{\Wcal}v_k }\ \! a_{kl}\ \! \inner{g_l}{g_j},
	\end{split}
\end{equation}
where the last equality is due to the fact that $ \inner{u}{\Pi_{\Wcal}v }= \inner{\Pi_{\Wcal}u}{\Pi_{\Wcal}v }$, for any $v,u\in\Hcal$.
Accordingly, we have
\begin{equation}
	\left[(\hatK   g_j)(\vcx_{i-1})\right]_{i=1,j=1}^{\nS,\!\ \nG} = \mxW_{\Vcal}\mxA\mxG. 
\end{equation}
Then, following same steps of calculations as in the proof of Theorem \ref{thm:Tikhonov_reg_case}, one can show that $\mxA$ can be obtained by solving convex program \eqref{eqn:opt_finite_Tikhonov_reg_case_PiW}.
\end{proof}
\fi 
In the next theorem, we discuss the situation where the dimension of $\Wcal$ is finite. This result is employed in Section \ref{sec:edmd} to provide the connection between the discussed Koopman learning probelm and
the extended dynamic mode decomposition (EDMD) method \cite{koopman_datadrivenapproximation_edmd}. 
\begin{theorem}	\label{thm:Tikhonov_reg_case_W_part_2}
	Let the hypotheses of Theorem~\ref{thm:Tikhonov_reg_case_W_part_1} hold,  $w_1,\ldots,w_{\nW}\in\Hcal$ be linear independent vectors such that $\Wcal = \linspan\{w_1,\ldots,w_{\nW}\}$, matrix $\mxP_{\Wcal}$ be defined as ${\mxP_{\Wcal}=[w_l(\vcx_{k-1})]_{k=1,l=1}^{\nS,\!\ \nW}}$, and, $\mxW$ be the Gramian matrix of vectors $w_1,\ldots,w_{\nW}$. 
	Then, one can represent the unique solution of \eqref{eqn:reg_learning_koopman_Tikhonov_W} as 
	\begin{equation}\label{eqn:hatK_W}
		\hatK_{\Wcal} = \sum_{j=1}^{\nW} \sum_{l=1}^{\nG} c_{jl}\ \! w_j \otimes g_l,	
	\end{equation}
	where $\mxC=[c_{jl}]_{j=1,l=1}^{\nW,\!\ \nG}$ is the solution of the following optimization problem
	\begin{equation}\label{eqn:opt_finite_Tikhonov_reg_case_W}
		\min_{\mxC\in \Rbb^{\nW\times\nG}}\ \|\mxP_{\Wcal}\mxC\mxG-\mxY\|_{\fro}^2 \ \!+ \ \! \lambda \ \!  \|\mxW^{\frac{1}{2}}\mxC\mxG^{\frac{1}{2}}\|^2.
	\end{equation}
\end{theorem}
\begin{proof}
For each 
$k\in[\nS]$, 
there exist $q_{k1},\ldots,q_{k\nW}$ such that
\begin{equation}\label{eqn:PiW_vk_minimization_norm}
\Pi_{\Wcal} v_k= \sum_{j=1}^{\nW} q_{kj}w_j = \argminOp_{w\in\Wcal} \frac{1}{2}\|v_k-w\|^2, 
\end{equation}
Define matrix $\mxQ$ as $\mxQ = [q_{kj}]_{k=1,j=1}^{\nS,\ \!\nW}$, and let the row vectors $\vcq_k$ and $\vcp_k$ be respectively defined as 
$\vcq_k =[q_{kj}]_{j=1}^{\nW}$ and $\vcp_k =[\inner{v_k}{w_j}]_{j=1}^{\nW}$.
From \eqref{eqn:PiW_vk_minimization_norm}, we can easily see  that $\vcq_k = \vcp_k \mxW^{-1}$.
Also, due to the definition of vectors $v_1,\ldots,v_{\nS}$, we know that $\mxP_{\Wcal} = [p_1^\tr,\ldots,p_{\nS}^\tr\,]^\tr$, which implies $\mxQ=\mxP_{\Wcal}\mxW^{-1}$.
From \eqref{eqn:PiW_vk_minimization_norm}, which says that $\Pi_{\Wcal} v_k= \sum_{j=1}^{\nW} q_{kj}w_j$,  
one can see that $\hatK_{\Wcal}$ in \eqref{eqn:hatK_PiW} has a representation as in \eqref{eqn:hatK_W}. Moreover, we have
\begin{equation*}
	\hatK_{\Wcal} 
	\!=\! 
	\sum_{j=1}^{\nW} \sum_{l=1}^{\nG}(\sum_{k=1}^{\nS} q_{kj}a_{kl})w_j\otimes g_l
	\!=\! 
	\sum_{j=1}^{\nW} \sum_{l=1}^{\nG}[\mxQ^\tr\mxA]_{(j,l)}w_j\otimes g_l,	
\end{equation*}
which yields $\mxC=\mxQ^\tr\mxA$.
Also, for each
$k,i\in[\nS]$, 
one has
\begin{equation}
\begin{split}
[\mxW_{\Vcal}]_{(k,i)} 
&=
\inner{\Pi_{\Wcal}v_k}{\Pi_{\Wcal}v_i} = 
\inner{\sum_{j=1}^{\nW} q_{kj}w_j}{\sum_{l=1}^{\nW} q_{il}w_l} 
\\&= 
\sum_{j=1}^{\nW}\sum_{l=1}^{\nW}q_{kj} \inner{w_j}{ w_l} q_{il} = 
[\mxQ\mxW\mxQ^\tr]_{(k,i)},
\end{split}
\end{equation}
and, in consequence, $\mxW_{\Vcal}=\mxQ\mxW\mxQ^\tr$. We thus get $\mxW_{\Vcal}=\mxP_{\Wcal}\mxW^{-1}\mxP_{\Wcal}^\tr\,$.
Subsequently, we have
\begin{equation}
\mxP_{\Wcal}\mxC = 
\mxP_{\Wcal}\mxQ^\tr\mxA =  
\mxP_{\Wcal}\mxW^{-1}\mxP_{\Wcal}^\tr\,\mxA =
\mxW_{\Vcal}\mxA,
\end{equation}
which gives $\|\mxP_{\Wcal}\mxC\mxG-\mxY\|_{\fro}^2=\|\mxW_{\Vcal}\mxA\mxG-\mxY\|_{\fro}^2$.
Furthermore, from $\mxA^\tr\mxW_{\Vcal}\mxA=\mxA^\tr\mxQ\mxW\mxQ^\tr\mxA=\mxC^\tr\mxW\mxC$, we know that 
\begin{equation*}
\begin{split}
\|\mxW^{\frac12}\mxC\mxG^{\frac12}\|^2 
&=
\sup_{\vc{z}\in\Rbb^{\nG}}
\frac
{\vc{z}^\tr\mxG^{\frac12}\mxC^\tr\mxW\mxC\mxG^{\frac12}\vc{z}}
{\|\vc{z}\|^2}
\\&=
\sup_{\vc{z}\in\Rbb^{\nG}}
\frac
{\vc{z}^\tr\mxG^{\frac12}\mxA^\tr\mxW_{\Vcal}\mxA\mxG^{\frac12}\vc{z}}
{\|\vc{z}\|^2}
=
\|\mxW^{\frac{1}{2}}_{\Vcal}\mxA
\mxG^{\frac{1}{2}}\|^2.
\end{split}
\end{equation*}
Therefore, using the mentioned change of variable 
in \eqref{eqn:opt_finite_Tikhonov_reg_case_PiW}, we get the convex program \eqref{eqn:opt_finite_Tikhonov_reg_case_W}. This concludes the proof.
\end{proof}
\begin{remark}
From the proof of Theorem~\ref{thm:Tikhonov_reg_case_W_part_2}, one can see that $\mxC =\mxW^{-1}\mxP_{\Wcal}\mxA$ and $\mxW_{\Vcal}=\mxP_{\Wcal}\mxW^{-1}\mxP_{\Wcal}^\tr\,$. One may employ these identities as a change-of-variable, and subsequently, obtain more tractable optimization problems.
\end{remark}
\begin{corollary}\label{cor:hatK=hatK_W=tildeK_W}
If $\Vcal:=\linspan\{v_1,\ldots,v_{\nS}\}\subseteq\Wcal$, 
then we have 
\begin{equation}
\tildeK_{\Wcal}=\hatK_{\Wcal}=\hatK.
\end{equation}
\end{corollary}
The approach discussed in Theorem \ref{thm:Tikhonov_reg_case} for learning the Koopman operator $\hatK$ demands the knowledge of $v_1,\ldots,v_{\nS}$.
This issue can be addressed by Theorem \ref{thm:Tikhonov_reg_case_W_part_1} when $\Pi_{\Wcal}v_1,\ldots,\Pi_{\Wcal}v_{\nS}$ are known.
Note that this knowledge is not sufficient for the scheme proposed in Theorem \ref{thm:Tikhonov_reg_case_approx} for approximating $\hatK$ in $\Lcal_{\Wcal}$, where indeed, the knowledge of $v_1,\ldots,v_{\nS}$ is again required to first solve problem \eqref{eqn:reg_learning_koopman_Tikhonov}, and then derive the approximation.
On the other hand, in the case of finite dimensional space $\Wcal$, when vectors $w_1,\ldots,w_{\nW}\in\Hcal$ are given such that $\Wcal =\linspan\{w_1,\ldots,w_{\nW}\}$, for solving the learning problem \eqref{eqn:reg_learning_koopman_Tikhonov_W} it is enough to know $\{w_m(\vcx_{k-1})\}_{k=1,m=1}^{\nS,\nG}$. In this situation, the knowledge of $v_1,\ldots,v_{\nS}$ is not required.
Also, from Corollary \ref{cor:hatK=hatK_W=tildeK_W}, we can see that if the set of vectors $w_1,\dots,w_{\nW}$ is rich enough in the sense that  $\Vcal\subseteq\Wcal=\linspan\{w_1,\ldots,w_{\nW}\}$,
then, the learning problems  \eqref{eqn:reg_learning_koopman_Tikhonov},
\eqref{eqn:reg_learning_koopman_Tikhonov_W}, and
\eqref{eqn:opt_finite_Tikhonov_reg_case_W} admit same solution.
Hence, under the condition $\Vcal\subseteq\Wcal=\linspan\{w_1,\ldots,w_{\nW}\}$, when $\{w_m(\vcx_{k-1})\}_{k=1,m=1}^{\nS,\nG}$ is given,
we can solve each of the above-mentioned learning problems without the knowledge of $v_1,\ldots,v_{\nS}$.

%% file: sec_06_edmd.tex
\section{Connection to the Extended Dynamic Mode Decomposition}\label{sec:edmd}
In this section, we consider the case where  $\Gcal=\linspan\{g_1,\ldots,g_{\nG}\}$ is invariant for the learned Koopman operator, 
i.e., based on the notations introduced in Section \ref{sec:learning}, 
we have $\hatK \in \Lcal_{\Gcal}$.
Without loss of generality, we assume $g_1,\ldots,g_{\nG}$ are linearly independent.

In the extended dynamic mode decomposition (EDMD) method, the Koopman operator is approximated by a finite dimensional linear map $\mxU:\Gcal\to\Gcal$, and then, the observation data $\Dscr$ is employed to estimate this map \cite{koopman_datadrivenapproximation_edmd}. 
Since the dimension of $\Gcal$ is finite, the map $\mxU$ admits a matrix representation in the basis $\{g_1,\ldots,g_{\nG}\}$.
More precisely, there exists matrix $\mxM\in\Rbb^{\nG\times\nG}$ such that $\mxU g_l = \sum_{j=1}^{\nG} [\mxM]_{(j,l)}g_j$, for $l=1,\ldots,\nG$. The matrix $\mxM$ is estimated by minimizing the empirical loss $\Ecal_{\mxM}$ defined as
\begin{equation*}
	\Ecal_{\mxU}(\mxM):=
	\|\mxP_{\Gcal}\mxM-\mxY\|_{\fro}^2 =
	\sum_{k=1}^{\nS} 
	\sum_{l=1}^{\nG} 
	\big((\mxU g_l)(\vcx_{k-1})-g_l(\vcx_k)\big)^2\!,
\end{equation*}
where ${\mxP_{\Gcal}:=[g_l(\vcx_{k-1})]_{k=1,l=1}^{\nS,\!\ \nG}}$, which is assumed to be full column rank, i.e., 
$\rank(\mxP_{\Gcal})= \nG$. Accordingly, that empirical loss $\Ecal_{\mxM}$ has a unique minimizer $\mxMstar=\mxP_{\Gcal}^\dagger\mxY$, where $\mxP_{\Gcal}^\dagger :=
(\mxP_{\Gcal}^\tr\mxP_{\Gcal})^{-1}\mxP_{\Gcal}^\tr$ is the 
Moore-Penrose pseudoinverse
of $\mxP_{\Gcal}$.

\begin{theorem}\label{thm:edmd_lifting}
Define matrix $\mxC_{\mxMstar}$ as $\mxMstar\,\mxG^{-1}\!$, 
and, let operator $\hatK_{\mxU}\in\Lcal(\Hcal)$ be defined as following
	\begin{equation}\label{eqn:hatK_U}
		\hatK_{\mxU} = \sum_{j=1}^{\nG} \sum_{l=1}^{\nG}\ \! [\mxC_{\mxMstar}]_{(j,l)}\ \! g_j \otimes g_l.	
	\end{equation}
	Then, $\hatK_{\mxU}$ is a solution of $\min_{\mxK\in\Lcal_{\Gcal}} \Ecal(\mxK)$.
	Moreover, the EDMD map $\mxU$ coincides with the restriction of $\hatK_{\mxU}$ to $\Gcal$, i.e., $\mxU = \hatK_{\mxU}{\big|_{\Gcal}}$. 
\end{theorem}
\begin{proof}
Similar to the proof of Theorem \ref{thm:Tikhonov_reg_case_W_part_1}, one can show that $\Ecal(\mxK)=\Ecal(\Pi_{\Gcal}\mxK\Pi_{\Gcal})$, for each $\mxK\in\Lcal_{\Gcal}$.
Accordingly, if $\min_{\mxK\in\Lcal_{\Gcal}} \Ecal(\mxK)$ admits a solution, it has also a solution in the set of operators $\Lcal(\Gcal)$ which can be characterized as  $\{\mxK_{\mxC}:=\sum_{j=1}^{\nG} \sum_{l=1}^{\nG} c_{jl}\ \! g_j \otimes g_l\ \!|\!\ \mxC\in\Rbb^{\nG\times\nG}\}$.
Therefore, it is enough to consider the problem  $\min_{\mxK\in\Lcal(\Gcal)} \Ecal(\mxK)$.
Given $\mxC\in\Rbb^{\nG\times\nG}$, we define $\mxM=\mxC\mxG$, where, based on same steps as in the proof of Theorem \ref{thm:Tikhonov_reg_case_W_part_2}, we can show that $\Ecal(\mxK_{\mxC})=\Ecal_{\mxU}(\mxM)$. 
Hence, there is a bijection between the value and solutions of $\min_{\mxK\in\Lcal(\Gcal)} \Ecal(\mxK)$  and $\min_{\mxM\in\Rbb^{\nG\times\nG}} \Ecal_{\mxU}(\mxM)$.  
Since $\mxMstar$ is a solution for the latter problem, then $\mxK_{\mxC_{\mxMstar}}$, which coincides with $\hatK_{\mxU}$, is a solution of $\min_{\mxK\in\Lcal(\Gcal)} \Ecal(\mxK)$, which concludes the proof of first part. 
For 
$i\in[\nG]$, 
due to \eqref{eqn:hatK_U}, we have
\begin{equation}
\begin{split}
\hatK_{\mxU}g_i &= 
\sum_{j=1}^{\nG} \sum_{l=1}^{\nG} [\mxC_{\mxMstar}]_{(j,l)} g_j  \inner{g_l}{g_i}
\\&=	
\sum_{j=1}^{\nG}  [\mxC_{\mxMstar}\mxG]_{(j,l)} g_j = \sum_{j=1}^{\nG}  [\mxMstar]_{(j,l)} g_j,	
\end{split}
\end{equation}
where the last equality is due to $\mxMstar=\mxC_{\mxMstar}\mxG$. This concludes the proof.
\end{proof}
	
\begin{theorem}\label{thm:lim_K_lambda}
For $\lambda>0$, let $\hatK_{\Gcal,\lambda}$ be the unique solution of  
\begin{equation}\label{eqn:reg_learning_koopman_G_lambda}
	\minOp_{\mxK\in\Lcal_{\Gcal} } \Jcal_{\lambda}(\mxK):= \Ecal(\mxK) + \lambda \|\mxK\|^2.
\end{equation}
Then, $\lim_{\lambda \downarrow 0} \hatK_{\Gcal,\lambda} = \hatK_{\mxU}$ and $\lim_{\lambda \to \infty} \hatK_{\Gcal,\lambda} = 0$, both in operator norm topology.
\end{theorem}
\begin{proof}
From Theorem~\ref{thm:Tikhonov_reg_case_W_part_1} and Theorem~\ref{thm:Tikhonov_reg_case_W_part_2}, we know that, for each $\lambda>0$, 
\eqref{eqn:reg_learning_koopman_G_lambda} admits a unique solution 
$\hatK_{\Gcal,\lambda} = \sum_{j=1}^{\nG} \sum_{l=1}^{\nG} [\mxC_{\lambda}]_{(j,l)}\ \! g_j \otimes g_l$, where $\mxC_{\lambda}$ is defined as
\begin{equation}\label{eqn:opt_finite_Tikhonov_reg_case_G_lambda}
	\mxC_{\lambda} = \argminOp_{\mxC\in \Rbb^{\nW\times\nG}}\ \|\mxP_{\Gcal}\mxC\mxG-\mxY\|_{\fro}^2 \ \!+ \ \! \lambda \ \!  \|\mxG^{\frac{1}{2}}\mxC\mxG^{\frac{1}{2}}\|^2.
\end{equation}
By the definition of $\mxC_{\lambda}$, we have
\begin{equation*}
\|\mxP_{\Gcal}\mxC_{\lambda}\mxG-\mxY\|_{\fro}^2 \le 
\|\mxP_{\Gcal}\mxC_{\lambda}\mxG-\mxY\|_{\fro}^2 + \lambda  \|\mxG^{\frac{1}{2}}\mxC_{\lambda}\mxG^{\frac{1}{2}}\|^2\le \|\mxY\|_{\fro}^2,
\end{equation*}
which implies that $\|\mxP_{\Gcal}\mxC_{\lambda}\mxG\|_{\fro}\le 2 \|\mxY\|_{\fro}$.
Using Lemma \ref{lem:fro_norm_ineq} in Appendix \ref{sec:appendix_lemmas}, we have
\begin{equation}
\begin{split}\!\!\!\!
\|\mxC_{\lambda}\|_{\fro} &= 
\|(\mxP_{\Gcal}^\tr\mxP_{\Gcal})^{-1}\mxP_{\Gcal}^\tr(\mxP_{\Gcal}\mxC_{\lambda}\mxG)\mxG^{-1}\|_{\fro}\\
&\!\!\!\!\!\!\le \|\mxP_{\Gcal}^\dagger\| \|\mxP_{\Gcal}\mxC_{\lambda}\mxG\|_{\fro}\|\mxG^{-1}\|
\le 2\|\mxP_{\Gcal}^\dagger\| \|\mxY\|_{\fro}\|\mxG^{-1}\|.
\end{split}	
\end{equation}
Accordingly, \eqref{eqn:opt_finite_Tikhonov_reg_case_G_lambda} in equivalent to the following program
\begin{equation}\label{eqn:opt_finite_Tikhonov_reg_case_G_lambda_constraint}
	\!\!\!
	\argminOp_{\mxC\in\Mcal} J_{\lambda}(\mxC):=\|\mxP_{\Gcal}\mxC\mxG-\mxY\|_{\fro}^2+  \lambda \ \!  \|\mxG^{\frac{1}{2}}\mxC\mxG^{\frac{1}{2}}\|^2\!\!.\!
\end{equation}
where $\Mcal:=\big\{\mxC\in \Rbb^{\nW\times\nG}\ \! \big|\ \! \|\mxC_{\lambda}\|\le2\|\mxP_{\Gcal}^\dagger\| \|\mxY\|_{\fro}\|\mxG^{-1}\|\big\}$,
which is a convex and compact set.
Let $\Ccal_{\lambda}$ be the solution set of \eqref{eqn:opt_finite_Tikhonov_reg_case_G_lambda_constraint} for $\lambda\ge 0$, which is a singleton due to the strong convexity of the objective function.
Moreover, we know that function $J_{\lambda}$ is continuous with respect $(\mxC,{\lambda})$. Hence, from Maximum Theorem \cite{Infinitedimensionalanalysis2006}, it follows that set-valued map $\lambda\mapsto\Ccal_{\lambda}$ is upper hemicontinuous with non-empty and compact values, which implies that $\lim_{\lambda\downarrow 0 } \Ccal_{\lambda}= \Ccal_{0}$, i.e., $\lim_{\lambda\downarrow 0 } \mxC_{\lambda}= \mxC_{\mxMstar}$.
Accordingly, due to the structure of $\hatK_{\lambda}$ and $\hatK_{\mxU}$, we have
$\lim_{\lambda \downarrow 0} \hatK_{\Gcal,\lambda} = \hatK_{\mxU}$ in operator norm topology. 
On the other hand, since $0\in\Lcal_{\Gcal}$ is feasible for the problem and due to the definition of $\hatK_{\Gcal,\lambda}$, we know that
\begin{equation*}
\lambda\|\hatK_{\Gcal,\lambda}\|^2
\le \Jcal_{\lambda}(\hatK_{\Gcal,\lambda}) 
\le \Jcal_{\lambda}(0) =  \|\mxY\|_{\fro}^2,
\end{equation*}
which implies $\|\hatK_{\Gcal,\lambda}\|
\le \frac{1}{\sqrt{\lambda}} \|\mxY\|_{\fro}$. Therefore, 
we have $\lim_{\lambda\to\infty}\|\hatK_{\Gcal,\lambda}\|=0$, and subsequently, it follows that $\lim_{\lambda \to +\infty} \hatK_{\Gcal,\lambda} = 0$, in operator norm topology. This concludes the proof.
\end{proof}

\begin{remark} \normalfont
For problem \eqref{eqn:reg_learning_koopman_Tikhonov}, one can see as $\lambda\to\infty$, the unique solution introduced in Theorem \ref{thm:Tikhonov_reg_case} converges to zero, in operator norm topology. 
The same claim holds for the unique solution of  \eqref{eqn:reg_learning_koopman_Tikhonov_W}.
\end{remark}

%% file: sec_07_generalization.tex
\section{Generalized Representer Theorem and Cases of Interest}
\label{sec:gen}
In this section, we generalize the learning problems and the main results discussed in Section \ref{sec:learning}, and then, we apply this result to various cases of  interest.

Given  a loss function  $L:\Rbb\to\Rbb$, the \emph{empirical loss} introduced in \eqref{eqn:empirical_loss} can be generalized to $\Ecal_L:\Lcal(\Hcal)\to\Rbb$ defined as following
\begin{equation}\label{eqn:empirical_loss_L}
	\Ecal_L(\mxK) := 
	\sum_{k=1}^{\nS}
	\sum_{l=1}^{\nG}  
	L(y_{kl}-(\mxK g_l)(\vcx_{k-1})).
\end{equation}
One can see that in \eqref{eqn:empirical_loss} the chosen $L$ is the quadratic function $L(e)=e^2$.
For being less sensitive to the outliers, $L$ can be the \emph{Huber loss} function $L_\rho$ defined as
\begin{equation}\label{eqn:Huber}
L_\rho(e) = 
\begin{cases}
\frac{1}{2}e^2, &\text{ if } |e|\le \rho,\\
\rho(|e|-\frac{\rho}{2}), &\text{ otherwise, }\\
\end{cases}
\end{equation}
where $\rho\in\Rbb_+$, or it can be the \emph{pseudo-Huber loss} defined as
\begin{equation}\label{eqn:smoothed-Huber}
	L_\rho(e) = (e^2+\rho^2)^{\frac{1}{2}}-\rho^2,
\end{equation}
which is the smoothed version of Huber loss \eqref{eqn:Huber} \cite{maronna2019robust,hastie2009elements}.
Moreover, to formulate a \emph{robust optimization} problem \cite{el1997robust,ben1998robust} for the learning Koopman operator, one can define the empirical loss function as
\begin{equation}\label{eqn:empirical_loss_robust}
	\Ecal_{\Ucal_{\rho}}(\mxK) := 
	\max_{\Delta\in\Ucal_{\rho}} 
	\sum_{k=1}^{\nS}
	\sum_{l=1}^{\nG}   
	L\big(y_{kl}-(\mxK g_l)(\vcx_{k-1})-\delta_{k,l}\big),
\end{equation}
where 
$\Delta$ is the matrix $\Delta:=[\delta_{k,l}]_{k=1,l=1}^{\nS,\nG}$, $\Ucal_{\rho}\subseteq\Rbb^{\nS\times \nG}$ is a given \emph{uncertainty set}, and, $\rho\in\Rbb_+$ is a real scalar characterizing $\Ucal_{\rho}$. 
For example,  one may employ the following uncertainty set 
\begin{equation}
\Ucal_{\rho} :=\Big\{\Delta\in\Rbb^{\nS\times \nG}\ \!\Big|\ \! \|\Delta\|_{\fro}\le \rho\Big\},
\end{equation}
for which
the empirical loss function \eqref{eqn:empirical_loss_robust} is simplified to
\begin{equation*}\label{eqn:empirical_loss_robust_simp}
	\Ecal_{\Ucal_{\rho}}(\mxK) = 
	\Big[
	\rho + 
	\Big(
	\sum_{k=1}^{\nS}
	\sum_{l=1}^{\nG}   
	\big(y_{kl}-(\mxK g_l)(\vcx_{k-1})\big)^2\Big)^{\frac12}\Big]^2\!\!\!,
\end{equation*}
when $L$ is the quadratic loss.
Similarly, for a \emph{distributionally robust} formulation of the learning problem \cite{esfahani2018data}, one may define the empirical loss as 
\begin{equation*}\label{eqn:empirical_loss_dist_robust}
	\Ecal_{\Pcal_{\rho}}(\mxK)  := \!
	\!\!\sup_{\Delta\sim\PP\in\Pcal_{\rho}} \!\EE_{\PP}\!
	\left[\,
	\sum_{k=1}^{\nS}\!
	\sum_{l=1}^{\nG}  
	L\big(y_{kl}-(\mxK g_l)(\vcx_{k-1})-\delta_{k,l}\big)\!
	\right]\!\!, \!
\end{equation*}
where  $\Pcal_{\rho}$ is a given \emph{ambiguity set} for the probability distributions.
One can see that each of the these empirical loss functions is convex when the loss function $L$ is convex.
Given the set of indices $\Ical\subseteq\{1,\ldots,\nS\}\times\{1,\ldots,\nG\}$ and function $\ell:\Rbb^{|\Ical|}\times \Rbb^{|\Ical|}\to \Rbb_+$, the empirical loss can be generally defined as 
\begin{equation}\label{eqn:empirical_loss_ell}
	\Ecal_{\ell}(\mxK)=\ell\Big(\big[(\mxK g_l)(\vcx_{k-1})\big]_{(k,l)\in\Ical},\big[y_{kl}\big]_{(k,l)\in\Ical}\Big),
\end{equation}
which is a convex function when $\ell(\cdot,\mxY_{\Ical}):\Rbb^{|\Ical|}\to\Rbb$ is convex for $\mxY_{\Ical}:=\big[y_{kl}\big]_{(k,l)\in\Ical}\in \Rbb^{|\Ical|}$.
Note that \eqref{eqn:empirical_loss_ell} generalizes all of the empirical loss functions introduced above.

Given a generic regularization function $\Rcal:\Lcal(\Hcal)\to \Rbb_+\cup\{+\infty\}$ and $\lambda\in\Rbb_+$, the learning problem for the Koopman operator can be defined in the most general form as following
\begin{equation}\label{eqn:reg_learning_koopman_generalized}
	\begin{array}{cl}
		\minOp_{\mxK\in\Fcal} &  \Jcal_{\ell}(\mxK):= \Ecal_{\ell}(\mxK) + \lambda \Rcal(\mxK)\\
		\mathrm{s.t.} &
		\mxK\in\Ccal,	
	\end{array}	
\end{equation}
where $\Fcal$ either denotes $\Lcal(\Hcal)$, or $\Lcal_{\Wcal}$, for a closed subspace $\Wcal\subseteq\Hcal$, and $\Ccal$ is a subset of $\Hcal$.
In the following, we characterize when this learning problem is tractable, i.e., we provide suitable conditions under which a representer theorem holds for \eqref{eqn:reg_learning_koopman_generalized}.

Define $\overline{\Rcal}:\Lcal(\Hcal)\to \Rbb\cup\{+\infty\}$ as 
$\overline{\Rcal} := \lambda \Rcal+\delta_{\Ccal}$,
and let $\Pi_{\Gcal}$ be the projection operator on $\Gcal$ introduced in \eqref{eqn:Gcal}.
For brevity, we unify the notations for both cases of $\Fcal$.
Indeed, when $\Fcal$ is $\Lcal_{\Wcal}$, 
we set $\nZ$ as $\nS$, $z_k$ as $z_k=\Pi_{\Wcal}v_k$, for $k=1,\ldots,\nZ$, and $\Zcal$ as the subspace $\Zcal = \linspan\{z_1,\ldots,z_{\nZ}\}$. 
Moreover, we denote by $\Pi_{\Zcal}$ and $\mxZ$ respectively as the projection operator on $\Zcal$ and the Gramian matrix of vectors $z_1,\ldots,z_{\nZ}$.
By abuse of notation, we use the same letters for $\nS$, $v_1,\ldots,v_{\nS}$, $\Vcal$, $\Pi_{\Vcal}$, and, $\mxV$, for the case $\Fcal=\Lcal(\Hcal)$.
Also, in order to provide arguments analogous to the case of finite dimensional subspace $\Wcal=\linspan\{w_1,\ldots,w_{\nW}\}$ as in Theorem \ref{thm:Tikhonov_reg_case_W_part_2}, we adopt the same notational convention for $\nW$,  $w_1,\ldots,w_{\nW}$, $\Wcal$, $\Pi_{\Wcal}$, and, $\mxW$.
In order to provide a generalized representer theorem for \eqref{eqn:reg_learning_koopman_generalized}, we need the following assumption.
\begin{assumption}\label{ass:Rbar}
		\normalfont
	For any $\mxS\in\Lcal(\Hcal)$, we have 

\begin{equation}\label{eqn:PiFcal_S_PiGcal_le_S}
\overline{\Rcal}(\Pi_{\Zcal}\!\ \mxS\!\ \Pi_{\Gcal}) \le \barRcal(\mxS).
\end{equation}
\end{assumption}
Based on the discussions in Section \ref{sec:learning} for learning the Koopman operator with Tikhonov regularization, one can see that the property \eqref{eqn:PiFcal_S_PiGcal_le_S} holds for \eqref{eqn:reg_learning_koopman_Tikhonov}.
\begin{theorem}[Generalized Representer Theorem]\label{thm:gen_reg_case}
	Let Assumptions \ref{ass:e_xk_bounded} and \ref{ass:Rbar} hold, and $v_1,\ldots,v_{\nS}\in\Hcal$ be the vectors defined in Theorem \ref{thm:Tikhonov_reg_case}. 
	\iftrue
	Suppose that the optimization problem \eqref{eqn:reg_learning_koopman_generalized} admits a  solution. Then,  
	\eqref{eqn:reg_learning_koopman_generalized} has a solution in the following form 
	\begin{equation}\label{eqn:hatK_Z}
		\hatK = \sum_{k=1}^{\nZ} \sum_{l=1}^{\nG} a_{kl}\ z_k \otimes g_l,	
	\end{equation}
	where $a_{kl}\in\Rbb$, for $k=1,\ldots,\nZ$ and $l=1,\ldots,\nG$.
	\else
	Suppose that the optimization problem \eqref{eqn:reg_learning_koopman_generalized} admits a  solution. Then, there exist $a_{kl}$, for $k=1,\ldots,\nZ$ and $l=1,\ldots,\nG$,  such that \eqref{eqn:reg_learning_koopman_generalized} has a solution in following form 
	\begin{equation}\label{eqn:hatK_general}
		\hatK = \sum_{k=1}^{\nS} \sum_{l=1}^{\nG} a_{kl}\ v_k \otimes g_l,	
	\end{equation}
	when  $\Fcal=\Lcal(\Hcal)$, and a solution as
	\begin{equation}\label{eqn:hatK_PiW_general}
		\hatK_{\Wcal} = \sum_{k=1}^{\nS} \sum_{l=1}^{\nG} a_{kl}\ \!(\Pi_{\Wcal}v_k) \otimes g_l,	
	\end{equation}	
	when $\Fcal=\Lcal_{\Wcal}$.
	\fi	
	Moreover, when $\Dcal:=\Fcal\cap\dom(\Rcal)\cap\Ccal$ is a non-empty, closed and convex set, 
	$\ell(\cdot,\mxY_{\Ical}):\Rbb^{|\Ical|}\to\Rbb$ is a convex function for each $\mxY_{\Ical}:=\big[y_{kl}\big]_{(k,l)\in\Ical}\in \Rbb^{|\Ical|}$, 
	$\Rcal$ is convex and lower semicontinuous, and $\barRcal$ is coercive, then, \eqref{eqn:reg_learning_koopman_generalized} admits at least one solution, with the parametric representation in \eqref{eqn:hatK_Z}.
	Additionally, if $\Rcal$ is strictly convex on $\Dcal$, then the solution of \eqref{eqn:reg_learning_koopman_generalized} is unique.
\end{theorem}
\begin{proof}
Define function $\barJcal_{\!\ell}:\Lcal(\Hcal)\to\Rbb\cup\{+\infty\}$ as
\begin{equation}\label{eqn:barJcal}
\barJcal_{\!\ell}(\mxK):= 
\Ecal_{\ell}(\mxK) + \lambda \Rcal(\mxK)+\delta_{\Ccal}(\mxK)+\delta_{\Fcal}(\mxK), 
\end{equation}
for each $\mxK\in\Lcal(\Hcal)$.	
One can see that  the learning problem \eqref{eqn:reg_learning_koopman_generalized} is equivalent to $\min_{\mxK\in\Lcal(\Hcal)}\barJcal_{\!\ell}(\mxK)$.
Let $\mxS$ be a solution of  \eqref{eqn:reg_learning_koopman_generalized} which is clearly a solution for  this optimization problem as well. 
Consider the case $\Fcal$ is $\Lcal(\Hcal)$ and let operator $\hatK$ be defined as $\hatK=\Pi_{\Vcal}\mxS\Pi_{\Gcal}$.
Note that in this case, the last term in \eqref{eqn:barJcal} is zero for $\mxS$ and $\hatK$.
Moreover, similar to the proof of Theorem~\ref{thm:Tikhonov_reg_case}, one can show that  $(\hatK g_l)(\vcx_{k-1}) = (\mxS g_l)(\vcx_{k-1})$, for each 
$k\in[\nS]$ and $l\in[\nG]$. 
Hence, by the definition of $\Ecal_{\ell}$ in \eqref{eqn:empirical_loss_ell}, we see that $\Ecal_{\ell}(\hatK) = \Ecal_{\ell}(\mxS)$. According to Assumption \ref{ass:Rbar}, it follows that
\begin{equation}
\barJcal_{\!\ell}(\hatK)
=
\Ecal_{\ell}(\hatK) + \barRcal(\hatK)
\le
\Ecal_{\ell}(\mxS) + \barRcal(\mxS)
= \barJcal_{\!\ell}(\mxS).
\end{equation}
Therefore, $\hatK$ is a solution of \eqref{eqn:reg_learning_koopman_generalized} as well, and we have $\barJcal_{\!\ell}(\hatK) = \barJcal_{\!\ell}(\mxS)$.
Due to linearity of $\hatK$, one can see that $\hatK$ has a parametric form as in \eqref{eqn:hatK_Z}.
For the case $\Fcal=\Lcal_{\Wcal}$, define operator $\hatK_{\Wcal}$ as $\hatK_{\Wcal}=\Pi_{\Wcal}\mxS\Pi_{\Gcal}$. 
By similar arguments, we can show that  $\hatK_{\Wcal}$ is a solution of \eqref{eqn:reg_learning_koopman_generalized} for which we have
the given parametric representation.

Due to the convexity property of $\ell$, we know that $\ell(\cdot,\mxY_{\Ical}):\Rbb^{|\Ical|}\to\Rbb$ is a continuous function, for $\mxY_{\Ical}=\big[y_{kl}\big]_{(k,l)\in\Ical}\in \Rbb^{|\Ical|}$.
Furthermore, we know that
\begin{equation}
	\Ecal_{\ell}(\mxK)=\ell\Big(\big[
	\inner{\mxK g_l}{\vcv_k} \big]_{(k,l)\in\Ical}, \big[y_{kl}\big]_{(k,l)\in\Ical}\Big),
\end{equation}
which implies that $\Ecal_{\ell}:\Hcal\to\Rbb_+$ is a proper, convex and continuous function.
Moreover, $\Rcal$ is convex and lower semicontinuous.
Note that we have $\dom(\barJcal_{\!\ell})=\Fcal\cap\dom(\Rcal)\cap\Ccal=\Dcal$, which is 
a non-empty, closed and convex set. 
This implies that  $\barJcal_{\!\ell}$ is a proper function, and also $\delta_{\Dcal}$ is  proper, convex and lower semicontinuous.
One can see that $\barJcal_{\!\ell}(\mxK)=\Ecal_{\ell}(\mxK)+\Rcal(\mxK)+\delta_{\Dcal}(\mxK)$, for each $\mxK\in\Hcal$.
Therefore, $\barJcal_{\!\ell}$ is proper, convex and lower semicontinuous. Moreover, it is strictly convex, when $\Rcal$ is a strictly convex function.
From non-negativity of $\Ecal_{\ell}$ and $\delta_{\Dcal}$, and, due to $\barJcal_{\!\ell}(\mxK)=\Ecal_{\ell}(\mxK)+\barRcal(\mxK)+\delta_{\Dcal}(\mxK)$, for each $\mxK\in\Hcal$, it follows that $\barJcal_{\!\ell}$ is coercive.
Therefore, \eqref{eqn:reg_learning_koopman_generalized} admits at least one solution, which is unique when $\Rcal$ is strictly convex 
\cite{peypouquet2015convex}.
\end{proof}	
\begin{remark}\label{rem:variational_principle}
\normalfont
Let $\Dcal:=\Fcal\cap\dom(\Rcal)\cap\Ccal$, and, define $\Jcal_{\!\ell}:\Dcal\to\Rbb$ as the restriction of function $\barJcal_{\!\ell}$ introduced \eqref{eqn:barJcal} to $\Dcal$, i.e., $\Jcal_{\!\ell} = \barJcal_{\!\ell}|_{\Dcal}$.
Due to \emph{variational principle} \cite{brezis2010functional}, if $\Dcal$ is a non-empty and weakly sequentially closed set, and  $\Jcal_{\!\ell}:\Dcal\to\Rbb$ is weakly sequentially lower semicontinuous and coercive, then,  there exists $\mxK\in\Dcal$ such that $\Jcal_{\!\ell}(\mxK)=\inf_{\mxS\in\Dcal}\Jcal_{\!\ell}(\mxS)$, i.e., \eqref{eqn:reg_learning_koopman_generalized} admits at least one solution. 
Note that here no convexity assumption is required, and hence, these conditions are more general.  
\end{remark}
\begin{theorem}\label{thm:extending_ass_2}
{\rm (i)}	Let $\lambda\in\Rbb_+$,  $\Rcal:\Lcal(\Hcal)\to \Rbb_+\cup\{+\infty\}$ be a regularization function, and $\Ccal_{\alpha}\subseteq\Hcal$, for $\alpha\in\Acal$, be given sets. Define $\Ccal:=\cap_{\alpha\in\Acal}\Ccal_{\alpha}$ and $\barRcal_{\alpha}:=\lambda\Rcal+\delta_{\Ccal_{\alpha}}$, for $\alpha\in\Acal$. 
If Assumption \ref{ass:Rbar} is satisfied by  $\barRcal_{\alpha}$, for each $\alpha\in\Acal$, then $\barRcal:=\lambda\Rcal+\delta_{\Ccal}$ also satisfies Assumption \ref{ass:Rbar}.\\
{\rm (ii)} Let $\Rcal_1,\ldots,\Rcal_m:\Lcal(\Hcal)\to \Rbb_+\cup\{+\infty\}$ be given regularization functions, $\lambda_1,\ldots,\lambda_m\in\Rbb_+$ and $\Ccal$ be a subset of $\Hcal$.  
If $\barRcal_{i}=\lambda_i\Rcal_i+\delta_{\Ccal}$ satisfies Assumption \ref{ass:Rbar}, for $i=1,\ldots,m$, then $\barRcal:=\sum_{i=1}^m\lambda_i\Rcal_i+\delta_{\Ccal}$ also satisfies Assumption \ref{ass:Rbar}.
\end{theorem}	
\begin{proof}
{\rm (i)} For $\mxS\in\Lcal(\Hcal)$, when $\mxS\notin\Ccal$ or $\lambda\Rcal(\mxS)=+\infty$, Assumption \ref{ass:Rbar} holds trivially. Hence, we assume that $\lambda\Rcal(\mxS)$ is finite and $\mxS\in\Ccal$, i.e., $\delta_{\Ccal}(\mxS)=0$. Therefore, for each $\alpha$, we know that $\mxS\in\Ccal_\alpha$,  and subsequently, $\delta_{\Ccal_{\alpha}}(\mxS)=0$. Since Assumption \ref{ass:Rbar} holds for $\barRcal_{\alpha}$, we have
\begin{equation*}
\lambda\Rcal(\Pi_{\Zcal}\mxS\Pi_{\Gcal}) + \delta_{\Ccal_{\alpha}}(\Pi_{\Zcal}\mxS\Pi_{\Gcal})
\le
\lambda\Rcal(\mxS) + \delta_{\Ccal_{\alpha}}(\mxS) = \lambda\Rcal(\mxS) ,	
\end{equation*}
which implies that $\Pi_{\Zcal}\mxS\Pi_{\Gcal} \in\Ccal_{\alpha}$ and $\lambda\Rcal(\Pi_{\Zcal}\mxS\Pi_{\Gcal})\le \lambda\Rcal(\Pi_{\Zcal}\mxS\Pi_{\Gcal})$.
Therefore, we have $\Pi_{\Zcal}\mxS\Pi_{\Gcal} \in\Ccal$, and $\delta_{\Ccal}(\Pi_{\Zcal}\mxS\Pi_{\Gcal})=0$. Subsequently, it follows that $\barRcal(\Pi_{\Zcal}\mxS\Pi_{\Gcal})\le \barRcal(\mxS)$. Hence, Assumption \ref{ass:Rbar} is satisfied by $\barRcal$. 
\\
{\rm (ii)} It is easy to check that
\begin{equation*}
\barRcal(\mxS)=\sum_{i=1}^m\lambda_i\Rcal_i(\mxS)+\delta_{\Ccal}(\mxS)=\sum_{i=1}^m\left(\lambda_i\Rcal_i(\mxS)+\delta_{\Ccal}(\mxS)\right),
\end{equation*}
for any $\mxS\in\Lcal(\Hcal)$. 
Based on this fact, the proof is straightforward.
\end{proof}	
\begin{remark}
\normalfont
Theorem~\ref{thm:extending_ass_2} allows utilizing the result of Theorem~\ref{thm:gen_reg_case} to the case where different constraints and regularization terms are considered together. 	
\end{remark}
In the following, we provide applications of the above theorems.
\subsection{Learning Koopman Operator with Frobenius Norm Regularization}
\iftrue
\else
\ \\\ \\\ \\\ \\\ \\\ \\ \\\ \\\ \\\ \\\ \\
\ \\\ \\\ \\\ \\\ \\\ \\ \\\ \\\ \\\ \\\ \\
\ \\\ \\\ \\\ \\\ \\\ \\ \\\ \\\ \\\ \\\ \\
\ \\\ \\\ \\\ \\\ \\\ \\ \\\ \\\ \\\ \\\ \\
\fi 
In Section \ref{sec:learning}, the regularization function utilized is based on the operator norm.
On the other hand, one may propose employing the Hilbert-Schmidt or Frobenius norm of the Koopman operator to define the regularization term, i.e., $\Rcal:\Lcal(\Hcal)\to\Rbb_+\cup\{+\infty\}$ is defined such that, for each $\mxK\in\Lcal(\Hcal)$, we have $\Rcal(\mxK) = \|\mxK\|_{\fro}^2:=\trace(\mxK^*\mxK)$, where $\mxK^*$ is the adjoint of $\mxK$ and $\trace(\cdot)$ denotes the trace function.
Accordingly, the learning problem is formulated as following
\begin{equation}\label{eqn:reg_learning_koopman_Frobenius}
	\minOp_{\mxK\in\Fcal}\   
	\Ecal(\mxK)
	+ \lambda \|\mxK\|_{\fro}^2,\\
\end{equation}
where $\Ecal(\mxK)$ is the empirical loss defined in \eqref{eqn:empirical_loss}.
Before we proceed further, we recall that, given an orthonormal basis $\{b_k\}_{k=1}^\infty$ for $\Hcal$, the trace of operator $\mxS\in \Lcal_{\Wcal}$ is defined as 
\begin{equation}\label{eqn:def_trace}
	\trace(\mxS) := \sum_{k=1}^{\infty}\inner{b_k}{\mxS b_k},
\end{equation}
when the summation converges \cite{brezis2010functional}.
Based on \eqref{eqn:def_trace}, we have
\begin{equation}\label{eqn:def_fro_norm}
	\Rcal(\mxK) = \|\mxK\|_{\fro}^2= \sum_{k=1}^{\infty}\inner{b_k}{\mxK^*\mxK b_k} = \sum_{k=1}^{\infty}\|\mxK b_k\|^2.
\end{equation}
Note that the left-hand sides of \eqref{eqn:def_trace} and \eqref{eqn:def_fro_norm} are independent of the choice of orthonormal basis  $\{b_k\}_{k=1}^\infty$.

The next theorem characterizes the solution of learning problem \eqref{eqn:reg_learning_koopman_Frobenius}.
\begin{theorem}
	\label{thm:HS_reg_case}
	Let Assumption \ref{ass:e_xk_bounded} hold and $\lambda>0$.
	Then, the optimization problem \eqref{eqn:reg_learning_koopman_Frobenius} has a \emph{unique} solution with parametric representation as in \eqref{eqn:hatK_Z},
	where $\mxA=[a_{kl}]_{k=1,l=1}^{\nZ,\!\ \nG}\in\Rbb^{\nZ\times\nG}$ is the solution of the following quadratic program
	\begin{equation}\label{eqn:fro_ref_finite}
		\min_{\mxA\in \Rbb^{\nZ\times\nG}}\ \|\mxZ\mxA\mxG-\mxY\|_{\fro}^2 \ + \ \lambda \|\mxZ^{\frac{1}{2}}\mxA\mxG^{\frac{1}{2}}\|_{\fro}^2.
	\end{equation}
\end{theorem}
\begin{proof}
We know that $\mxK=0$ is a feasible point for \eqref{eqn:reg_learning_koopman_Frobenius}. Therefore, for the optimal solution of \eqref{eqn:reg_learning_koopman_Frobenius}, we need to have
\begin{equation}
\lambda\|\hatK\|_{\fro}^2\le 
\Ecal(\hatK) + \lambda \|\hatK\|_{\fro}^2 \le 
\Ecal(0) + \lambda \|0\|_{\fro}^2 = \|\mxY\|_{\fro}^2.
\end{equation} 
Accordingly, \eqref{eqn:reg_learning_koopman_Frobenius} is equivalent to the following problem
\begin{equation}\label{eqn:reg_learning_koopman_Frobenius_C}
	\begin{array}{cl}
		\minOp_{\mxK\in\Fcal} & \Ecal(\mxK)+ \lambda \|\mxK\|_{\fro}^2,\\
		\mathrm{s.t.} &
		\mxK\in\Ccal,	
	\end{array}	
\end{equation}
where $\Ccal$ is defined as 
\begin{equation}\label{eqn:Ccal_fro_lambda}
	\Ccal:=\Big\{\mxS\in\Lcal(\Hcal)\ \!\Big|\ \!\|\mxS\|_{\fro}\le \frac{1}{\sqrt{\lambda}}\|\mxY\|_{\fro}\Big\}.
\end{equation}
Let $\mxS\in\Lcal(\Hcal)$, and $\{b_k\}_{k=1}^\infty$ be an orthonormal basis for $\Hcal$ such that $\Gcal = \linspan\{b_1,\ldots,b_{\barnG}\}$, where $\barnG\le \nG$. 
If $k\le\barnG$, we have $\|\Pi_{\Zcal}\mxS\Pi_{\Gcal}b_k\| = \|\Pi_{\Zcal}\mxS b_k\|$, and since $\|\Pi_{\Zcal}\|\le 1$, it follows that $\|\Pi_{\Zcal}\mxS\Pi_{\Gcal}b_k\| \le \| \mxS b_k\|$.
If $k>\barnG$, then $\|\Pi_{\Zcal}\mxS\Pi_{\Gcal} b_k\| = 0$.
Therefore, due to the definition of Frobenius norm, we have 
\begin{equation*}
\begin{split}
	\|\Pi_{\Zcal}\!\ \mxS\!\ \Pi_{\Gcal}\|_{\fro}^2 
	&=
	\sum_{k=1}^\infty \|\Pi_{\Zcal}\!\ \mxS\!\ \Pi_{\Gcal}\!\ b_k\|^2
	\le 
	\sum_{k=1}^\infty \|\mxS\!\ b_k\|^2
	=
	\|\mxS\|_{\fro}^2.
\end{split}
\end{equation*}
Thus, for any $\mxS\in\Lcal(\Hcal)$, we have 
$\Rcal(\Pi_{\Zcal}\mxS\Pi_{\Gcal})=\|\Pi_{\Zcal}\mxS\Pi_{\Gcal}\|_{\fro}^2\le\|\mxS\|_{\fro}^2=\Rcal(\mxS)$.
Moreover, for each $\mxS\in\Ccal$, one can see that
\begin{equation}
\|\Pi_{\Zcal}\mxS\Pi_{\Gcal}\|_{\fro}\le\|\mxS\|_{\fro}
\le
\frac{1}{\sqrt{\lambda}}\|\mxY\|_{\fro},
\end{equation}
which implies that $\Pi_{\Zcal}\mxS\Pi_{\Gcal}\in \Ccal$.
Therefore, $\delta_{\Ccal}(\Pi_{\Zcal}\mxS\Pi_{\Gcal})\le\delta_{\Ccal}(\mxS)$, and hence, Assumption \ref{ass:Rbar} holds.
By definition, we know that $\Ccal\subset\dom(\Rcal)$ and $0\in\Dcal:=\Fcal\cap\dom(\Rcal)\cap\Ccal=\Fcal\cap\Ccal$.
Since $\|\cdot\|_{\fro}$ is a norm on $\dom(\Rcal)$ and $\Ccal\subset\dom(\Rcal)$,  we have that $\Ccal$ is a convex set. 
Let $\{\mxS_n\}_{n\in\Nbb}\subset\Lcal(\Hcal)$ be a sequence such that 
$\lim_{n\to\infty}\mxS_n=\mxS\in\Lcal(\Hcal)$ in norm topology.
For any $k\in\Nbb$, we know that $\lim_{n\to\infty}\|\mxS_nb_k\|^2=\|\mxS b_k\|^2$.
From Fatou's lemma, it follows that
\begin{equation}\label{eqn:fatuo_lemma_fro}
\begin{split}
\|\mxS\|_{\fro}^2=\sum_{k=1}^\infty&\|\mxS b_k\|^2 =\sum_{k=1}^\infty \liminf_{n}\|\mxS_n b_k\|^2 
\\&\le 
\liminf_{n}\sum_{k=1}^\infty \|\mxS_n b_k\|^2 
=
\liminf_{n}\|\mxS_n\|_{\fro}^2. 
\end{split}
\end{equation} 
Therefore, we have $\Rcal(\mxS)\le\liminf_{n}\Rcal(\mxS_n)$, and $\Rcal$ is lower semicontinuous.
Moreover, if  $\{\mxS_n\}_{n\in\Nbb}\subset\Ccal$, then $\|\mxS_n\|_{\fro}\le \frac{1}{\sqrt{\lambda}}\|\mxY\|_{\fro}$, for each $n$, and subsequently, due to \eqref{eqn:fatuo_lemma_fro}, we have $\|\mxS\|_{\fro}\le \frac{1}{\sqrt{\lambda}}\|\mxY\|_{\fro}$ which implies that $\mxS\in\Ccal$.
Hence, $\Ccal$ and $\Dcal=\Fcal\cap\Ccal$ are non-empty, closed and convex sets.
Due to $\|\mxS\|\le\|\mxS\|_{\fro}$, we know that $\Rcal$ as well as $\barRcal$ are coercive.  
Furthermore, from Lemma \ref{lem:Fro_norm_T_strongly_convex}, we have that
$\Rcal$ is strictly convex.
Therefore, due to Theorem \ref{thm:gen_reg_case}, \eqref{eqn:reg_learning_koopman_Frobenius_C} admits a unique solution with the parametric form in \eqref{eqn:hatK_Z}.
This solution coincides with the unique solution of \eqref{eqn:reg_learning_koopman_Frobenius} due to the equivalency of the corresponding programs.   
Since for any $h_1,h_2\in\Hcal$, we have $(h_1\otimes h_2)^*=h_2\otimes h_1$, 
for $\hatK$ in the given parametric form, we know that 
$\hatK^* = \sum_{k=1}^{\nZ} \sum_{l=1}^{\nG} a_{kl}\ \! g_l \otimes z_k$.
One can easily see that
\begin{equation}\label{eqn:h1xh2_h3xh4}
(h_1\otimes h_2)(h_3\otimes h_4) = \inner{h_2}{h_3} (h_1\otimes h_4),
\end{equation}
for any $h_1,h_2,h_3,h_4\in\Hcal$.
Accordingly, we have
\begin{equation}\label{eqn:T*T}
\begin{split}	
	\hatK^*\hatK
	&=
	\sum_{k=1}^{\nZ} \sum_{j=1}^{\nG}
	\sum_{l=1}^{\nG}\sum_{i=1}^{\nZ} 
	a_{kl}\ a_{ij}\ (g_l \otimes z_k)(z_i \otimes g_j)
	\\
	&=
	\sum_{j=1}^{\nG}\sum_{l=1}^{\nG}(g_l \otimes g_j)
	\sum_{k=1}^{\nZ} \sum_{i=1}^{\nZ} 
	a_{kl}\ \! \inner{z_k}{z_i}\ \! a_{ij}
	\\
	&=
	\sum_{j=1}^{\nG}\sum_{l=1}^{\nG}(g_l \otimes g_j)
	[\mxA^\tr\mxZ\mxA]_{(l,j)}.
\end{split}
\end{equation}
From $\|\hatK\|_{\fro}^2 = \trace(\hatK^*\hatK)$, and due to $\trace(g_l \otimes g_j)=\inner{g_l}{g_j}$, for each 
$j,l\in[\nG]$, 
it follows from \eqref{eqn:T*T} that
\begin{equation}
\begin{split}
&\|\hatK\|_{\fro}^2 = \sum_{j=1}^{\nG}\sum_{l=1}^{\nG}\inner{g_l}{g_j}
[\mxA^\tr\mxZ\mxA]_{(l,j)} 
\\&\
= \trace(\mxG\mxA^\tr\mxZ\mxA)		
=\trace(\mxG^{\frac{1}{2}}\mxA^\tr\mxZ^{\frac{1}{2}}\mxZ^{\frac{1}{2}}\mxA\mxG^{\frac{1}{2}})=\|\mxZ^{\frac{1}{2}}\mxA\mxG^{\frac{1}{2}}\|_{\fro}^2.\!\!\!\!\!\!
\end{split}
\end{equation}
Similar to the proof of Theorem \ref{thm:Tikhonov_reg_case}, we can show that $\Ecal(\hatK)=\|\mxZ\mxA\mxG-\mxY\|_{\fro}^2$. Replacing these terms, we obtain optimization \eqref{eqn:fro_ref_finite}. This concludes the proof.
\end{proof}
Define $J_{\fro}:\Rbb^{\nZ\times\nG}\to\Rbb_+$ as the objective function in \eqref{eqn:fro_ref_finite}, i.e., for any $\mxA\in\Rbb^{\nZ\times\nG}$, we have 
\begin{equation}\label{eqn:J_fro}
J_{\fro}(\mxA) = \|\mxZ\mxA\mxG-\mxY\|_{\fro}^2 \ + \ \lambda \|\mxZ^{\frac{1}{2}}\mxA\mxG^{\frac{1}{2}}\|_{\fro}^2. 	
\end{equation}
The first derivative of $J_{\fro}$ is
\begin{equation}\label{eqn:grad_J_fro}
\frac{1}{2}\nabla_{\!\mxA}J_{\fro}(\mxA) = \mxZ^2\mxA\mxG^2 +\lambda\mxZ\mxA\mxG-\mxZ\mxY\mxG.
\end{equation}
To solve \eqref{eqn:fro_ref_finite}, we can use the first order necessary condition $\nabla_{\!\mxA}J_{\fro}(\mxA)=0$, which is a linear system of equation with respect to $\mxA$. Indeed, $\nabla_{\!\mxA}J_{\fro}(\mxA)=0$ is a \emph{generalized Sylvester equation}, which can be solved efficiently. Also, using \eqref{eqn:grad_J_fro}, one can employ iterative schemes such as BFGS.
 
\begin{remark}\normalfont
According to the discussion above, one can see that employing Frobenius norm for the regularization is less computationally demanding compared to the case of regularization with operator norm.
Accordingly, one may prefer learning Koopman operator based on \eqref{eqn:reg_learning_koopman_Frobenius} rather than \eqref{eqn:reg_learning_koopman_Tikhonov}.
\end{remark}
\begin{remark} \label{rem:lim_K_lambda_Frobenius}
\normalfont
For $\lambda>0$ and $\Fcal=\Lcal_{\Gcal}$, let $\hatK_{\Gcal,\lambda}$ denotes the unique solution of the learning problem \eqref{eqn:reg_learning_koopman_Frobenius}.
Then, similar to Theorem \ref{thm:lim_K_lambda}, one can show that $\lim_{\lambda \downarrow 0} \hatK_{\Gcal,\lambda} = \hatK_{\mxU}$ and $\lim_{\lambda \to \infty} \hatK_{\Gcal,\lambda} = 0$, both in Frobenius norm and operator norm topology.
\end{remark}

\subsection{Learning Koopman Operator with Rank Constraint}
Learning a low rank operator can be relevant when a reduced version of Koopman operator is of interest, possibly for a model reduction of the system  \cite{peitz2019koopman,klus2020data}.
Accordingly, one may introduce a rank constraint in the learning problem as 
\begin{equation}\label{eqn:Ccal_rank}
    \Ccal:=\Big\{\mxS\in\Lcal(\Hcal)
    \ \!\Big|\ \! 
    \rank(\mxS):=\dim(\mxS(\Hcal))\le r\Big\},	
\end{equation}
where $r\in\Zbb_+$ is a given bound on the rank of Koopman operator.
The resulting learning problem is as following
\begin{equation}\label{eqn:reg_learning_koopman_rank}
\begin{array}{cl}
	\minOp_{\mxK\in\Fcal} & \Ecal(\mxK)\\
	\mathrm{s.t.} &
	\mxK\in\Ccal,	
\end{array}	
\end{equation}
where $\Ecal(\mxK)$ is the empirical loss defined in \eqref{eqn:empirical_loss}.
\begin{theorem}\label{thm:rank_const_case}
	Under Assumption \ref{ass:e_xk_bounded}, if the optimization problem \eqref{eqn:reg_learning_koopman_rank} admits a solution,  it has a solution with parametric form given in \eqref{eqn:hatK_Z},
	where $\mxA=[a_{kl}]_{k=1,l=1}^{\nZ,\!\ \nG}\in\Rbb^{\nZ\times\nG}$ is the solution of the following problem
	\begin{equation}\label{eqn:rank_const_finite}
		\begin{array}{cl}	
			\minOp_{\mxA\in \Rbb^{\nZ\times\nG}}&\ \|\mxZ\mxA\mxG-\mxY\|_{\fro}^2,\\
			\mathrm{s.t.} & \ \rank(\mxZ\mxA\mxG)\le r.
		\end{array}
	\end{equation}
\end{theorem}
\begin{proof}
	We know that \eqref{eqn:reg_learning_koopman_rank} is a special case of \eqref{eqn:reg_learning_koopman_generalized} where $\Rcal\equiv 0$, and $\Ccal$ is given in \eqref{eqn:Ccal_rank}. Accordingly, we have $\barRcal = \delta_{\Ccal}$.
	For any operator $\mxS\in\Hcal$, we know that
	\begin{equation*}\begin{split}
			\rank(\Pi_{\Zcal}\mxS\Pi_{\Gcal})
			&\le \min\Big\{\rank(\Pi_{\Zcal}),\rank(\mxS),\rank(\Pi_{\Gcal})\Big\}
			\\&\le\ \rank(\mxS).		
		\end{split}
	\end{equation*}
	Hence, $\mxS\in\Ccal$ implies that $\Pi_{\Zcal}\mxS\Pi_{\Gcal}\in\Ccal$.
    Subsequently, we have 
	\begin{equation}
	\barRcal(\Pi_{\Zcal}\mxS\Pi_{\Gcal}) = 	\delta_{\Ccal}(\Pi_{\Zcal}\mxS\Pi_{\Gcal})\le \delta_{\Ccal}(\mxS) = \barRcal(\mxS).
	\end{equation}
   	Therefore, the Assumption \ref{ass:Rbar} holds for \eqref{eqn:reg_learning_koopman_rank}.
	Accordingly, due to Theorem \ref{thm:gen_reg_case}, if \eqref{eqn:reg_learning_koopman} admits a solution, it has also a solution $\hatK$ with parametric form given in \eqref{eqn:hatK_Z}. 
	By taking orthonormal bases for $\Zcal$ and $\Gcal$, one can easily show that $\rank(\hatK) = \rank(\mxZ\mxA\mxG)$.
	More precisely, let $\{b_1,\ldots,b_{\barnG}\}$ and $\{p_1,\ldots,p_{\barnZ}\}$ be two sets of orthonormal vectors such that $\Gcal=\linspan\{b_1,\ldots,b_{\barnG}\}$ and $\Zcal=\linspan\{p_1,\ldots,p_{\barnZ}\}$.
	Then, for each 
    $l\in[\nG]$ and $k\in [\nZ]$, 
    there exist $e_{l1},\ldots, e_{l\barnG}$ and $q_{k1},\ldots, q_{k\barnZ}$ such that we have
	$g_l=\sum_{j=1}^{\barnG} e_{lj}b_j$ and $z_k=\sum_{i=1}^{\barnG} q_{ki}p_i$.
	Accordingly, one can see that
	\begin{equation}\label{eqn:hatK_expansion_orthonormal}
		\begin{split}	
		\hatK
			&=
			\sum_{i=1}^{\barnZ} \sum_{j=1}^{\barnG}
			\left(
			\sum_{k=1}^{\nZ}\sum_{l=1}^{\nG} 
			q_{ki}\ \! a_{kl}\ \! e_{lj}
			\right)
			(p_i \otimes b_j)
			\\
			&=
			\sum_{i=1}^{\barnZ} \sum_{j=1}^{\barnG}
			[\mxQ^\tr\mxA\mxE]_{(i,j)}
			(p_i \otimes b_j),
		\end{split}
	\end{equation}
	where matrices $\mxE$ and $\mxQ$ are defined respectively as $\mxE=[e_{lj}]_{l=1,j=1}^{\nG,\barnG}$ and
	$\mxQ=[q_{ki}]_{k=1,i=1}^{\nZ,\barnZ}$.
	From \eqref{eqn:hatK_expansion_orthonormal}, we know that $\rank(\hatK)=\rank(\mxQ^\tr\mxA\mxE)$.
	Moreover, we can write the Gramian matrices $\mxG$ and $\mxZ$ respectively as $\mxG=\mxE\mxE^\tr$ and $\mxZ=\mxQ\mxQ^\tr$.
	Accordingly, due to Lemma \ref{lem:rank_ABBT} in Appendix \ref{sec:appendix_lemmas}, we have
	\begin{equation}
	\begin{split}
	\rank(\hatK)&=\rank(\mxQ^\tr\mxA\mxE)
	=
	\rank(\mxQ^\tr\mxA\mxE\mxE^\tr)
	\\&=
	\rank(\mxQ\mxQ^\tr\mxA\mxE\mxE^\tr)=
	\rank(\mxZ\mxA\mxG).	
	\end{split}
	\end{equation}
	Based on similar arguments as in the proof of Theorem \ref{thm:Tikhonov_reg_case}, one can show that $\Ecal(\hatK)=\|\mxZ\mxA\mxG-\mxY\|_{\fro}^2$. Replacing these terms, we obtain optimization \eqref{eqn:rank_const_finite}. This concludes the proof. 
\end{proof}	
\begin{remark} \normalfont
	For the learning problem \eqref{eqn:reg_learning_koopman_rank}, the conditions provided in	Theorem~\ref{thm:gen_reg_case} or Remark~\ref{rem:variational_principle} for guaranteeing the existence or uniqueness of the solution are not satisfied, e.g., function $\barRcal$ is not coercive.
\end{remark}

By the change of variable $\mxB= \mxZ\mxA\mxG$, the optimization problem \eqref{eqn:rank_const_finite} can be modified to
\begin{equation}\label{eqn:rank_const_finite_B}
	\begin{array}{cl}	
		\minOp_{\mxB\in \Rbb^{\nZ\times\nG}}&\ \|\mx{B}-\mxY\|_{\fro}^2,\\
		\mathrm{s.t.} & \ \rank(\mx{B})\le \tilde{r},
	\end{array}
\end{equation}
where  $\tilde{r}$ is defined as $\tilde{r} = \min\{r,\rank(\mxG),\rank(\mxZ)\}$.
Following this, one can use Eckart-Young-Mirsky theorem \cite{eckart1936approximation} to solve \eqref{eqn:rank_const_finite_B}. 
More precisely, let  $\mxY = \underline{\mxU}\Sigma\overline{\mxU}^\tr$ be the singular value decomposition of matrix $\mxY$, $\Sigma_1=\diag(\sigma_1,\ldots,\sigma_{\tilde{r}})$
be the diagonal matrix containing first $\tilde{r}$ largest singular values of $\mxY$, and,  $\underline{\mxU}_1$ and $\overline{\mxU}_1$ be respectively the matrices containing first $\tilde{r}$ columns of $\underline{\mxU}$ and $\overline{\mxU}$. Then, $\mxB=\underline{\mxU}\Sigma_1\overline{\mxU}^\tr$ is a solution of \eqref{eqn:rank_const_finite_B}, and subsequently, we can obtain $\mxA$ by solving $\mxZ\mxA\mxG = \underline{\mxU}\Sigma_1\overline{\mxU}^\tr$ for $\mxA$.

\subsection{Learning Koopman Operator with Nuclear Norm Regularization}
In various learning problems such as collaborative filtering, nuclear norm regularization is employed to penalize the complexity of the model \cite{ji2009accelerated}.
Indeed, the nuclear norm is interpreted as a convex relaxation of rank \cite{recht2010guaranteed}.
Similar to the matrices, the nuclear norm of operator $\mxS\in\Lcal(\Hcal)$ is defined as 
$\|\mxS\|_*=\trace(|\mxS|)$, where $|\mxS|$ denotes the square root of $\mxS$, i.e., $|\mxS|$ is a non-negative operator such that $|\mxS|^2=\mxS^*\mxS$. 
Therefore, due to \eqref{eqn:def_trace}, given an orthonormal basis  $\{b_k\}_{k=1}^\infty$, we have
\begin{equation}\label{eqn:def_trace_norm}
	\|\mxS\|_* := \sum_{k=1}^{\infty}\inner{b_k}{|\mxK| b_k},
\end{equation}
when the summation in \eqref{eqn:def_trace_norm} converges. Furthermore, it is known that
\begin{equation}\label{eqn:def_trace_norm_infC}
\|\mxK\|_* = \sup\Big\{|\trace(\mxC\mxK)|\ \! \Big|\ \!  \mxC\in\Kcal(\Hcal),\|\mxC\|\le 1\Big\},
\end{equation} 
where $\Kcal(\Hcal)$ denotes the space of compact operators on $\Hcal$ \cite{conway2019course}.
Considering nuclear norm of the Koopman operator as the regularization term, we have the following learning problem
\begin{equation}\label{eqn:reg_learning_koopman_nuc}
		\minOp_{\mxK\in\Fcal} \ \Ecal(\mxK)+\lambda\|\mxK\|_*,\\
\end{equation}
where $\lambda>0$ and $\Ecal(\mxK)$ is the empirical loss defined in \eqref{eqn:empirical_loss}.
\begin{theorem}
	\label{thm:Tr_reg_case}
	Under Assumption \ref{ass:e_xk_bounded}, the optimization problem \eqref{eqn:reg_learning_koopman_nuc} admits a solution  
	$\hatK$  with parametric form \eqref{eqn:hatK_Z},
	where $\mxA=[a_{kl}]_{k=1,l=1}^{\nZ,\!\ \nG}$ is the solution of  following convex program
	\begin{equation}\label{eqn:nuc_reg_finite}
	\min_{\mxA\in \Rbb^{\nZ\times\nG}}\ \|\mxZ\mxA\mxG-\mxY\|_{\fro}^2 + \lambda \|\mxZ^{\frac12}\mxA\mxG^{\frac12}\|_*\,.
	\end{equation}
\end{theorem}
\iftrue
\begin{proof}
See Appendix \ref{sec:appendix_proof_thm_12}.
\end{proof}
\else
\begin{proof}
For program \eqref{eqn:reg_learning_koopman_nuc}, $\mxK=0$ is a feasible solution. Therefore, for the optimal solution of \eqref{eqn:reg_learning_koopman_nuc}, the following inequality holds
\begin{equation}\label{eqn:proof_thm_Tr_reg_case_01}
	\lambda\|\hatK\|_*\le 
	\Ecal(\hatK) + \lambda \|\hatK\|_* \le 
	\Ecal(0) + \lambda \|0\|_* = \|\mxY\|_{\fro}^2.
\end{equation} 
By virtue of \eqref{eqn:proof_thm_Tr_reg_case_01}, we define $\Ccal$ as 
\begin{equation}\label{eqn:Ccal_nuc_lambda}
	\Ccal:=\Big\{\mxS\in\Lcal(\Hcal)\ \!\Big|\ \!\|\mxS\|_*\le \frac{1}{\lambda}\|\mxY\|_{\fro}\Big\},
\end{equation}
and, introduce the following constrained problem
\begin{equation}\label{eqn:reg_learning_koopman_nuc_C}
	\begin{array}{cl}
		\minOp_{\mxK\in\Fcal} & \Ecal(\mxK)+ \lambda \|\mxK\|_*,\\
		\mathrm{s.t.} &
		\mxK\in\Ccal,	
	\end{array}	
\end{equation}
which is equivalent to \eqref{eqn:reg_learning_koopman_nuc}. 
Let $\mxS\in\Lcal(\Hcal)$, and $\{b_k\}_{k=1}^\infty$ be an orthonormal basis for $\Hcal$ such that $\Gcal = \linspan\{b_1,\ldots,b_{\barnG}\}$.
Accordingly, for compact operator $\mxC$ with $\|\mx{C}\|\le 1$, we have
\begin{equation}
\begin{split}
	\big|\trace(\mx{C}\Pi_{\Zcal}\!\ \mxS\!\ \Pi_{\Gcal})\big|
	&= 
	\bigg|\sum_{k=1}^\infty \inner{\mx{C}\Pi_{\Zcal}\!\ \mxS\!\ \Pi_{\Gcal}\!\ b_k}{b_k}\bigg|
	\\&=
	\bigg|\sum_{k=1}^{\barnG} \inner{\mx{C}\Pi_{\Zcal}\!\ \mxS\!\  b_k}{b_k}\bigg|.
\end{split}
\end{equation}
Note that $\mx{C}\Pi_{\Zcal}$ is a compact operator and $\|\mx{C}\Pi_{\Zcal}\|\le 1$.
Therefore, due to \eqref{eqn:def_trace_norm_infC}, we have
\begin{equation}
	\begin{split}
		\|\Pi_{\Zcal}\!\ \mxS\!\ \Pi_{\Gcal}\|_* 
		&\le 
		\sup_{\|\mx{C}\|\le 1, \mx{C}\in\Kcal(\Hcal)}
		\bigg|\sum_{k=1}^{\barnG} \inner{\mx{C} \mxS\!\  b_k}{b_k}\bigg|
		\\&=
		\sup_{\|\mx{C}\|\le 1, \mx{C}\in\Kcal(\Hcal)}
		\bigg|\sum_{k=1}^{\infty} \inner{\Pi_{\Gcal}\mx{C} \mxS\!\  b_k}{b_k}\bigg|
	\end{split}
\end{equation}
Since $\Pi_{\Gcal}\mx{C}$ is a compact operator and $\|\Pi_{\Gcal}\mx{C}\|\le 1$,
from \eqref{eqn:def_trace_norm_infC}, we have
\begin{equation}\begin{split}
		\|\Pi_{\Zcal}\!\  \mxS\!\ \Pi_{\Gcal}\|_* 
		&\le
		\sup_{\|\mx{C}\|\le 1, \mx{C}\in\Kcal(\Hcal)}
		\bigg|\sum_{k=1}^{\infty} \inner{\mx{C} \mxS\!\  b_k}{b_k}\bigg| 
		\\&=
		\sup_{\|\mx{C}\|\le 1, \mx{C}\in\Kcal(\Hcal)}
		\big|\trace(\mx{C}\mxS)\big| = \|\mxS\|_*. 
\end{split}\end{equation}
Therefore, for any $\mxS\in\Lcal(\Hcal)$, one can see that 
$\Rcal(\Pi_{\Zcal}\mxS\Pi_{\Gcal})=\|\Pi_{\Zcal}\mxS\Pi_{\Gcal}\|_*\le\|\mxS\|_*=\Rcal(\mxS)$.
Moreover, for $\mxS\in\Ccal$, we have
\begin{equation}
	\|\Pi_{\Zcal}\mxS\Pi_{\Gcal}\|_*\le\|\mxS\|_*
	\le
	\frac{1}{\lambda}\|\mxY\|_{\fro},
\end{equation}
and, subsequently, it follows that $\Pi_{\Zcal}\mxS\Pi_{\Gcal}\in \Ccal$.
Therefore, $\delta_{\Ccal}(\Pi_{\Zcal}\mxS\Pi_{\Gcal})\le\delta_{\Ccal}(\mxS)$, and hence, Assumption \ref{ass:Rbar} holds for $\barRcal=\lambda\Rcal+\delta_{\Ccal}$.
From the definition of $\Ccal$ and $\Rcal$, it follows that  $\dom(\Rcal)\cap\Ccal=\Ccal$,
and subsequently, we have $\Dcal:=\Fcal\cap\dom(\Rcal)\cap\Ccal=\Fcal\cap\Ccal$.
One can easily see that $0\in\Dcal$.
Since $\|\cdot\|_*$ is a norm on $\dom(\Rcal)$ and $\Ccal\subset\dom(\Rcal)$,  we know that $\Ccal$ is a convex set, and also, $\Rcal$ is a convex function.
For $\mxS\in\Lcal(\Hcal)$, we have the inequality $\|\mxS\|\le\|\mxS\|_*$, which implies the coercivity of $\Rcal$. Hence, $\barRcal=\Rcal+\delta_{\Ccal}$ is a coercive function.
Let $\{\mxS_n\}_{n\in\Nbb}\subset\Lcal(\Hcal)$ be a sequence such that $\lim_{n\to\infty}\mxS_n=\mxS\in\Lcal(\Hcal)$ in norm topology.
Subsequently, we know that  $\lim_{n\to\infty}|\mxS_n|=|\mxS|$ in norm topology.
Hence, for any $k\in\Nbb$, we have that $\lim_{n\to\infty}\inner{b_k}{|\mxS_n|b_k}=\inner{b_k}{|\mxS|b_k}$.
Moreover, since $\{|\mxS_n|\}_{n\in\Nbb}$ and $\mxS$ are non-negative operators, we know that $\inner{b_k}{|\mxS_n|b_k}\ge 0$ and $\inner{b_k}{|\mxS|b_k}\ge 0$, for any $n,k\in\Nbb$.
Hence, from Fatou's lemma \cite{folland1999real}, it follows that
\begin{equation}\label{eqn:fatuo_lemma_nuc}
\begin{split}
\|\mxS\|_*&=\sum_{k=1}^\infty\inner{b_k}{|\mxS|b_k} =\sum_{k=1}^\infty \liminf_{n}\inner{b_k}{|\mxS_n|b_k} 
\\&\le 
\liminf_{n}\sum_{k=1}^\infty \inner{b_k}{|\mxS_n|b_k} 
=
\liminf_{n}\|\mxS_n\|_*. 
\end{split}
\end{equation} 
Accordingly, we know that $\Rcal(\mxS)\le\liminf_{n}\Rcal(\mxS_n)$, and $\Rcal$ is lower semicontinuous.
Also, if  $\{\mxS_n\}_{n\in\Nbb}\subset\Ccal$, then $\|\mxS_n\|_*\le \frac{1}{\lambda}\|\mxY\|_{\fro}$, for each $n$. Therefore, due to \eqref{eqn:fatuo_lemma_nuc}, we know that $\|\mxS\|_*\le \frac{1}{\lambda}\|\mxY\|_{\fro}$, and, subsequently, we have $\mxS\in\Ccal$.
Hence, $\Ccal$ and $\Dcal=\Fcal\cap\Ccal$ are non-empty, closed and convex sets.
Accordingly, Theorem \ref{thm:gen_reg_case} implies that \eqref{eqn:reg_learning_koopman_nuc_C} admits solution $\hatK$ with the parametric form in \eqref{eqn:hatK_Z}, which is also a solution for \eqref{eqn:reg_learning_koopman_nuc} due to the equivalency of the corresponding programs.  
Let $\{b_n\}_{n\in\Nbb}$ be an orthonormal basis such that $\Gcal=\linspan\{b_1,\ldots,b_{\barnG}\}$.
Accordingly, for any $l\in\{1,\ldots,\nG\}$, there exist $e_{l1},\ldots, e_{l\barnG}$ such that we have $g_l=\sum_{j=1}^{\barnG} e_{lj}b_j$.
From \eqref{eqn:T*T}, we have
\begin{equation}\label{eqn:T*T_E}\!\!\!
	\begin{split}	
		\hatK^*\hatK
		&=
		\sum_{j=1}^{\nG}\sum_{l=1}^{\nG}
		\big(\sum_{i=1}^{\barnG} e_{li}b_i\big)\! \otimes\! \big(\sum_{k=1}^{\barnG} e_{jk}b_k\big)
		[\mxA^\tr\mxZ\mxA]_{(l,j)}
		\\&=
		\sum_{i=1}^{\barnG}\sum_{k=1}^{\barnG} 
		\Big( 
		\sum_{j=1}^{\nG}\sum_{l=1}^{\nG} 
		e_{li} [\mxA^\tr\mxZ\mxA]_{(l,j)} e_{jk}
		\Big)		
		(b_i\otimes b_k) 
		\\&=
		\sum_{i=1}^{\barnG}\sum_{k=1}^{\barnG} 
		[\mxE^\tr\mxA^\tr\mxZ\mxA\mxE]_{(i,k)}
		(b_i\otimes b_k), 
	\end{split}
\end{equation}
where $\mxE$ is the matrix defined as $\mxE=[e_{lj}]_{l=1,j=1}^{\nG,\barnG}$.
One can easily see that $\mxG=\mxE\mxE^\tr$.
Due to \eqref{eqn:T*T_E}, we know that there exist $r_{lj}$, $l,j=1,\ldots,\barnG$, such that $|\hatK^*\hatK| = \sum_{i=1}^{\barnG}\sum_{k=1}^{\barnG}r_{ik}(b_i\otimes b_k)$.
From \eqref{eqn:h1xh2_h3xh4}, we have 
\begin{equation}\label{eqn:|T*T|^2_E}\!\!\!
	\begin{split}	
		|\hatK^*\hatK|^2
		&=
		\sum_{i=1}^{\barnG}\sum_{j=1}^{\barnG}r_{ij}(b_i\otimes b_j)
		\sum_{l=1}^{\barnG}\sum_{k=1}^{\barnG}r_{lk}(b_l\otimes b_k)
		\\&=
		\sum_{i=1}^{\barnG}\sum_{k=1}^{\barnG}
		\sum_{j=1}^{\barnG}\sum_{l=1}^{\barnG}
		r_{ij}r_{lk}\inner{b_j}{b_l} (b_i\otimes b_k)
		\\&=
		\sum_{i=1}^{\barnG}\sum_{k=1}^{\barnG}
		[\mxR^2]_{(i,k)}(b_i\otimes b_k), 
	\end{split}
\end{equation}
where $\mxR$ is the matrix defined as $\mxR=[r_{ik}]_{i=1,k=1}^{\barnG,\barnG}$.
From \eqref{eqn:T*T_E} and \eqref{eqn:|T*T|^2_E}, it follows that 
\begin{equation}\label{eqn:R2}
\mxR^2=\mxE^\tr\mxA^\tr\mxZ\mxA\mxE=(\mxZ^{\frac12}\mxA\mxE)^\tr(\mxZ^{\frac12}\mxA\mxE).
\end{equation}
Note that we have
\begin{equation}\label{eqn:inner_h1_h2xh3_h1}
	\inner{h_1}{(h_2\otimes h_3)h_1 } = \inner{h_1}{h_2} \inner{h_1}{h_3},
\end{equation}
for any $h_1,h_2,h_3,h_4\in\Hcal$.
Therefore, we have 
\begin{equation}\label{eqn:trace_R}
	\begin{split}	
		\trace(|\hatK^*\hatK|) 
		&=
		\sum_{j=1}^{\infty}\inner{b_j}{\sum_{i=1}^{\barnG}\sum_{k=1}^{\barnG}r_{ik}(b_i\otimes b_k)b_j }
		\\&=
		\sum_{j=1}^{\infty}\sum_{i=1}^{\barnG}\sum_{k=1}^{\barnG}r_{ik}
		\inner{b_j}{b_i }\inner{b_j}{b_k } 
		\\&= \sum_{i=1}^{\barnG}r_{ii}=\trace(\mxR).
	\end{split}
\end{equation}
Due to \eqref{eqn:trace_R} and \eqref{eqn:R2}, we know that 
$\|\hatK\|_* = \|\mxZ^{\frac12}\mxA\mxE\|_*$.
Note that, for any matrix $\mxM$, we have
$\|\mxM\|_* = \|\mxM^\tr\|_* = \trace\big((\mxM^\tr\mxM)^{\frac12}\big) = \trace\big((\mxM\mxM^\tr)^{\frac12}\big)$. 
From this fact and $\mxG^{\frac12}\mxG^{\frac12}=\mxG=\mxE\mxE^\tr$, it follows that
\begin{equation}
\begin{split}
\|\hatK\|_* &
= \trace\big((\mxZ^{\frac12}\mxA\mxE)(\mxE^\tr\mxA^\tr\mxZ^{\frac12})\big)
\\&=\trace\big((\mxZ^{\frac12}\mxA\mxG^{\frac12})(\mxG^{\frac12}\mxA^\tr\mxZ^{\frac12})\big)
\\&=\|\mxG^{\frac12}\mxA^\tr\mxZ^{\frac12}\|_*=\|\mxZ^{\frac12}\mxA\mxG^{\frac12}\|_*.
\end{split}
\end{equation}
Also, one can show that $\Ecal(\hatK)=\|\mxZ\mxA\mxG-\mxY\|_{\fro}^2$ using arguments similar to the proof of Theorem \ref{thm:Tikhonov_reg_case}. Substituting these terms in \eqref{eqn:reg_learning_koopman_nuc}, we obtain the optimization problem \eqref{eqn:rank_const_finite}. This concludes the proof. 
\end{proof}
\fi
\begin{remark}
	\normalfont
	The regularization function $\Rcal(\cdot):=\|\cdot\|_*$, employed in \eqref{eqn:reg_learning_koopman_nuc}, is not strictly convex.
	Therefore, Theorem~\ref{thm:gen_reg_case} can not guarantee the uniqueness of the solution.
	However, if we only consider the minimum norm solution of \eqref{eqn:reg_learning_koopman_nuc}, then it is unique and with the parametric form \eqref{eqn:hatK_Z}.
\end{remark}
\begin{remark} \label{rem:lim_K_lambda_nuc}
\normalfont
Let $\hatK_{\Gcal,\lambda}$ be the unique minimum norm solution of the learning problem \eqref{eqn:reg_learning_koopman_nuc}, for $\lambda>0$ and $\Fcal=\Lcal_{\Gcal}$.
Similar to Theorem \ref{thm:lim_K_lambda}, we can show that $\lim_{\lambda \downarrow 0} \hatK_{\Gcal,\lambda} = \hatK_{\mxU}$, and $\lim_{\lambda \to \infty} \hatK_{\Gcal,\lambda} = 0$, in nuclear norm topology which implies convergence in Frobenius norm and operator norm as well. 
\end{remark}

%% file: sec_08_stable.tex
\subsection{Learning Stable Koopman Operator}
Let $\xeq = [x_{\mathrm{eq},1},\ldots,x_{\mathrm{eq},n}]^\tr\in\Xcal$ be an equilibrium point for dynamics \eqref{eqn:dyn_f}.
In this section, we assume the Hilbert space of observables is a reproducing kernel Hilbert space $\Hcal_{\kernel}$ endowed with kernel $\kernel:\Xcal\times\Xcal$ for which we have 
\begin{equation}\label{eqn:kernel_x0_0x_0}
\kernel(\vcx,\xeq)=\kernel(\xeq,\vcx)=0, \quad\forall\vcx\in\Xcal.
\end{equation}
Indeed, given kernel $\kernelh:\Xcal\times\Xcal$, we can define $\kernel$ as
\begin{equation*}
	\kernel(\vcx,\vcy) = \kernelh(\vcx,\vcy) -\kernelh(\vcx,\xeq) -\kernelh(\xeq,\vcy)+\kernelh(\xeq,\xeq), 
\end{equation*}
for each $\vcx,\vcy\in\Xcal$.
Then, one can see that $\kernel$  satisfies the property \eqref{eqn:kernel_x0_0x_0}.
Note that when $\kernel$ has this property, then for each observable 
$g\in\Hcal_{\kernel}$, 
we have 
\begin{equation}
g(\xeq)=\inner{\kernel(\xeq,\cdot)}{g} = 0.	
\end{equation}

\begin{assumption}\label{ass:Hcal_separation}
	There exist $h_1,\ldots,h_{\nX}\in\Hcal_{\kernel}$ and positive scalars $L_1,\ldots,L_{\nX},\alpha_1,\ldots,\alpha_{\nX}$ such that, for each $\vcx=[x_1,\ldots,x_{\nX}]^\tr\in\Xcal$, we have	
	\begin{equation}\label{eqn:hj_assumption}
	|h_j(\vcx)|\ge L_j |x_j-x_{\mathrm{eq},j}|^{\alpha_j}, \quad \forall j\in\{1,\ldots,{\nX}\}.
	\end{equation}
\end{assumption}
One can see if quadratic function $h(\vcx)=\|\vcx-\xeq\|^2$ belongs to $\Hcal_{\kernel}$, then Assumption \ref{ass:Hcal_separation} is satisfied. Indeed, inequality \eqref{eqn:hj_assumption} holds for $h_j=h$, $L_j=1$, and $\alpha_j=2$, for $j=1,\ldots,{\nX}$, .

\begin{theorem} \label{thm:stability}
Let Assumption \ref{ass:Hcal_separation} hold for $\Hcal_{\kernel}$,  $\sup_{\vcx\in\Xcal}\kernel(\vcx,\vcx)<\infty$, and there exist $\epsilon>0$ such that $\|\mxK\|\le 1-\epsilon$. Then, $\xeq$ is a globally stable equilibrium point.  	
\end{theorem}
\begin{proof}
Consider a trajectory of system \eqref{eqn:dyn_f} as $\{\vcx_n\}_{n\in\Nbb}$, and let $g\in\Hcal_{\kernel}$.
From the definition of Koopman operator, the reproducing property of kernel, and  Cauchy-Schwartz inequality, it follows that
\begin{equation}
\begin{split}
|g(\vcx_n)| &= |(\mxK^n g)(\vcx_0)| 
\\&= |\inner{\kernel(\vcx_0,\cdot)}{\mxK^n g}|\le \|\kernel(\vcx_0,\cdot)\|\|\mxK^n g\|.	
\end{split}
\end{equation}
Since $\|\kernel(\vcx_0,\cdot)\|=\kernel(\vcx_0,\vcx_0)^{\frac12}$ and $\|\mxK\|\le 1-\epsilon$, we have
\begin{equation}
	|g(\vcx_n)| \le(1-\epsilon)^n\kernel(\vcx_0,\vcx_0)^{\frac12} \|g\|.
\end{equation}
Due to \eqref{eqn:hj_assumption} and by replacing $g$ with $h_j$, it follows that
\begin{equation*}
	L_j \left|x_{n,j}-x_{\mathrm{eq},j}\right|^{\alpha_j} \le |h_j(\vcx_n)| \le(1-\epsilon)^n\kernel(\vcx_0,\vcx_0)^{\frac12} \|h_j\|,
\end{equation*}
where $x_{n,j}$ is the $j^\nth$ coordinate of $\vcx_n$, for $j\in[\nX]$.
Accordingly, for each $j$, we have
\begin{equation*}
	\left|x_{n,j}-x_{\mathrm{eq},j}\right| \le (1-\epsilon)^{\frac{n}{\alpha}}
	\max_{1\le j\le\nX}\bigg[
	\frac{\sup_{\vcx\in\Xcal}\kernel(\vcx,\vcx)^{\frac12}	
		\|h_j\|}{L_j}
	\bigg]^{\frac{1}{\alpha_j}}\!\!,
\end{equation*}
where $\alpha=\max\{\alpha_1,\ldots,\alpha_{\nX}\}$. Hence, we have
\begin{equation}
	\lim_{n\to\infty}\vcx_n=\xeq,
\end{equation} 
where the convergence is uniform and with exponential rate. This concludes the proof.
\end{proof}
Motivated by Theorem \ref{thm:stability}, we can include in the learning problem the side-information on the stability of equilibrium point  \eqref{eqn:reg_learning_koopman_generalized} as following
\begin{equation}\label{eqn:reg_learning_koopman_generalized_stability}
	\begin{array}{cl}
		\minOp_{\mxK\in\Fcal} & \Ecal_{\ell}(\mxK) + \lambda \Rcal(\mxK)\\
		\mathrm{s.t.} &
		\mxK\in\Ccal,\ 
		\|\mxK\|\le 1-\epsilon,	
	\end{array}	
\end{equation}
where $\epsilon\in(0,1)$.

\begin{theorem}
Under the hypotheses of Theorem~\ref{thm:gen_reg_case},  the existence, the uniqueness and the parametric representation \eqref{eqn:hatK_Z} for the solution of learning problem \eqref{eqn:reg_learning_koopman_generalized_stability} are implied.  	
\end{theorem}
\begin{proof}
We know that $\|\Pi_{\Zcal}\mxS\Pi_{\Gcal}\|\le\|\mxS\|$, for all $\mxS\in\Lcal(\Hcal_{\kernel})$.
Therefore, $\|\mxS\|\le 1-\epsilon$ implies that $\|\Pi_{\Zcal}\mxS\Pi_{\Gcal}\|\le 1-\epsilon$. 
Subsequently, we have $\delta_{\Ccal_{\epsilon}}(\Pi_{\Zcal}\mxS\Pi_{\Gcal})\le\delta_{\Ccal_{\epsilon}}(\mxS)$,
where $\Ccal_{\epsilon}$ is the non-empty, closed and convex set defined as $$\Ccal_{\epsilon}:=\{\mxS\in\Lcal(\Hcal_{\kernel})|\|\mxS\|\le 1-\epsilon\}.$$ 
Accordingly, $\barRcal_{\epsilon}:=\delta_{\Ccal_{\epsilon}}$ satisfies Assumption \ref{ass:Rbar}. Following this, the proof of theorem is concluded directly from 
Theorem \ref{thm:extending_ass_2} and Theorem~\ref{thm:gen_reg_case}.
\end{proof}
Suppose that we have employed the empirical loss function 
introduced in \eqref{eqn:empirical_loss}, and the regularization function is either $\Rcal(\mxK)=\|\mxK\|^2$, or, $\Rcal(\mxK)=\|\mxK\|_{\fro}^2$.
Therefore, due to Theorem~\ref{thm:stability}, we know that the learning problem \eqref{eqn:reg_learning_koopman_generalized_stability} admits a unique solution $\hatK$ as in \eqref{eqn:hatK_Z}. 
Define matrix $\mxB$ as $\mxB=\mxZ^{\frac{1}{2}}\mxA\mxG^{\frac{1}{2}}$.
Based on a  discussion similar to Section~\ref{ssec:opt}, one can see that if $\Rcal(\mxK)=\|\mxK\|^2$, 
$\mxB$ can be obtained as the solution of following convex optimization
\begin{equation}\label{eqn:Tikhonov_reg_finite_stable}
	\begin{array}{cl}
		\minOp_{\mxB\in \Rbb^{\nZ\times\nG},\ \! \beta\in\Rbb}
		&
		\ \|\mxZ^{\frac{1}{2}}\mxB\mxG^{\frac{1}{2}}-\mxY\|_{\fro}^2 
		\ \!+ \ \! 
		\lambda \beta^2\\
		\mathrm{s.t.}
		&
		\ \begin{bmatrix}
			\beta \eye_{\nG} & \mxB\\\mxB^\tr &  \beta \eye_{\nZ}
		\end{bmatrix} \succeq 0,
		\\ & \ \beta \le 1-\epsilon,
	\end{array}	
\end{equation}
and, when $\Rcal(\mxK)=\|\mxK\|_{\fro}^2$,
we can obtain $\mxB$ by solving following convex program
\begin{equation}\label{eqn:Frobenius_reg_finite_stable}
	\begin{array}{cl}
		\minOp_{\mxB\in \Rbb^{\nZ\times\nG},\ \! \beta\in\Rbb}
		&
		\ \|\mxZ^{\frac{1}{2}}\mxB\mxG^{\frac{1}{2}}-\mxY\|_{\fro}^2 \ \!+ \ \! \lambda \|\mxB\|_{\fro}^2\\
		\mathrm{s.t.}
		&
		\ \begin{bmatrix}
			\beta \eye_{\nG} & \mxB\\\mxB^\tr &  \beta \eye_{\nZ}
		\end{bmatrix} \succeq 0,
		\\ & \ \beta \le 1-\epsilon.
	\end{array}	
\end{equation}

%% file: sec_09_numerics.tex
\section{Numerical Experiments and Examples}
\begin{figure}[t]
	\centering
	\includegraphics[width=0.485\textwidth,trim={9mm 7mm 5cm 7mm},clip]{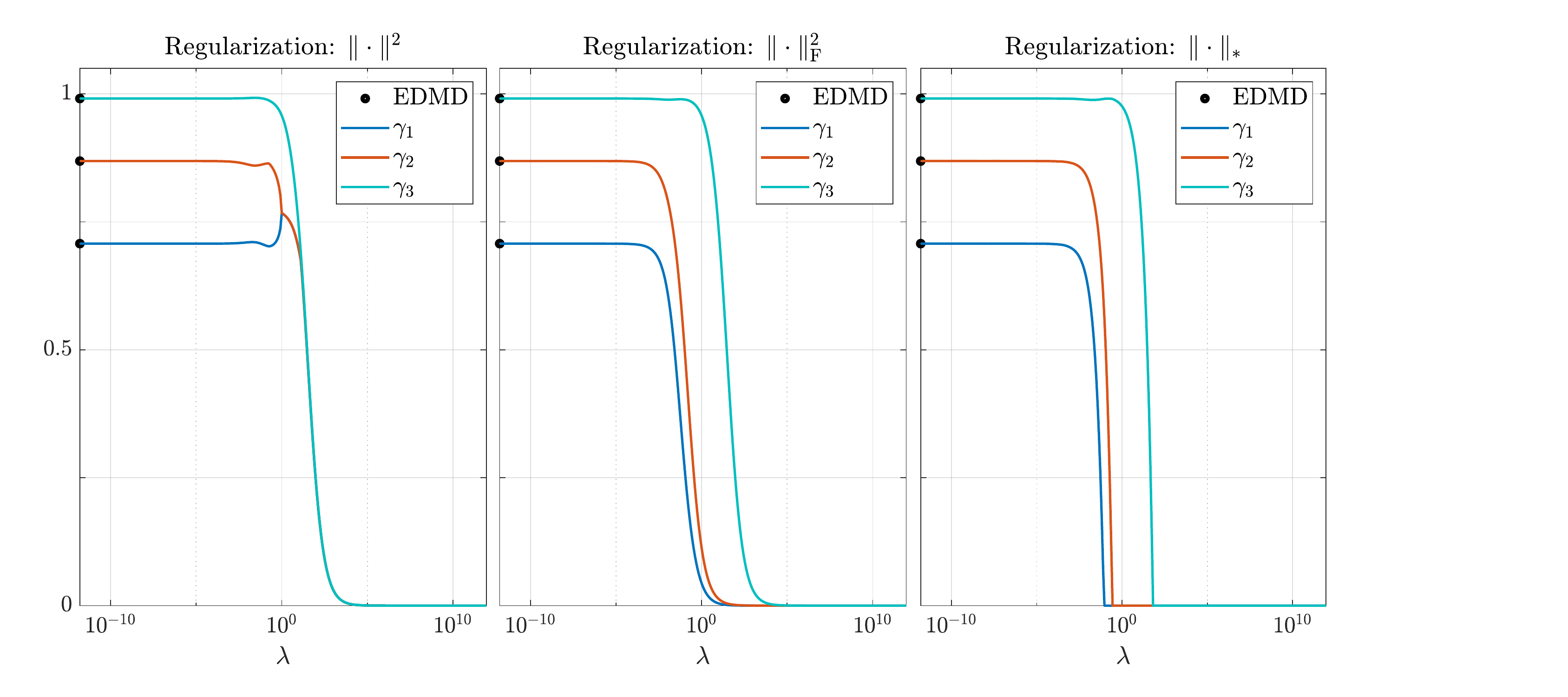}
	\caption{The figure shows the magnitude of eigenvalues derived from the proposed scheme converge to the ones obtained from EDMD as the weight of regularization, $\lambda$, goes to $0$.
	We can see that convergence has different behavior depending on the choice of regularization.
	}
	\label{fig:lambda_sweep}
\end{figure}
\begin{figure}[t]
	\centering
	\includegraphics[width=0.495\textwidth,trim={7mm 4mm 5mm 5mm}, clip]{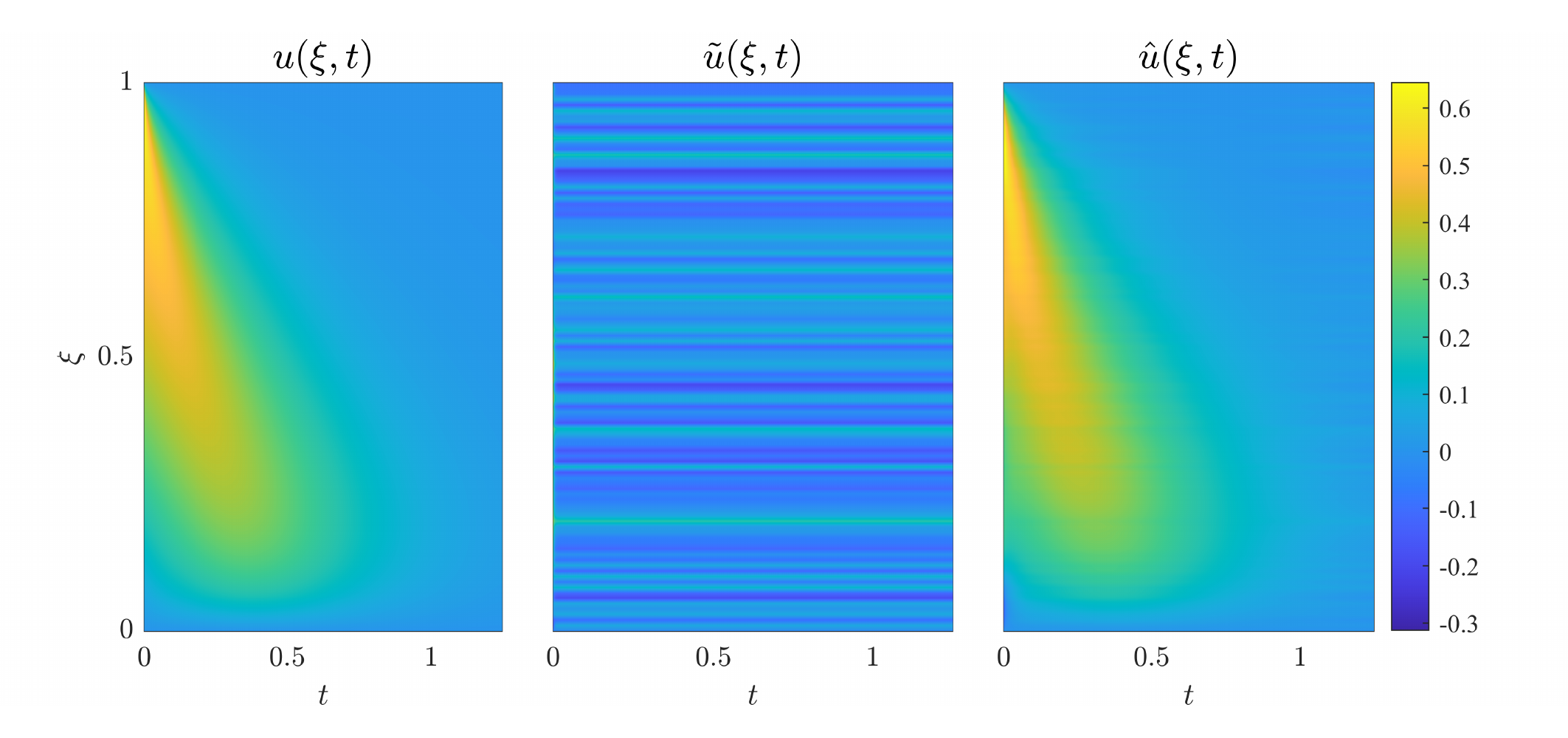}
	\caption{The figure shows the exact solution (left) of PDE \eqref{eqn:PDE_AD} with initial condition  $u(\xi,0)=1-\mathrm{e}^{-\xi}$, , and the approximate solutions derived from Koopman estimations, one obtained by the EDMD approach (middle) and the other via the learning method with Frobenius norm regularization (right).
	}
	\label{fig:PDE}
\end{figure}

In this section, we provide numerical examples elaborating the presented results.
Throughout this section, the Hilbert space of observables is specified as the RKHS with kernel $\kernel:\Xcal\times\Xcal\to\Rbb$, and observables $\{g_l\}_{l\in[\nG]}$ are defined according to Remark~\ref{rem:RKHS}, i.e.,  
$g_l(\cdot):=\kernel(\vcp_l,\cdot)$, for $l\in[\nG]$,
where $\Pcal:=\{\vcp_l\}_{l=1}^{\nG}$ is a finite set in $\Xcal$. 
Furthermore, for tuning hyperparameters such as the regularization weight, we employ a cross-validation scheme implemented through a Bayesian optimization procedure.
\begin{example}\label{exm:lambda_sweep}\normalfont
The connection between EDMD and learning problems \eqref{eqn:reg_learning_koopman_Tikhonov}, 
\eqref{eqn:reg_learning_koopman_Frobenius}, 
and,
\eqref{eqn:reg_learning_koopman_nuc} is discussed respectively in Theorem~\ref{thm:lim_K_lambda}, Remark~\ref{rem:lim_K_lambda_Frobenius}, and Remark~\ref{rem:lim_K_lambda_nuc}. 
To illustrates this feature, we consider the following stable nonlinear dynamics
\begin{equation}\label{eqn:dyn_example_01}
\begin{split}
x_{1,k+1} &= \mu_1 x_{1,k},\\
x_{2,k+1} &= \mu_2 x_{2,k}+(\mu_1^2-\mu_2)x_{1,k}^2,\\
\end{split}
\end{equation}
where $\mu_1=0.95$ and $\mu_2 = 0.75$ \cite{koopman_invariant_Naoya}.
Let the Hilbert space of observables be a RKHS endowed with the Gaussian kernel defined as following
\begin{equation}\label{eqn:Gaussian_kernel}
    \kernel(\vcx,\vcy) = \mathrm{e}^{-\frac1{2\ell_{\kernel}^2}\|\vcx-\vcy\|^2}, \quad \forall, \vcx,\vcy\in\Rbb^2,
\end{equation}
where $\ell_{\kernel}=1$, and let $\mxK:\Hcal_{\kernel}\to\Hcal_{\kernel}$ denote the Koopman operator corresponding to the nonlinear dynamical system \eqref{eqn:dyn_example_01}.
Suppose the system is initialized at $\vcx_0=[1\ \ \!0]^\tr$, and then generated a trajectory of length $\nS=60$.
Also, let $g_1$, $g_2$, and $g_3$ be three observables  defined as 
\begin{equation}
g_l(\vcx):=\mathrm{e}^{-\frac1{2\ell_{\kernel}^2}\|\vcx-\vc{p}_l\|^2},	\quad l=1,2,3,
\end{equation}
where $\vc{p}_1 = [1,0]^\tr$, $\vc{p}_2 = [1, 1]^\tr$ and $\vc{p}_3 = [0,1]^\tr$.
Suppose that we have the values of these observables along the trajectory of the system, i.e., $g_l(\vcx_{k})$ is given for $l=1,2,3$ and $k=0,\ldots,60$.  
\begin{figure*}[!t] 
    \centering
  \subfloat{
       \includegraphics[width=0.24\textwidth,trim={1mm 1mm 3mm 0mm},clip]{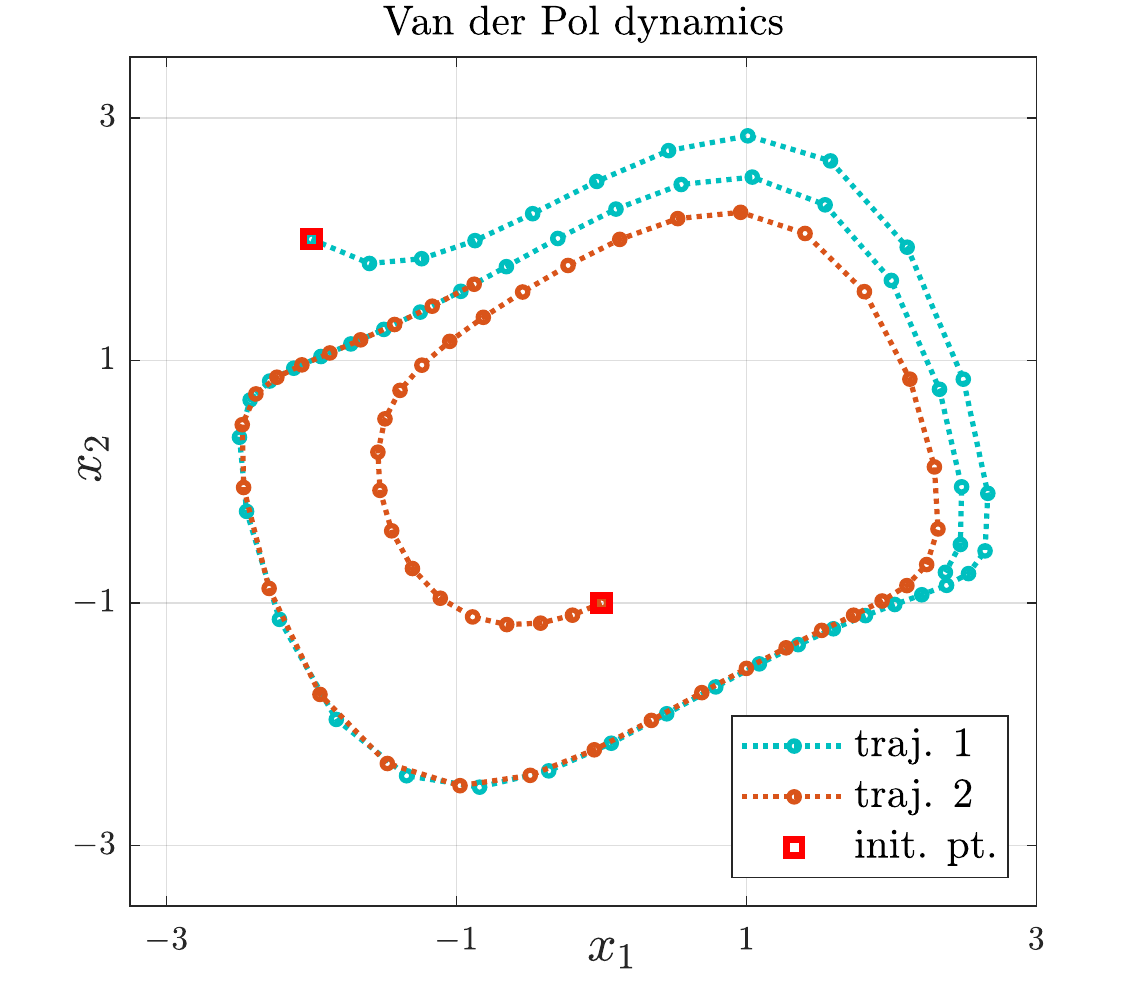}
       }
    \hfill \ 
  \subfloat{
    	\includegraphics[width=0.735\textwidth,trim={3mm 6mm 2mm 1mm},clip]{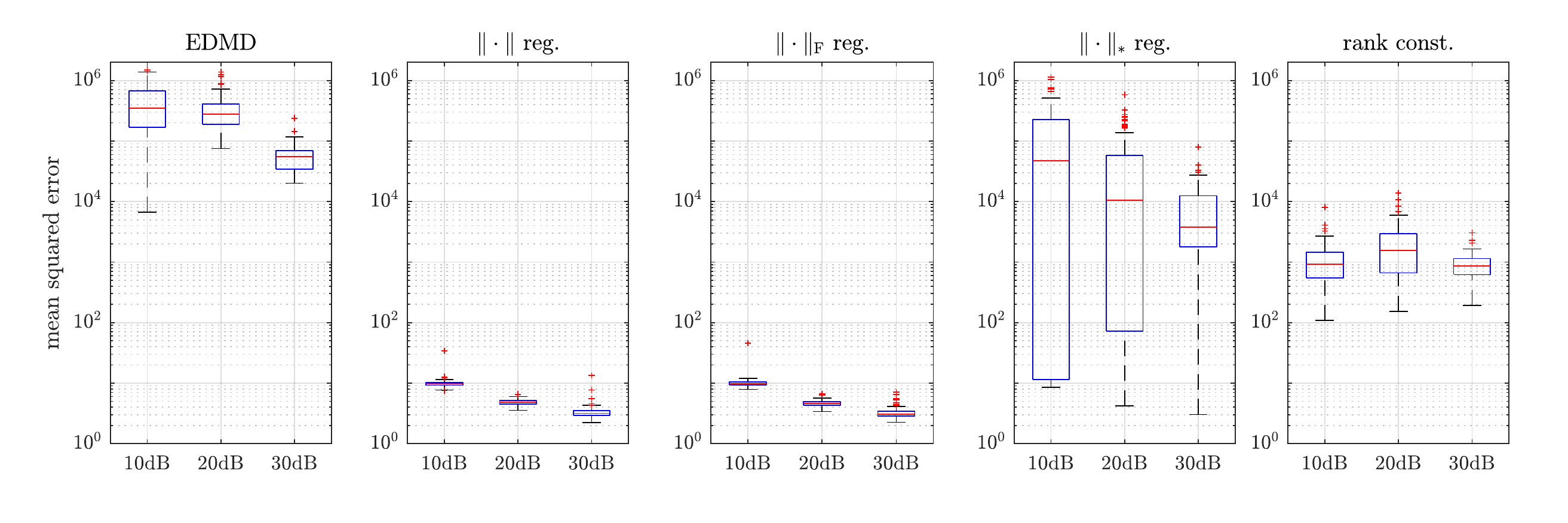}
        	}
	\caption{Left: two trajectories of Van der Pol dynamics \eqref{eqn:VanderPol}. Right: the box-plots for the mean squared error of different Koopman learning methods.}    
    \label{fig:vanderpol}    
\end{figure*}
Given this data, we can apply EDMD method discussed in Section \ref{sec:edmd}, and also, the proposed scheme introduced in Theorem \ref{thm:Tikhonov_reg_case_W_part_2}, for different values of $\lambda$ and, $w_l:=g_l$, for $l=1,2,3$.
Each of these methods provides an estimation of the eigenvalues of the Koopman operator.
From Theorem \ref{thm:lim_K_lambda}, we expect that as the weight of regularization, $\lambda$, goes to $0$, the magnitude of eigenvalues derived from the proposed scheme, denoted here by $\gamma_1$, $\gamma_2$ and $\gamma_3$, converge to the ones obtained from EDMD.
Moreover, we know that as $\lambda\to+\infty$, the solution of \eqref{eqn:reg_learning_koopman_G_lambda} goes to $0$. Accordingly, we expect that $\gamma_1$, $\gamma_2$ and $\gamma_3$ converge to $0$.
Based on Remark \ref{rem:lim_K_lambda_Frobenius}, and Remark \ref{rem:lim_K_lambda_nuc},
we expect to observe similar results for the case of $\Rcal(\mxK)=\|\mxK\|_{\fro}^2$ and
$\Rcal(\mxK)=\|\mxK\|_{*}$.
%
%
%
%
Figure \ref{fig:lambda_sweep} demonstrates these asymptotic phenomena. 
One can see that depending on the choice of regularization functions, we have different forms of convergence for  $\gamma_1$, $\gamma_2$ and $\gamma_3$.
Furthermore, comparing the cases $\Rcal(\mxK) = \|\mxK\|^2$ and
$\Rcal(\mxK) = \|\mxK\|_{\fro}^2$, when the regularization term is based on the Frobenius norm, we observe that as $\lambda\to\infty$, the convergence of  $\gamma_1,\gamma_2,\gamma_3$ to $0$ is with higher rate. 
Moreover, when nuclear norm is employed for the regularization term, $\gamma_1$, $\gamma_2$, and $\gamma_3$ converge to $0$ faster than the two previous cases.
These phenomena can be explained based on the inequality between the operator norm, Frobenius norm and nuclear norm, i.e., 
$$\|\mxS\| 
\le \|\mxS\|_{\fro} 
\le \|\mxS\|_* \le \infty,$$ 
which holds for all $\mxS\in\Lcal(\Hcal_{\kernel})$.
Additionally, we can see that when $\Rcal(\mxK) = \|\mxK\|_*$, the rank of operator $\hatK$ drops as $\lambda$ increases.
The is due to the nature of nuclear norm which leads to promoting low-rank solutions.
\xqed{$\triangle$}\end{example}
\begin{example}\label{exm:comparison}\normalfont
The Van der Pol oscillator is described as 
\begin{equation}\label{eqn:VanderPol}
\begin{split}
\dot{x}_1 & = x_2,\\
\dot{x}_2 &= \mu (1-x_1^2) x_2 - x_1 + u,\\
\end{split}
\end{equation}
where $\mu>0$ is the damping coefficient and $u$ is the input. To obtain a discrete-time system as in \eqref{eqn:dyn_f}, we set $\mu=0.5$ and $u=0$ in \eqref{eqn:VanderPol}, and employ forward Euler method with step size $\Delta t = 0.2$s. 
We consider trajectories starting from initial points $\vcx_0^{{\stiny{\text{$(1)$}}}} = [-2,2]^\tr$ and $\vcx_0^{{\stiny{\text{$(2)$}}}} = [0,-1]^\tr$ respectively with length $\nS^{{\stiny{\text{$(1)$}}}}=50$ and  $\nS^{{\stiny{\text{$(2)$}}}}=50$ (see Figure~\ref{fig:vanderpol}). We employ Matern kernel $\kernel$ with parameter $\nu_{\kernel}=\frac52$ and length scale $\ell_{\kernel}=1$ \cite{GPR}, and define observables $\{g_l\}_{l=1}^{\nG}$ as the sections of kernel $\kernel$ at points $\Pcal =\{0.5(i,j)|i,j=-5,...,5\}$.  
To perform a Monte Carlo experiment, we corrupt the trajectories data points with zero mean white Gaussian additive noise. More precisely, we consider different variance values to include low, medium, and high signal-to-noise ratio (SNR) levels, namely with $10\,$dB, $20\,$dB, and $30\,$dB, and concerning each of these SNR levels, 120 noise realization is generated for being added to the trajectories.
Given the data, we can estimate the Koopman operator using the EDMD approach and the regularized learning methods mentioned in the previous sections by solving their  equivalent finite-dimensional optimization problems.
%
To compare the performance of these methods, we employ test observables $\{\overline{g}_l\}_{l=1}^{\nGb}$ specified as the sections of kernel $\kernel$ at points $\overline{\Pcal} =\{0.1(i,j)|i,j=-5,...,5\}$, and the mean squared error  for the estimated Koopman operator $\hatK$ defined as 
\begin{equation}\label{eqn:MSE}
    \text{MSE}(\hatK) = 
    \frac{1}{\nGb} \sum_{l=1}^{\nGb} \int_{R}
    \big((\hatK\circ g_l)(\vcx)-g_l(F(\vcx))\big)^2\drm \vcx,
\end{equation}
where $F$ denotes the right-hand side of Van der Pol dynamics \eqref{eqn:VanderPol}, $R = [-2.5,2.5]\times[-2.5,2.5]$, and the integral is calculated numerically using a grid with $\Delta\vcx = (0.025,0.025)$. 
%
Figure~\ref{fig:vanderpol} illustrates and compares estimation performance outcomes. One can observe that the introduced learning schemes outperform the EDMD method, i.e., the incorporation of regularization terms and constraints results in a more accurate estimation of the Koopman operator. Furthermore, learning techniques with regularization terms defined based on the operator and Frobenius norms show improved estimation performances than those with nuclear norm regularization. The observed outperformance can be due to the fact that the nuclear norm dominates the operator and Frobenius norms, and consequently, they have more effective impacts. The same arguments hold when the impact of rank constraint is compared with nuclear norm regularization. Moreover, we observe that the estimation performances generally improve as the SNR level increases, which is an expected phenomenon. The minor exception is the case of including rank constraint, which can be due to its non-convexity. Finally, we can see that nuclear norm regularization and rank constraint are not as effective as the regularization terms specified by the operator and Frobenius norm, indicating that the original Koopman operator is probably not low rank here.
\xqed{$\triangle$}\end{example}
\begin{figure}[t]
	\centering
	\includegraphics[width=0.45\textwidth,trim={4mm 4mm 9mm 3mm},clip]{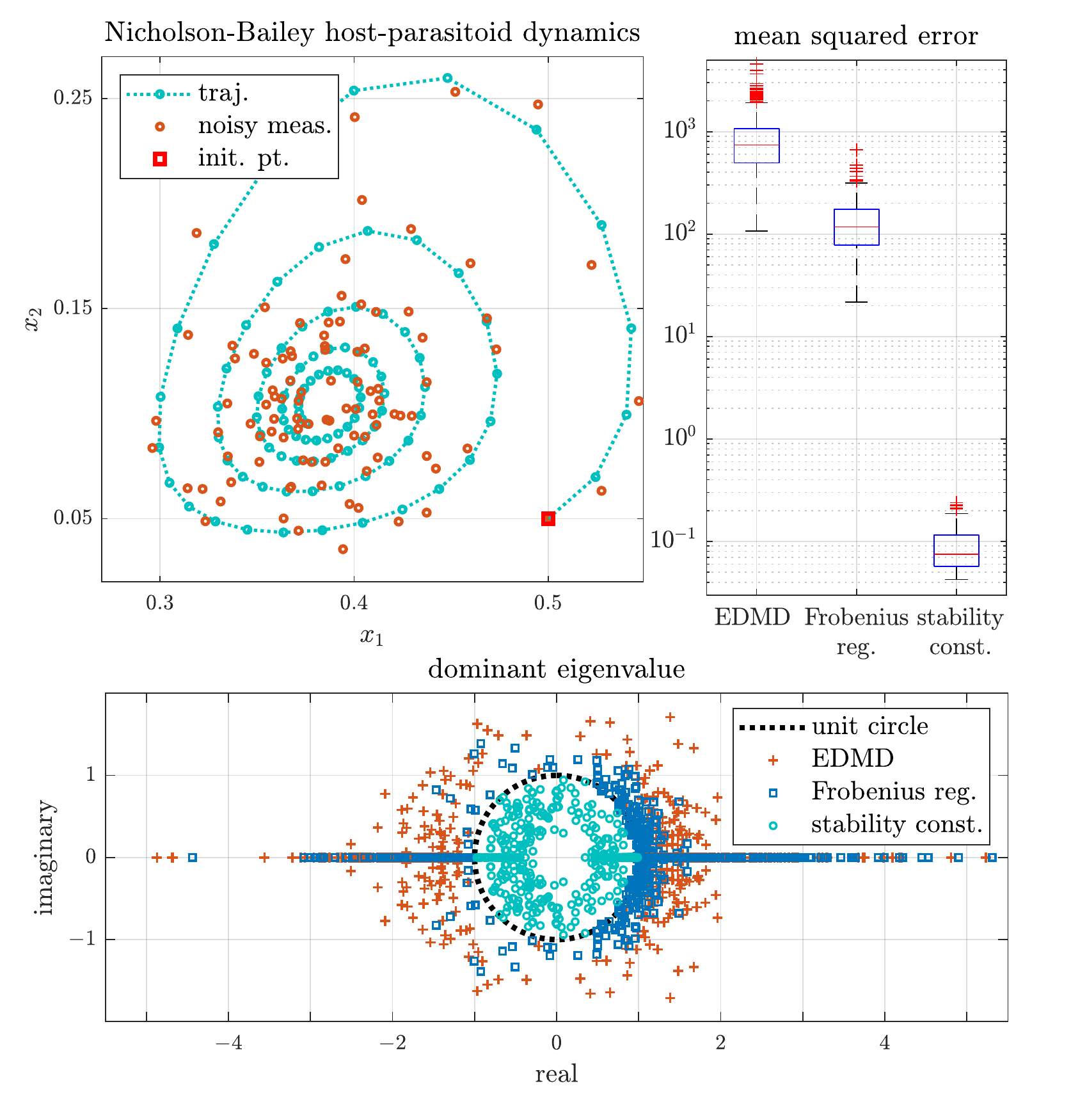}
	\caption{Top-left: A trajectory of dynamics \eqref{eqn:HP} and its noisy measurement with $30\,$dB SNR. Top-right: performance comparison for Koopman operator estimate using the EDMD approach, the learning scheme with Frobenius norm regularization, and its modified version with the stability-inducing constraint. Bottom: the dominant eigenvalues of estimated Koopman operators.
	}
	\label{fig:HP}
\end{figure}
\begin{example}\label{exm:stability_SI}\normalfont
The incorporation of potentially available side-information about the Koopman operator can improve the learning accuracy by shrinking the hypothesis space and rejecting spurious solution candidates.
To demonstrate this feature, we employ a scaled version of the Nicholson-Bailey model for host-parasitoid dynamics described as 
\begin{equation}\label{eqn:HP}
\begin{split}
x_{1,k+1} &=  R_0\, x_{1,k}\, (1+2x_{2,k})^{-\frac12},\\
x_{2,k+1} &=  c\,   x_{1,k}\, \big(1-(1+2x_{2,k})^{-\frac12}\big),\\
\end{split}
\end{equation}
where $R_0=1.1$ and $c=3$. 
We consider the trajectory starting from the initial point $\vcx_0 = [0.5,0.05]^\tr$ and with length $\nS=100$ (see Figure~\ref{fig:HP}). We employ Gaussian kernel \eqref{eqn:Gaussian_kernel} with $\ell_{\kernel}=0.1$ \cite{GPR}.
The observables $\{g_l\}_{l=1}^{\nG}$ are defined as the sections of kernel $\kernel$ at points $\Pcal =\{(0.3,0.05) + 0.025(i,j)|i,j=0,...,8\}$. 
According to the behavior of the system shown in Figure~\ref{fig:HP}, one can conclude that the system is stable. 
Meanwhile, as we estimate the Koopman operator using the observables data and through the EDMD approach and the learning method with Frobenius norm regularization with $\lambda=10^{-6}$, we observe that the magnitude of the dominant eigenvalues of the estimated Koopman operators is less than one. This observation confirms the discussed stability side-information.
Following this, we perform a Monte Carlo experiment by corrupting the trajectory data with realizations of zero mean white Gaussian additive noise. The noise variance is chosen such that the resulting SNR is $30\,$dB. We generate 450 noise realizations for being added to the trajectory data, and subsequently, the observables are evaluated on the noisy trajectory.
Given the observables data, we repeat previous Koopman operator estimation schemes. Furthermore, to integrate the stability side-information, we modify the learning method with Frobenius norm regularization by including the stability-inducing constraint $\|\mxK\|\le 1-\epsilon$ with $\epsilon = 10^{-5}$.
Figure~\ref{fig:HP} (bottom) demonstrates the dominant eigenvalues of the estimated Koopman operators. 
We can see from Figure~\ref{fig:HP} that when the stability-inducing constraint is used, the dominating eigenvalues are inside the unit circle, whereas they are mainly located outside the unit circle when the other techniques are implemented. Comparing the EDMD results, it can be seen that the dominant eigenvalues have a smaller magnitude when Frobenius norm regularization is used, which can be due to the inequality a<b. Indeed, the Frobenius norm regularization partially incorporates the stability side-information.
Similar to the previous example, we quantitatively compare the Koopman operator estimation results using
test observables $\{\overline{g}_l\}_{l=1}^{\nGb}$ defined as the sections of kernel $\kernel$ at points $\overline{\Pcal} =\{(0.3,0.05) + 0.01(i,j)|i,j=0,...,20\}$ and the mean squared error defined in \eqref{eqn:MSE} and calculated on the region $R = [0.3,0.5]\times[0.05,0.25]$. Figure~\ref{fig:HP} (top-right) compares the performance of discussed Koopman operator estimation schemes. One can see that the inclusion of stability side-information results in improved learning and estimation accuracy. 
\xqed{$\triangle$}\end{example}
\begin{example}\label{exm:PDE}\normalfont
Consider the following convection-diffusion PDE
\begin{equation}\label{eqn:PDE_AD}
    \frac{\partial u}{\partial t}(\xi,t)
    =
    a
    \frac{\partial u}{\partial \xi}(\xi,t)
    +
    b
    \frac{\partial^2 u}{\partial \xi^2}(\xi,t),
\end{equation}
where $(\xi,t)\in[0,1]\times [0,\infty)$, $a = 1$, and $b = 0.1$. 
We discretize $\xi$ and $t$ domains respectively with $\Delta\xi=10^{-2}$ and $\Delta t=10^{-4}$ to obtain discrete-time dynamics $\vcx^+=F(\vcx)$ with state dimension $\nX=101$.
Due to the definition of the Koopman operator, we know that $g (\vcx_n) = \mxK^n(\vcx_0)$, for any $n\in\Nbb$.
Thus, when $\vcx$ can be derived using the value of observables at $\vcx$, one can employ the estimation of the Koopman operator to obtain the solution of the system for an arbitrarily given initial condition. The precision of this solution indicates the accuracy of the estimated Koopman operator. 
Accordingly, we consider the solution of the discrete-time system that corresponds to the initial condition $u(\xi,0)=\sin(\pi\xi)$.
Also, we employ kernel $\kernel(\vcx,\vcy) = 1+\vcx^\tr\vcy$, and specify $\nG=2\nX$ observables $\{g_l\}_{l=1}^{\nG}$ as the sections of kernel $\kernel$ at points $\Pcal =\{\vcp_l\}_{l=1}^{\nG}$ which are randomly chosen from the standard normal distribution in $\Rbb^{\nX}$.
We estimate the Koopman operator using data of the observables, and subsequently, we obtain the solution corresponding to the initial condition  $u(\xi,0)=1-\mathrm{e}^{-\xi}$ based on the discussion above. 
To estimate the Koopman operator, we employ the EDMD approach and the learning method with Frobenius norm regularization, and denote the resulting approximated solutions respectively by $\tilde{u}$ and $\hat{u}$.  
To quantitatively compare these approximate solutions, we define solution mismatches as $\Delta\tilde{u}:=\tilde{u}-u$ and $\Delta\hat{u}:=\hat{u}-u$, and subsequently their $\Lscr^2$-norm is calculated on the region $[0,1]\times [0,1.25]$, which results in $\|\Delta\tilde{u}\|_2 = 2.543 \times 10^{-1}$ and  $\|\Delta\hat{u}\|_2 = 1.004 \times 10^{-2}$. In Figure~\ref{fig:PDE}, the approximate solutions are compared to the exact solution $u$. Figure~\ref{fig:PDE} shows that $\hat{u}$ is nearly identical to the exact solution, whereas $\tilde{u}$ appears to be considerably different.
The quantitative evaluation and graphical comparison support that the Koopman operator estimation obtained by the learning method with Frobenius norm regularization is significantly more accurate than the one derived from the EDMD approach.
\xqed{$\triangle$}\end{example}

%% file: sec_97_conc.tex
\section{Conclusion}
In this paper, we investigated the problem of learning Koopman operator of a discrete-time autonomous system. The learning problem was formulated as a constrained regularized optimization over the infinite-dimensional space of linear operators.
We showed that a representer theorem holds for the learning problem under certain but general conditions.
This allows a finite-dimension reformulation of the problem without any approximation and precision loss.
Moreover, we investigated the incorporation of various forms of regularization and constraint in the Koopman operator learning problem, including the operator norm, the Frobenius norm, and rank.  
For each of these cases, we derived the corresponding finite-dimensional problem.

%% file: sec_99_myappendix.tex
\subsection{Learning with Multiple Trajectories}
\label{ssec:multi_traj}
We can extended the learning problem \eqref{eqn:reg_learning_koopman_Tikhonov} to the case of multiple trajectories. Let $\nT\in\Nbb$, and, for $i=1,\ldots,\nT$, $\vcx_0^{\itrajs},\vcx_1^{\itrajs},\ldots,
\vcx_{\nS^{\itrajs}}^{\itrajs}$  
be a trajectory of dynamical system \eqref{eqn:dyn_f}, where $\nS^{\itrajs}\in\Nbb$.
Suppose that the set of data $\Dscr_{\nT}$ defined as 
\begin{equation*}\label{eqn:D_multiple_traj}
	\Dscr_{\nT}:=\Big\{y_{kl}^{\itrajs}  := g_l(\vcx_k^{\itrajs})\Big|  
	1\le i\le\nT,0\le k\le\nS^{\itrajs},1\le l\le\nG\Big\},	
\end{equation*}
is given. 
One may define $y_{kl}^{\itrajs}$ as  $y_{kl}^{\itrajs}= g_l(\vcx_k^{\itrajs})+\varepsilon_{k,l}^{\itrajs}$
to consider the uncertainty in the values of observables.
Similar to before, the learning problem is defined as 
\begin{equation}\label{eqn:reg_learning_koopman_Tikhonov_multiple_traj}
	\minOp_{\mxK\in\Lcal(\Hcal)}  
	\sumOp_{i=1}^{\nT}
	\sumOp_{k=1}^{\nS^{\itrajs}}
	\sumOp_{l=1}^{\nG} 
	\big(y_{kl}^{\itrajs} -(\mxK g_l)(\vcx_{k-1}^{\itrajs})\big)^2 + \lambda \|\mxK\|^2\!\!,
\end{equation}
where $\lambda>0$.
For $i=1,\ldots,\nT$ and $k=1,\ldots,\nS^{\itrajs}$, let the evaluation operator at  $\vcx_{k-1}^{\itrajs}$ be bounded. 
Then, the optimization problem \eqref{eqn:reg_learning_koopman_Tikhonov_multiple_traj} has a unique solution $\hatK$ as
\begin{equation}\label{eqn:hatK_multi}
	\hatK = \sum_{i=1}^{\nT}\sum_{k=1}^{\nS^{\itrajs}} \sum_{l=1}^{\nG} a_{kl}^{(i)}\ v_k^{(i)} \otimes g_l,
\end{equation}
where  $\{v_k^{\itrajs}|1\le i\le\nT,1\le k\le\nS^{\itrajs}\}$ are vectors in $\Hcal$ such that $g(\vcx_{k-1}^{\itrajs})= \inner{v_{k}^{\itrajs}}{g}$, for any $g\in\Hcal$.
Let  $\nS=\sum_{i=1}^{\nT}\nS^{\itrajs}$ and, matrices $\mathrm{\mathbf{A}}$ and $\mathrm{\mathbf{Y}}$ be respectively defined as
\begin{equation}
\mathrm{\mathbf{A}} := \left[ \mxA^{\itrajsOp{1}}{}^\tr\!\ldots\ \! \mxA^{\itrajsOp{\nT}}{}^\tr \right]^\tr \in \Rbb^{\nS\times\nG},
\end{equation}
and
\begin{equation}
\mathrm{\mathbf{Y}} := \left[ \mxY^{\itrajsOp{1}}{}^\tr\!\ldots\ \! \mxY^{\itrajsOp{\nG}}{}^\tr \right]^\tr \in \Rbb^{\nS\times\nG},	
\end{equation}
where $\mxA^{\itrajs}:=[a_{kl}^{\itrajs}]_{k=1,l=1}^{\nS^{\itrajs},\nG}$
and 
$\mxY^{\itrajs}:=[y_{kl}^{\itrajs}]_{k=1,l=1}^{\nS^{\itrajs},\nG}$,
for $i=1,\ldots,\nT$.
Also, let $\mathrm{\mathbf{V}}\in\Rbb^{\nS\times\nS}$ be the following matrix
\begin{equation}
\mathrm{\mathbf{V}}\in\Rbb^{\nS\times\nS}
=
\begin{bmatrix}
\mxV^{\itrajsOp{1,1}}&\ldots&\mxV^{\itrajsOp{1,\nT}}\\
\vdots&\ldots&\vdots\\
\mxV^{\itrajsOp{\nT,1}}&\ldots&\mxV^{\itrajsOp{\nT,\nT}}\\
\end{bmatrix}
\end{equation}
where, for any $i,j=1,\ldots,\nT$, matrix $\mxV^{\itrajsOp{i,j}}$ is defined such that $[\mxV^{\itrajsOp{i,j}}]_{(k,l)} = \inner{\vcv_{k}^{\itrajsOp{i}}}{\vcv_{l}^{\itrajsOp{j}}}$ for $k=1,\ldots,\nS^{\itrajsOp{i}}$ and $l=1,\ldots,\nS^{\itrajsOp{j}}$.
Then, in order to find the coefficients in \eqref{eqn:hatK_multi}, it is enough to solve the following convex program
\begin{equation}\label{eqn:opt_finite_Tikhonov_reg_case_multi_traj}
	\min_{\mathrm{\mathbf{A}}\in \Rbb^{\nS\times\nG}}\ \|\mathrm{\mathbf{V}}\mathrm{\mathbf{A}}\mxG-\mathrm{\mathbf{Y}}\|_{\fro}^2 \ \!+ \ \! \lambda \ \!  \|\mathrm{\mathbf{V}}^{\frac{1}{2}}\mxA\mxG^{\frac{1}{2}}\|^2,
\end{equation}
where $\mxG$ is the Gramian matrix of $g_1,\ldots,g_{\nG}$. 
Similar to Section \ref{ssec:opt}, we can define $\mathrm{\mathbf{B}}=\mathrm{\mathbf{V}}^{\frac{1}{2}}\mathrm{\mathbf{A}}\mxG^{\frac{1}{2}}$ and derive following equivalent convex program
\begin{equation}\label{eqn:opt_finite_Tikhonov_reg_case_multi_traj_LMI}
	\begin{array}{cl}
		\minOp_{\mathrm{\mathbf{B}}\in \Rbb^{\nS\times\nG},\ \! \beta\in\Rbb}
		&
		\ \|\mathrm{\mathbf{V}}^{\frac{1}{2}}\mathrm{\mathbf{B}}\mxG^{\frac{1}{2}}-\mathrm{\mathbf{Y}}\|_{\fro}^2 \ \!+ \ \! \lambda \ \!  \beta^2\\
		\mathrm{s.t.}
		&
		\ \begin{bmatrix}
			\beta \eye_{\nG}        & \mathrm{\mathbf{B}}\\
			\mathrm{\mathbf{B}}^\tr &  \beta \eye_{\nS}
		\end{bmatrix} \succeq \zeromx,
	\end{array}	
\end{equation}
\subsection{Proof of Theorem \ref{thm:Tikhonov_reg_case_W_part_1}}\label{sec:appendix_proof_thm_3}
The existence and uniqueness of the solution $\hatK_{\Wcal}$ follows from the same lines of arguments as in the proof of Theorem~\ref{thm:Tikhonov_reg_case}.
Define the linear subspaces  $\tildeWcal$ as 
\begin{equation}\label{eqn:tildeWcal}
	\tildeWcal:=\linspan\{\Pi_{\Wcal}v_1,\ldots,\Pi_{\Wcal}v_{\nS}\},
\end{equation}
and, let $\Pi_{\tildeWcal}$ be the projection operator on $\tildeWcal$.
For 
$k\in[\nS]$, 
we know that $\Pi_{\Wcal}v_k \in\Wcal$, and therefore, $\tildeWcal$ is a subspace of $\Wcal$.
Define operator $\mxS$ as $\mxS:=\Pi_{\tildeWcal}\ \!\hatK_{\Wcal}\ \!\Pi_{\Gcal}$ which belongs to $\Lcal_{\Wcal}$ due to $\tildeWcal\subseteq\Wcal$.
From the definition of $\Pi_{\Gcal}$, we know that $\Pi_{\Gcal}g_l=g_l$, for 
$l\in[\nG]$.
Thus, for any $k$ and $l$, we have
\begin{equation}\label{eqn:inner_S_hatK_W}
	\begin{split}
		\inner{v_k}{\mxS g_l} 
		&=
		\inner{v_k}{\Pi_{\tildeWcal} \hatK_{\Wcal} \Pi_{\Gcal} g_l}\\ 
		&
		=
		\inner{v_k}{\Pi_{\tildeWcal}\hatK_{\Wcal} g_l}
		=\inner{\Pi_{\tildeWcal}^* v_k}{\hatK_{\Wcal} g_l},
	\end{split}
\end{equation}
where  $\Pi_{\tildeWcal}^*$ is the adjoint of $\Pi_{\tildeWcal}$.
Since $\tildeWcal$ is a finite dimensional subspace, it is closed and the projection operator $\Pi_{\tildeWcal}$ is self-adjoint, i.e., $\Pi_{\tildeWcal}^*=\Pi_{\tildeWcal}$.
Also, due to $\tildeWcal\subseteq\Wcal$, we know that  $\Wcal^\perp\subseteq\tildeWcal^\perp$, and subsequently, we have $\Pi_{\tildeWcal}\Pi_{\Wcal^\perp} = 0$.
Accordingly, one can see that
\begin{equation}
	\Pi_{\tildeWcal}-\Pi_{\tildeWcal}\Pi_{\Wcal} 
	= 
	\Pi_{\tildeWcal}(\eye-\Pi_{\Wcal})
	=
	\Pi_{\tildeWcal}\Pi_{\Wcal^\perp} = 0.
\end{equation}
Consequently, for each   
$k\in[\nS]$, 
we have 
\begin{equation}\label{eqn:Pi_tildeW*_vk}
\Pi_{\tildeWcal}^*v_k = \Pi_{\tildeWcal}v_k=\Pi_{\tildeWcal}\ \!\Pi_{\Wcal}v_k= \Pi_{\Wcal}v_k,	
\end{equation}
where the last equality is due to 
$\Pi_{\Wcal}v_k\in\tildeWcal$. 
Note that $\Wcal$ is a closed subspace, and subsequently, $\Pi_{\Wcal}$ is a self-adjoint operator, i.e., $\Pi_{\Wcal}^*=\Pi_{\Wcal}$.
Accordingly,  from \eqref{eqn:inner_S_hatK_W} and \eqref{eqn:Pi_tildeW*_vk},  we can see that 
\begin{equation}\label{eqn:inner_S_hatK_W_02}
\begin{split}
	\inner{v_k}{\mxS g_l}  &= 
	\inner{\Pi_{\Wcal} v_k}{\hatK_{\Wcal} g_l}\\ 
	&= 
	\inner{v_k}{\Pi_{\Wcal}^*\hatK_{\Wcal} g_l}
	=
	\inner{v_k}{\Pi_{\Wcal}\hatK_{\Wcal} g_l}, 
\end{split}
\end{equation}
for each   $k$ and $l$.
Due to the definition of $\Lcal_{\Wcal}$ in \eqref{eqn:L_W} and since $\hatK_{\Wcal}\in\Lcal_{\Wcal}$, we know that $\hatK_{\Wcal}g_l\in\Wcal$, and subsequently, one has $\Pi_{\Wcal}\hatK_{\Wcal} g_l=\hatK_{\Wcal} g_l$, 
for $l\in[\nG]$.
Therefore, from \eqref{eqn:inner_S_hatK_W_02}, it follows that
\begin{equation}\label{eqn:ERM_ERM_W}
	\begin{split}
		\Ecal(\mxS) &= \sum_{k=1}^{\nS}
		\sum_{l=1}^{\nG} 
		\big(  y_{kl}-\inner{v_k}{\mxS g_l}\big)^2
		\\&=
		\sum_{k=1}^{\nS}
		\sum_{l=1}^{\nG}  
		\big(y_{kl}-\inner{v_k}{\hatK_{\Wcal} g_l}\big)^2=\Ecal(\hatK_{\Wcal}).	
	\end{split}
\end{equation}
Similar to the proof of Theorem \ref{thm:Tikhonov_reg_case}, one can show that $\|\mxS\|^2 \le \|\hatK_{\Wcal}\|^2$, and subsequently, one can see that $\Ecal(\mxS)+\lambda\|\mxS\|^2\le \Ecal(\hatK_{\Wcal}) +\lambda\|\hatK_{\Wcal}\|^2$.
From the uniqueness of the solution of \eqref{eqn:reg_learning_koopman_Tikhonov_W}, we  have $\hatK_{\Wcal}=\mxS=\Pi_{\Vcal}\!\ \hatK_{\Wcal}\!\ \Pi_{\Gcal}$. 
Due to the linearity of operator $\hatK_{\Wcal}$, it follows that there exist $a_{kl}\in\Rbb$, 
for 
$k\in[\nS]$ and $l\in[\nG]$, 
such that 
$\hatK_{\Wcal} = \sum_{k=1}^{\nS} \sum_{l=1}^{\nG} a_{kl}  (\Pi_{\Wcal}v_k) \otimes g_l$, i.e., we have \eqref{eqn:hatK_PiW}.
Considering $\hatK_{\Wcal}$ in this parametric form and 
due to the linearity of inner product, for each    
$i\in[\nS]$ and $j\in[\nG]$, 
it follows that 
\begin{equation}
	\begin{split}
		(\hatK   g_j)(\vcx_{i-1})
		&=
		\sum_{k=1}^{\nS} \sum_{l=1}^{\nG}a_{kl}\ \! \inner{v_i}{\big((\Pi_{\Wcal}v_k) \otimes g_l\big)g_j}
		\\&=
		\sum_{k=1}^{\nS} \sum_{l=1}^{\nG}a_{kl}\ \! \inner{v_i}{\Pi_{\Wcal}v_k \inner{g_l}{g_j}}
		\\&= 
		\sum_{k=1}^{\nS} \sum_{l=1}^{\nG} \inner{\Pi_{\Wcal}v_i}{\Pi_{\Wcal}v_k }\ \! a_{kl}\ \! \inner{g_l}{g_j},
	\end{split}
\end{equation}
where the last equality is due to the fact that $ \inner{u}{\Pi_{\Wcal}v }= \inner{\Pi_{\Wcal}u}{\Pi_{\Wcal}v }$, for any $v,u\in\Hcal$.
Accordingly, we have
\begin{equation}
	\left[(\hatK   g_j)(\vcx_{i-1})\right]_{i=1,j=1}^{\nS,\!\ \nG} = \mxW_{\Vcal}\mxA\mxG. 
\end{equation}
Then, following same steps of calculations as in the proof of Theorem \ref{thm:Tikhonov_reg_case}, one can show that $\mxA$ can be obtained by solving convex program \eqref{eqn:opt_finite_Tikhonov_reg_case_PiW}.
This concludes the proof. 
\qed
\subsection{Proof of Theorem \ref{thm:Tr_reg_case}}
\label{sec:appendix_proof_thm_12}
For program \eqref{eqn:reg_learning_koopman_nuc}, $\mxK=0$ is a feasible solution. Therefore, for the optimal solution of \eqref{eqn:reg_learning_koopman_nuc}, the following inequality holds
\begin{equation}\label{eqn:proof_thm_Tr_reg_case_01}
	\lambda\|\hatK\|_*\le 
	\Ecal(\hatK) + \lambda \|\hatK\|_* \le 
	\Ecal(0) + \lambda \|0\|_* = \|\mxY\|_{\fro}^2.
\end{equation} 
By virtue of \eqref{eqn:proof_thm_Tr_reg_case_01}, we define $\Ccal$ as 
\begin{equation}\label{eqn:Ccal_nuc_lambda}
	\Ccal:=\Big\{\mxS\in\Lcal(\Hcal)\ \!\Big|\ \!\|\mxS\|_*\le \frac{1}{\lambda}\|\mxY\|_{\fro}\Big\},
\end{equation}
and, introduce the following constrained problem
\begin{equation}\label{eqn:reg_learning_koopman_nuc_C}
	\begin{array}{cl}
		\minOp_{\mxK\in\Fcal} & \Ecal(\mxK)+ \lambda \|\mxK\|_*,\\
		\mathrm{s.t.} &
		\mxK\in\Ccal,	
	\end{array}	
\end{equation}
which is equivalent to \eqref{eqn:reg_learning_koopman_nuc}. 
Let $\mxS\in\Lcal(\Hcal)$, and $\{b_k\}_{k=1}^\infty$ be an orthonormal basis for $\Hcal$ such that $\Gcal = \linspan\{b_1,\ldots,b_{\barnG}\}$.
Accordingly, for compact operator $\mxC$ with $\|\mx{C}\|\le 1$, we have
\begin{equation}
\begin{split}
	\big|\trace(\mx{C}\Pi_{\Zcal}\!\ \mxS\!\ \Pi_{\Gcal})\big|
	&= 
	\bigg|\sum_{k=1}^\infty \inner{\mx{C}\Pi_{\Zcal}\!\ \mxS\!\ \Pi_{\Gcal}\!\ b_k}{b_k}\bigg|
	\\&=
	\bigg|\sum_{k=1}^{\barnG} \inner{\mx{C}\Pi_{\Zcal}\!\ \mxS\!\  b_k}{b_k}\bigg|.
\end{split}
\end{equation}
Note that $\mx{C}\Pi_{\Zcal}$ is a compact operator and $\|\mx{C}\Pi_{\Zcal}\|\le 1$.
Therefore, from \eqref{eqn:def_trace_norm_infC}, it follows that
\begin{equation}
	\begin{split}
		\|\Pi_{\Zcal}\!\ \mxS\!\ \Pi_{\Gcal}\|_* 
		&\le 
		\sup_{\|\mx{C}\|\le 1, \mx{C}\in\Kcal(\Hcal)}
		\bigg|\sum_{k=1}^{\barnG} \inner{\mx{C} \mxS\!\  b_k}{b_k}\bigg|
		\\&=
		\sup_{\|\mx{C}\|\le 1, \mx{C}\in\Kcal(\Hcal)}
		\bigg|\sum_{k=1}^{\infty} \inner{\Pi_{\Gcal}\mx{C} \mxS\!\  b_k}{b_k}\bigg|.
	\end{split}
\end{equation}
Since $\Pi_{\Gcal}\mx{C}$ is a compact operator and $\|\Pi_{\Gcal}\mx{C}\|\le 1$,
from \eqref{eqn:def_trace_norm_infC}, we have
\begin{equation}\begin{split}
		\|\Pi_{\Zcal}\!\  \mxS\!\ \Pi_{\Gcal}\|_* 
		&\le
		\sup_{\|\mx{C}\|\le 1, \mx{C}\in\Kcal(\Hcal)}
		\bigg|\sum_{k=1}^{\infty} \inner{\mx{C} \mxS\!\  b_k}{b_k}\bigg| 
		\\&=
		\sup_{\|\mx{C}\|\le 1, \mx{C}\in\Kcal(\Hcal)}
		\big|\trace(\mx{C}\mxS)\big| = \|\mxS\|_*. 
\end{split}\end{equation}
Therefore, for any $\mxS\in\Lcal(\Hcal)$, one can see that 
$\Rcal(\Pi_{\Zcal}\mxS\Pi_{\Gcal})=\|\Pi_{\Zcal}\mxS\Pi_{\Gcal}\|_*\le\|\mxS\|_*=\Rcal(\mxS)$.
Moreover, for $\mxS\in\Ccal$, we have
\begin{equation}
	\|\Pi_{\Zcal}\mxS\Pi_{\Gcal}\|_*\le\|\mxS\|_*
	\le
	\frac{1}{\lambda}\|\mxY\|_{\fro},
\end{equation}
and, subsequently, it follows that $\Pi_{\Zcal}\mxS\Pi_{\Gcal}\in \Ccal$.
Therefore, $\delta_{\Ccal}(\Pi_{\Zcal}\mxS\Pi_{\Gcal})\le\delta_{\Ccal}(\mxS)$, and hence, Assumption \ref{ass:Rbar} holds for $\barRcal=\lambda\Rcal+\delta_{\Ccal}$.
From the definition of $\Ccal$ and $\Rcal$, it follows that  $\dom(\Rcal)\cap\Ccal=\Ccal$,
and subsequently, we have $\Dcal:=\Fcal\cap\dom(\Rcal)\cap\Ccal=\Fcal\cap\Ccal$.
One can easily see that $0\in\Dcal$.
Since $\|\cdot\|_*$ is a norm on $\dom(\Rcal)$ and $\Ccal\subset\dom(\Rcal)$,  we know that $\Ccal$ is a convex set, and also, $\Rcal$ is a convex function.
For $\mxS\in\Lcal(\Hcal)$, we have the inequality $\|\mxS\|\le\|\mxS\|_*$, which implies the coercivity of $\Rcal$. Hence, $\barRcal=\Rcal+\delta_{\Ccal}$ is a coercive function.
Let $\{\mxS_n\}_{n\in\Nbb}\subset\Lcal(\Hcal)$ be a sequence such that $\lim_{n\to\infty}\mxS_n=\mxS\in\Lcal(\Hcal)$ in norm topology.
Subsequently, we know that  $\lim_{n\to\infty}|\mxS_n|=|\mxS|$ in norm topology.
Hence, for any $k\in\Nbb$, we have that $\lim_{n\to\infty}\inner{b_k}{|\mxS_n|b_k}=\inner{b_k}{|\mxS|b_k}$.
Moreover, since $\{|\mxS_n|\}_{n\in\Nbb}$ and $\mxS$ are non-negative operators, we know that $\inner{b_k}{|\mxS_n|b_k}\ge 0$ and $\inner{b_k}{|\mxS|b_k}\ge 0$, for any $n,k\in\Nbb$.
Hence, from Fatou's lemma \cite{folland1999real}, it follows that
\begin{equation}\label{eqn:fatuo_lemma_nuc}
\begin{split}
\|\mxS\|_*&=\sum_{k=1}^\infty\inner{b_k}{|\mxS|b_k} =\sum_{k=1}^\infty \liminf_{n}\inner{b_k}{|\mxS_n|b_k} 
\\&\le 
\liminf_{n}\sum_{k=1}^\infty \inner{b_k}{|\mxS_n|b_k} 
=
\liminf_{n}\|\mxS_n\|_*. 
\end{split}
\end{equation} 
Accordingly, we know that $\Rcal(\mxS)\le\liminf_{n}\Rcal(\mxS_n)$, and $\Rcal$ is lower semicontinuous.
Also, if  $\{\mxS_n\}_{n\in\Nbb}\subset\Ccal$, then $\|\mxS_n\|_*\le \frac{1}{\lambda}\|\mxY\|_{\fro}$, for each $n$. Therefore, due to \eqref{eqn:fatuo_lemma_nuc}, we know that $\|\mxS\|_*\le \frac{1}{\lambda}\|\mxY\|_{\fro}$, and, subsequently, we have $\mxS\in\Ccal$.
Hence, $\Ccal$ and $\Dcal=\Fcal\cap\Ccal$ are non-empty, closed and convex sets.
Accordingly, Theorem \ref{thm:gen_reg_case} implies that \eqref{eqn:reg_learning_koopman_nuc_C} admits solution $\hatK$ with the parametric form in \eqref{eqn:hatK_Z}, which is also a solution for \eqref{eqn:reg_learning_koopman_nuc} due to the equivalency of the corresponding programs.  
Let $\{b_n\}_{n\in\Nbb}$ be an orthonormal basis such that $\Gcal=\linspan\{b_1,\ldots,b_{\barnG}\}$.
Accordingly, for each
$l\in[\nG]$, 
there exist $e_{l1},\ldots, e_{l\barnG}$ such that we have $g_l=\sum_{j=1}^{\barnG} e_{lj}b_j$.
From \eqref{eqn:T*T}, it follows that
\begin{equation}\label{eqn:T*T_E}\!\!\!
	\begin{split}	
		\hatK^*\hatK
		&=
		\sum_{j=1}^{\nG}\sum_{l=1}^{\nG}
		\big(\sum_{i=1}^{\barnG} e_{li}b_i\big)\! \otimes\! \big(\sum_{k=1}^{\barnG} e_{jk}b_k\big)
		[\mxA^\tr\mxZ\mxA]_{(l,j)}
		\\&=
		\sum_{i=1}^{\barnG}\sum_{k=1}^{\barnG} 
		\Big( 
		\sum_{j=1}^{\nG}\sum_{l=1}^{\nG} 
		e_{li} [\mxA^\tr\mxZ\mxA]_{(l,j)} e_{jk}
		\Big)		
		(b_i\otimes b_k) 
		\\&=
		\sum_{i=1}^{\barnG}\sum_{k=1}^{\barnG} 
		[\mxE^\tr\mxA^\tr\mxZ\mxA\mxE]_{(i,k)}
		(b_i\otimes b_k), 
	\end{split}
\end{equation}
where $\mxE$ is the matrix defined as $\mxE=[e_{lj}]_{l=1,j=1}^{\nG,\barnG}$.
One can easily see that $\mxG=\mxE\mxE^\tr$.
Due to \eqref{eqn:T*T_E}, we know that there exist $r_{ik}$, for
$i,k\in [\barnG]$, 
such that $|\hatK^*\hatK| = \sum_{i=1}^{\barnG}\sum_{k=1}^{\barnG}r_{ik}(b_i\otimes b_k)$.
From \eqref{eqn:h1xh2_h3xh4}, we have 
\begin{equation}\label{eqn:|T*T|^2_E}\!\!\!
	\begin{split}	
		|\hatK^*\hatK|^2
		&=
		\sum_{i=1}^{\barnG}\sum_{j=1}^{\barnG}r_{ij}(b_i\otimes b_j)
		\sum_{l=1}^{\barnG}\sum_{k=1}^{\barnG}r_{lk}(b_l\otimes b_k)
		\\&=
		\sum_{i=1}^{\barnG}\sum_{k=1}^{\barnG}
		\sum_{j=1}^{\barnG}\sum_{l=1}^{\barnG}
		r_{ij}r_{lk}\inner{b_j}{b_l} (b_i\otimes b_k)
		\\&=
		\sum_{i=1}^{\barnG}\sum_{k=1}^{\barnG}
		[\mxR^2]_{(i,k)}(b_i\otimes b_k), 
	\end{split}
\end{equation}
where $\mxR$ is the matrix defined as $\mxR=[r_{ik}]_{i=1,k=1}^{\barnG,\barnG}$.
From \eqref{eqn:T*T_E} and \eqref{eqn:|T*T|^2_E}, it follows that 
\begin{equation}\label{eqn:R2}
\mxR^2=\mxE^\tr\mxA^\tr\mxZ\mxA\mxE=(\mxZ^{\frac12}\mxA\mxE)^\tr(\mxZ^{\frac12}\mxA\mxE).
\end{equation}
Note that we have
\begin{equation}\label{eqn:inner_h1_h2xh3_h1}
	\inner{h_1}{(h_2\otimes h_3)h_1 } = \inner{h_1}{h_2} \inner{h_1}{h_3},
\end{equation}
for any $h_1,h_2,h_3,h_4\in\Hcal$.
Therefore, we have 
\begin{equation}\label{eqn:trace_R}
	\begin{split}	
		\trace(|\hatK^*\hatK|) 
		&=
		\sum_{j=1}^{\infty}\inner{b_j}{\sum_{i=1}^{\barnG}\sum_{k=1}^{\barnG}r_{ik}(b_i\otimes b_k)b_j }
		\\&=
		\sum_{j=1}^{\infty}\sum_{i=1}^{\barnG}\sum_{k=1}^{\barnG}r_{ik}
		\inner{b_j}{b_i }\inner{b_j}{b_k } 
		\\&= \sum_{i=1}^{\barnG}r_{ii}=\trace(\mxR).
	\end{split}
\end{equation}
Due to \eqref{eqn:trace_R} and \eqref{eqn:R2}, we know that 
$\|\hatK\|_* = \|\mxZ^{\frac12}\mxA\mxE\|_*$.
Note that, for any matrix $\mxM$, we have
$\|\mxM\|_* = \|\mxM^\tr\|_* = \trace\big((\mxM^\tr\mxM)^{\frac12}\big) = \trace\big((\mxM\mxM^\tr)^{\frac12}\big)$. 
From this fact and $\mxG^{\frac12}\mxG^{\frac12}=\mxG=\mxE\mxE^\tr$, it follows that
\begin{equation}
\begin{split}
\|\hatK\|_* &
= \trace\big((\mxZ^{\frac12}\mxA\mxE)(\mxE^\tr\mxA^\tr\mxZ^{\frac12})\big)
\\&=\trace\big((\mxZ^{\frac12}\mxA\mxG^{\frac12})(\mxG^{\frac12}\mxA^\tr\mxZ^{\frac12})\big)
\\&=\|\mxG^{\frac12}\mxA^\tr\mxZ^{\frac12}\|_*=\|\mxZ^{\frac12}\mxA\mxG^{\frac12}\|_*.
\end{split}
\end{equation}
Also, one can show that $\Ecal(\hatK)=\|\mxZ\mxA\mxG-\mxY\|_{\fro}^2$ using arguments similar to the proof of Theorem \ref{thm:Tikhonov_reg_case}. Substituting these terms in \eqref{eqn:reg_learning_koopman_nuc}, we obtain the optimization problem \eqref{eqn:rank_const_finite}. This concludes the proof.
\qed
\subsection{Supporting Lemmas}\label{sec:appendix_lemmas}
\begin{lemma}\label{lem:fro_norm_ineq}
	For matrices $\mxA$ and $\mxB$, we have $\|\mxA\mxB\|_{\fro}\le \|\mxA\|\|\mxB\|_{\fro}$. Moreover, if $\mxB$ is invertible, then  $\|\mxA\mxB\|_{\fro}\le \|\mxA\|_{\fro}\|\mxB\|$.
\end{lemma}

\begin{lemma}\label{lem:Fro_norm_T_strongly_convex}
The function $f:\Lcal(\Hcal)\to \Rbb_+$, defined as $f(\mxT)=\|\mxT\|_{\mathrm{F}}^2$, is strictly convex.
\end{lemma}	
\begin{proof}
Let $\{b_k\}_{k=1}^\infty$ be an orthonormal basis for $\Hcal$. Also, let $t\in(0,1)$ and $\mxT_1, \mxT_2\in\Lcal(\Hcal)$ such that $\mxT_1\ne \mxT_2$. 
Then, for each   $k$, we have 
$ 
\|\mxT_1 b_k\|^2 + \|\mxT_2 b_k\|^2 \ge 2\inner{\mxT_1b_k}{\mxT_2b_k}.
$ 
The equality holds if and only if ${\mxT_1b_k=\mxT_2b_k}$.
Since $\mxT_1\ne \mxT_2$, there exists $k$ such that this inequality is strict.
Now, multiplying both sides with $t(1-t)$ and replacing the terms, we have
\begin{equation*}
\begin{split}
&t\|\mxT_1 b_k\|^2 + (1-t)\|\mxT_2 b_k\|^2 
\\&\ \ \ \ge 
t^2\|\mxT_1 b_k\|^2 + (1-t)^2\|\mxT_2 b_k\|^2 + 2t(1-t)\inner{\mxT_1b_k}{\mxT_2b_k}
\\&\ \ \ =
\|(t\mxT_1+(1-t)\mxT_2) b_k\|^2.
\end{split}
\end{equation*}
By taking summation and due the definition of Frobenius norm, we have
\begin{equation}
t\|\mxT_1\|_{\fro}^2+(1-t)\|\mxT_2\|_{\fro}^2>
\|t\mxT_1+(1-t)\mxT_2\|_{\fro}^2.
\end{equation}
Accordingly, we have $tf(\mxT_1)+(1-t)f(\mxT_2)>f(t\mxT_1+(1-t)\mxT_2)$, which concludes the proof.
\end{proof}	

\begin{lemma}\label{lem:rank_ineq_fro}
	For matrices $\mxA\in\Rbb^{n\times m}$, $\mxB\in\Rbb^{m\times k}$ and $\mxC\in\Rbb^{k\times p}$, we have 
	\begin{equation}\label{eqn:rank_ineq_fro}
	\rank(\mxA\mxB)+\rank(\mxB\mxC)\le \rank(\mxB)+\rank(\mxA\mxB\mxC).
	\end{equation}
\end{lemma}

\begin{lemma}\label{lem:rank_ABBT}
	For matrices $\mxA\in\Rbb^{n\times m}$, $\mxB\in\Rbb^{m\times k}$ and $\mxC\in\Rbb^{k\times p}$, we have 
	$\rank(\mxA\mxB)=\rank(\mxA^\tr\mxA\mxB)$
	and
	$\rank(\mxB\mxC)=\rank(\mxB\mxC\mxC^{\tr})$.
\end{lemma}
\begin{proof}
	We know that $\rank(\mxA^\tr\mxA\mxB)\le \rank(\mxA\mxB)$. Therefore, due to \eqref{eqn:rank_ineq_fro}, we have
	\begin{equation}\label{eqn:rank_ABBT_proof_eq_01}
	\begin{split}
	\rank(\mxA^\tr\mxA)+\rank(\mxA\mxB)
	&\le 
	\rank(\mxA) + \rank(\mxA^\tr\mxA\mxB)
	\\&=
	\rank(\mxA) + \rank(\mxA\mxB).
	\end{split}
	\end{equation}
	From  $\rank(\mxA^\tr\mxA)=\rank(\mxA)$ and \eqref{eqn:rank_ABBT_proof_eq_01}, it follows that $\rank(\mxA\mxB)=\rank(\mxA^\tr\mxA\mxB)$.
	The other equality can be shown based on similar steps.
\end{proof}